\documentclass[fleqn,usenatbib]{mnras}

\usepackage{newtxtext,newtxmath}

\usepackage[T1]{fontenc}

\DeclareRobustCommand{\VAN}[3]{#2}
\let\VANthebibliography\thebibliography
\def\thebibliography{\DeclareRobustCommand{\VAN}[3]{##3}\VANthebibliography}

\usepackage{graphicx} 
\DeclareGraphicsExtensions{.jpg,.jpg,.pdf}
\usepackage{amsmath}	
\usepackage{multirow}
\usepackage{caption}
\usepackage{float}
\usepackage{pdflscape}
\usepackage{gensymb}
\usepackage{fix-cm}

\title[Small Planet Compositions]{Constraining Small Planet Compositions for Future Missions}

\author[Larissa Palethorpe et al.]{
Larissa Palethorpe$^{1,2,3}$\thanks{E-mail: larissa.palethorpe@bristol.ac.uk},
Annelies Mortier$^{4}$,
Jo Ann Egger$^{5,6,7}$,
Ken Rice$^{1,2}$,
Thomas G. Wilson$^{8}$,\newauthor
Andrew Vanderburg$^{9}$,
Aldo S. Bonomo$^{10}$,
Walter Boschin$^{11}$,
Andrew Collier Cameron$^{12}$,\newauthor
Yoshi Nike Emilia Eschen$^{8}$,
Avet Harutyunyan$^{11}$,
Luca Malavolta$^{13,14}$,
Aldo F. Martínez Fiorenzano$^{11}$,\newauthor
Alessandro Sozzetti$^{10}$,
Manu Stalport$^{15,16}$,
Vincent Van Eylen$^{17}$,
Christopher Allan Watson$^{18}$
\\
$^{1}$Institute for Astronomy, University of Edinburgh, Royal Observatory, Blackford Hill, Edinburgh, EH9 3HJ, UK\\
$^{2}$Centre for Exoplanet Science, University of Edinburgh, Edinburgh, EH9 3HJ, UK\\
$^{3}$HH Wills Physics Laboratory, University of Bristol, Tyndall Avenue, Bristol, BS8 1TL, UK\\
$^{4}$School of Physics \& Astronomy, University of Birmingham, Edgbaston, Birmingham, B15 2TT, UK\\
$^{5}$Weltraumforschung und Planetologie, Physikalisches Institut, University of Bern, Gesellschaftsstrasse 6, 3012 Bern, Switzerland\\
$^{6}$European Space Agency (ESA), European Space Research and Technology Centre (ESTEC), Keplerlaan 1, 2201 AZ Noordwĳk, The Netherlands\\
$^{7}$ESA research fellow\\
$^{8}$Department of Physics, University of Warwick, Gibbet Hill Road, Coventry CV4 7AL, UK\\
$^{9}$Center for Astrophysics, Harvard \& Smithsonian, 60 Garden Street, Cambridge, MA 02138, USA\\
$^{10}$INAF - Osservatorio Astrofisico di Torino, via Osservatorio 20, 10025 Pino Torinese, Italy\\
$^{11}$Fundaci\'on G. Galilei - INAF (Telescopio Nazionale Galileo), Rambla José Ana Fernandez Pérez 7, E-38712 Breña Baja, Tenerife, Spain\\
$^{12}$Centre for Exoplanet Science, SUPA School of Physics and Astronomy, University of St Andrews, North Haugh, St Andrews KY16 9SS, UK\\
$^{13}$Dipartimento di Fisica e Astronomia ``Galileo Galilei'', Università degli Studi di Padova, Vicolo dell'Osservatorio 3, 35122, Padova, Italy\\
$^{14}$INAF, Osservatorio Astronomico di Padova, Vicolo dell'Osservatorio 5, 35122, Padova, Italy\\
$^{15}$Space Sciences, Technologies and Astrophysics Research (STAR) Institute, Université de Liège, Allée du 6 Août 19C, 4000 Liège, Belgium\\
$^{16}$Astrobiology Research Unit, Université de Liège, Allée du 6 Août 19C, B-4000 Liège, Belgium\\
$^{17}$Mullard Space Science Laboratory, University College London, Holmbury St Mary, Dorking, Surrey, RH5 6NT, UK\\
$^{18}$Astrophysics Research Centre, School of Mathematics and Physics, Queen’s University Belfast, Belfast, BT7 1NN, UK\\
}

\date{Accepted 2026 March 06. Received ---; in original form ---}

\pubyear{2026}

\begin{document}
\label{firstpage}
\pagerange{\pageref{firstpage}--\pageref{lastpage}}
\maketitle

\begin{abstract}
Accurate mass and radius measurements of small transiting exoplanets are essential for probing their compositions, formation histories, and potential habitability. We present a uniform analysis of six planetary systems (each hosting at least one small transiting planet): K2-79, K2-106, K2-111, K2-222, K2-263, and TOI-1634. Our study combines new \textit{CHEOPS} transit observations with archival photometry from \textit{K2}, \textit{TESS}, and ground-based facilities, alongside new and archival radial velocity data from HARPS-N, HIRES, ESPRESSO, and others. For each system, we perform joint transit and RV modelling, achieving typical precisions better than 15\% and 5\% for mass and radius, respectively, and thus enabling precise bulk density determinations. These reveal a range of compositions, including rocky planets near the radius valley (e.g. K2-106\,b, TOI-1634\,b), intermediate-density planets requiring steam-rich or mixed volatile envelopes (e.g. K2-111\,b, K2-263\,b), and low-density regimes, consistent with gas dwarfs or water-worlds (e.g. K2-79\,b, K2-222\,b). Several systems show evidence of additional companions detectable via RVs but not seen in transit. The results highlight the value of coordinated \textit{CHEOPS} and HARPS-N observations in delivering some of the most precise bulk densities for small planets to date and support the preparation for future atmospheric characterisation missions.
\end{abstract}




\section{Introduction}

The detailed characterisation of small exoplanets is critical to understanding their formation, evolution, and potential habitability. While transit and RV surveys have revealed thousands of planets across a wide range of masses, radii, and orbital separations, the number of well-characterised small exoplanets ($1 < R_{p}< 4\,R_{\oplus}$) --- those with both masses and radii measured to better than 20\% and 5\%, respectively --- remains limited to fewer than 350 systems \citep{Otegi2020, ExoplanetArchive}. These planets exhibit remarkable diversity in composition and atmospheric properties \citep{Madhusudhan2019}, and their internal structures contain key information about formation pathways, accretion histories, and atmospheric evolution \citep{Mordasini2012, Dorn2015, Bean2021}; thus, characterising their internal structures and evolutionary histories remains a central goal for the exoplanet field.

A key step towards this goal lies in accurately determining these planets' bulk densities. This precision is particularly important in the context of upcoming and future missions. The European Space Agency's (ESA) \textit{PLAnetary Transits and Oscillations of stars} \citep[\textit{PLATO};][]{Mas2023} mission, scheduled for launch in 2027, is specifically designed to improve the accuracy of planetary radii, masses, and ages through long-baseline photometry combined with asteroseismology, \textit{Gaia} astrometry, and high-resolution spectroscopy, targeting precisions of $\sim5\%$ in radius and $\sim10\%$ in mass and age. Similarly the \textit{Earth2.0} space mission \citep[\textit{ET};][]{Ge2024} aims to detect and characterise Earth-sized planets around Sun-like stars, further expanding the sample of well-characterised small exoplanets by the end of the 2030s. Looking further ahead the \textit{Habitable Worlds Observatory} \citep[\textit{HWO};][]{HWO}, currently in the concept phase with a potential launch in the 2040s, will rely heavily on prior knowledge of planetary parameters. \textit{HWO} aims to characterise planets around nearby Sun-like stars through both direct imaging and spectroscopic observations, including reflected-light spectroscopy of directly imaged planets and, for transiting systems, transmission spectroscopy. Accurate knowledge of a planet's bulk density, derived from precise measurements of its mass and radius, will be critical for \textit{HWO}. Such measurements allow scientists to distinguish between rocky, water-rich, and volatile-dominated worlds before direct observation and are essential for interpreting atmospheric spectra in the context of interior composition \citep{Batalha2019, Salvador2024, Damiano2025}. This information also improves target selection and assessment of planetary habitability, when considered alongside additional factors such as orbital separation, stellar age, magnetic protection, and the irradiation environment of the host star \citep{Valente2024, Atkinson2024, Zhu2025, See2025}. In addition, accurate ephemerides, such as the transit centre time, are crucial to ensure that transits are successfully observed and valuable telescope time is not wasted. The exoplanet community's push towards detecting and characterising potentially habitable worlds thus hinges on refining the planetary parameters of validated small exoplanets.

The HARPS-N Collaboration has played a leading role in the characterisation of small transiting exoplanets, contributing more than a third of the current sample with well constrained masses and radii \citep{Pepe2013, Mortier2018, Mayo2019, Rice2019, Lacedelli2021, Rajpaul2021, Bonomo2023}. To expand this effort, we secured \textit{CHaracterising ExOPlanet Satellite} \citep[\textit{CHEOPS};][]{Benz2021} observations targeting six such systems, each originally identified by the \textit{K2} \citep{Howell2014} or the \textit{Transiting Exoplanet Survey Satellite} \citep[\textit{TESS};][]{Ricker2014} missions and actively monitored with HARPS-N under its GTO program. For each target, we obtained two transits with \textit{CHEOPS} to achieve two complementary goals: (1) refining planetary radius measurements, and (2) improving orbital ephemerides to tighten priors on RV modelling. This strategy enhances the precision of mass determinations by reducing phase uncertainties and improving transit timing predictions. By combining \textit{CHEOPS} and HARPS-N data with a range of archival photometry and spectroscopy, we present some of the most accurately constrained bulk densities currently available for small exoplanets, providing key insights into their interior structure and atmospheric evolution.

In this work, we present a comprehensive re-analysis of six planetary systems: K2-79, K2-106, K2-111, K2-222, K2-263, and TOI-1634. Each system hosts at least one small transiting planet with existing photometric and spectroscopic observations, and may exhibit compelling features such as planet multiplicity, low densities, and proximity to the radius valley \citep{owen_kepler_2013, Lopez_2013, Fulton_2017}. Table \ref{tab:stellar_params} summarises the stellar and planetary parameters for these targets with the values drawn from previous literature. We compile and uniformly model transit photometry from \textit{K2}, \textit{TESS}, \textit{CHEOPS}, and a range of ground-based instruments, alongside RV datasets from HARPS-N, HIRES, ESPRESSO, HARPS, FIES, PFS, IRD, and HDS, incorporating these literature stellar parameter values as priors in our analysis. The typical precision of these stellar parameters (2--5\% in mass; 1--4\% in radius) sets the baseline accuracy for the derived planetary masses, radii, and bulk densities, as these quantities scale directly with the host-star properties. Recent studies have demonstrated that modelling stellar spectral energy distributions can further improve stellar radii and masses, enabling planetary radii to be determined to $\sim2\%$ and masses to $\sim10\%$ precision \citep{Morell2026}. While we adopt literature stellar parameters in this work, continued improvements in stellar characteriation will directly translate into tighter constraints on planetary compositions. Our modelling approach incorporates both single- and multi-planet Keplerian fits, Gaussian Process (GP) noise modelling where appropriate, and a consistent treatment of systematics and uncertainties across instruments. This enables a robust derivation of planetary parameters and direct comparisons between systems.

The six systems studied here span a wide range of planetary properties and evolutionary pathways. Several planets, such as K2-106\,b and TOI-1634\,b, lie near the boundary between rocky super-Earths and volatile-rich sub-Neptunes, probing the effects of extreme irradiation and atmospheric loss \citep{Ginzburg_2018, Gupta_2020}. Others, like K2-79\,b and K2-222\,b, fall into the low-density regime where compositions are degenerate, highlighting the H/He vs water-rich envelopes \citep{Rogers2010, Zeng2021}. K2-263\,b, by contrast, likely formed beyond the snow line and now orbits at a cool enough distance to offer insights into the formation and inward migration of cold sub-Neptunes \citep{Bitsch2021}. Meanwhile, K2-111\,b orbits an $\alpha$-enhanced, iron-poor star, raising questions about the link between stellar composition and planet interiors \citep{Santos2017}. Together, these systems provide a valuable testbed for planet formation and atmospheric evolution models, and offer a diverse set of targets for future atmospheric characterisation with upcoming facilities such as \textit{HWO}.

The paper is structured as follows: Section \ref{sec:obs} describes the photometric and spectroscopic datasets used in our analysis. Section \ref{sec:modelling} outlines our modelling approach for each system, including transit and RV fitting methods. Section \ref{sec:int_struc} presents our results for each system, with updated interior structure modelling. In Section \ref{sec:HWO} we discuss implications for future study with upcoming missions. We conclude in Section \ref{sec:conclusion} with a summary and outlook for future work.

\begin{table*}
    \centering
    \caption{Stellar and planetary parameters from previous literature. Sources are listed at the bottom of the table. For TOI-1634\,b, planetary parameters from two sources are given due to significant differences in reported masses. In this case, * \,indicates \citet{Cloutier2021} whilst \dag \, indicates \citet{Hirano2021}. $\ddagger$ indicates parameters not reported in the literature. No planet c is known for K2-79 or K2-263. For K2-222\,c, the orbital period is reported, but no mass estimate or RV amplitude has been published. For TOI-1634, no confirmed planet c has been detected: \citet{Cloutier2021} and \citet{Hirano2021} both note possible longer-period RV trends, but neither reports a statistically significant or characterised second planet. `NT' indicates a non-transiting planet.}
    \resizebox{\linewidth}{!}{
    \begin{tabular}{lcccccc}
         \hline\hline 
         Parameter & \textbf{K2-79} & \textbf{K2-106} & \textbf{K2-111} & \textbf{K2-222} & \textbf{K2-263} & \textbf{TOI-1634} \\ 
         \hline
         \multicolumn{4}{l}{\textit{Astrometry}} \\
         \vspace{1.5mm}
         R.A. & 03h41m01.42s & 00h52m19.21s & 03h59m33.67s & 01h05m51.00s & 08h38m43.71s &  03h45m33.75s \\
         \vspace{1.5mm}
         Decl. & +13d31m09.07s & +10d47m40.94s & +21d17m54.69s & +11d45m13.38s & +15d40m50.17s &  +37d06m44.21s \\
         \vspace{1.5mm}
         $\pi$ [mas] & $3.841 \pm 0.015$ & $4.0603 \pm 0.0499$ & $4.9626 \pm 0.0674 $ & $10.0090 \pm 0.0608$ & $6.1262 \pm 0.0514$ & $28.512\pm 0.018$ \\
         \vspace{1.5mm}
         d [pc] & $256.011^{+2.349}_{-2.307}$ & $244.590^{+3.044}_{-2.972}$ & $200.394^{+2.763}_{-2.690}$ & $101.321^{+0.632}_{-0.625}$ & $162.481^{+1.380}_{-1.357}$ &  $35.2736^{+0.0527}_{-0.0526}$ \\
         \vspace{1.5mm}
         Spectral Type & G1  & G5 V & G2 & G0 & G9 V & M2 V \\
         \hline
         \multicolumn{4}{l}{\textit{Photometry}} \\
         \vspace{1.5mm}
         $B$ [mag] & $12.89 \pm 0.02 $  & $12.576 \pm 0.443$ & $11.80 \pm 0.03 $ & $10.04 \pm 0.03 $ & $12.35 \pm 0.03$ & $14.755\pm 0.060$\\
         \vspace{1.5mm}
         $V$ [mag] & $12.07 \pm 0.06$  & $12.101\pm 0.031$ & $11.14 \pm 0.04 $ & $9.54 \pm 0.03$ & $11.61 \pm 0.04$ & $13.24 \pm 0.04$ \\
         \vspace{1.5mm}
         $J$ [mag] & $10.36 \pm 0.02 $  & $10.770 \pm 0.023$ & $9.77 \pm 0.02$ & $8.42 \pm 0.02 $ & $10.22 \pm 0.02$ & $9.564\pm 0.021$ \\
         \vspace{1.5mm}
         $H$ [mag] & $10.00 \pm 0.02 $  & $10.454 \pm 0.026$ & $9.48 \pm 0.03 $ & $8.17 \pm 0.03$ & $9.81 \pm 0.02$ & $8.940\pm 0.021$ \\
         \vspace{1.5mm}
         $K$ [mag] & $9.91 \pm 0.02 $  &  $10.344 \pm 0.021$ & $9.38 \pm 0.02 $ & $8.11 \pm 0.02$ & $9.75 \pm 0.02$ & $8.699 \pm 0.014$ \\
         \hline
         \multicolumn{4}{l}{\textit{Atmospheric parameters}} \\
         \vspace{1.5mm}
         $T_{\rm eff}$ & $5897 \pm 118$ & $5532 \pm 74$ & $5775 \pm 60$ & $5492 \pm 119$ & $5368 \pm 44$ & $3472 \pm 70$ \\
         \vspace{1.5mm}
         [Fe/H] & $0.035 \pm 0.06$ & $0.10 \pm 0.05$ & $-0.46 \pm 0.05$ & $-0.315 \pm 0.06$ & $-0.08 \pm 0.03$ & $0.19 \pm 0.12$ \\
         \vspace{1.5mm}
         [Mg/H] & $0.10 \pm 0.03$ & $0.17 \pm  0.05$ & $-0.14 \pm 0.06$ & $-0.20 \pm 0.05$ & $-0.03 \pm 0.02$ &  $0.38 \pm 0.18$ \\
         \vspace{1.5mm}
         [Si/H] & $0.05 \pm 0.02$ & $0.12 \pm  0.10$ & $-0.27 \pm 0.04$ & $-0.21 \pm 0.02$ & $-0.04 \pm 0.02$ &  $0.77 \pm 0.31$ \\
         \hline
         \multicolumn{4}{l}{\textit{Stellar parameters}} \\
         \vspace{1.5mm}
         $t_*$ [Gyr] & $6.5 \pm 1.3$ & $7 \pm  2$ & $13.5_{-0.9}^{+0.4}$ & $7.1_{-1.7}^{+1.5}$ & $7 \pm 4$ &  $7.50^{+3.80}_{-4.00}$\\
         \vspace{1.5mm}
         $M_*$ [$M_{\odot}$] & $1.06 \pm 0.05$ & $0.94 \pm  0.03$ & $0.84 \pm 0.02$ & $0.94 \pm 0.05$ & $0.88 \pm 0.03$ &  $0.502 \pm 0.014$ \\
         \vspace{1.5mm}
         $R_*$ [$R_{\odot}$] & $1.269 \pm 0.051$ & $0.993 \pm  0.008$ & $1.25 \pm 0.02$ & $1.072 \pm 0.043$ & $0.85 \pm 0.02$ &  $0.450 \pm 0.013$ \\
         \vspace{1.5mm}
         $\rho_*$ [$\rho_{\odot}$] & $0.52 \pm 0.10$ & $1.03 \pm  0.05$ & $0.43 \pm 0.03$ & $0.76 \pm 0.13$ & $1.43 \pm 0.06$ &  $5.50 \pm 0.63$\\
         \vspace{1.5mm}
         $P_{\rm rot, max}$ [days] & $23.8 \pm 4.4$ & $18.6 \pm 2.8$ & $27.4_{-1.2}^{+1.4}$ & $17.5 \pm 2.8$ & $> 21.5 \pm 0.5$ &  $77_{-20}^{+26}$ \\
         \hline
         \multicolumn{4}{l}{\textit{Planetary parameters}} \\
         $T_{0,\rm b}$\,[BJD~2450000] & $7103.228 \pm 0.002$ & $7394.0091 \pm 0.0007$ & $7100.076 \pm 0.002$ & $7399.059 \pm 0.003$ & $8111.127 \pm 0.001$ & $8791.5147 \pm 0.0006$* \\ \vspace{1.5mm} &  &  &  &  &  & $8791.5150 \pm 0.0005$\dag \\
         $P_{\rm b}$\,[days] & $10.9947_{-0.0005}^{+0.0003}$ & $0.571313 \pm 0.000006$ & $5.3518 \pm 0.0004$ & $15.3886 \pm 0.0009$ & $50.81895 \pm 0.00009$ & $0.98934 \pm 0.00002$* \\ \vspace{1.5mm} &  &  &  &  &  & $0.989344 \pm 0.000002$\dag \\
         $M_{\rm b}$\,[$M_{\oplus}$] & $11.8 \pm 3.6$ & $7.80_{-0.70}^{+0.71}$ & $5.29_{-0.77}^{+0.76}$ & $6.0 \pm 1.9$ & $14.8 \pm 3.1$ & $4.91_{-0.70}^{+0.68}$* \\ \vspace{1.5mm} &  &  &  &  &  & $10.14 \pm 0.95 $\dag \\
         $R_{\rm b}$\,[$R_{\oplus}$] & $4.09_{-0.12}^{+0.17}$ & $1.676 \pm 0.037$ & $1.82_{-0.09}^{+0.11}$ & $2.35_{-0.07}^{0.08}$ & $2.41 \pm 0.12$ & $1.790_{-0.081}^{+0.080}$* \\ \vspace{1.5mm} &  &  &  &  &  & $1.749 \pm 0.079 $\dag \\
         $\rho_{\rm b}$\,[$\rho_{\oplus}$] & $0.17_{-0.05}^{+0.06}$ & $1.66_{-0.18}^{+0.19}$ & $0.87_{-0.17}^{+0.21}$ & $0.46_{-0.15}^{+0.16}$ & $1.05_{-0.25}^{+0.29}$ & $0.85_{-0.16}^{+0.18}$* \\ \vspace{1.5mm} &  &  &  &  &  & $1.90 \pm 0.31 $\dag \\
         $T_{\rm eq, b}$ [K] (A$_B$=0) & $1021_{-20}^{+21}$ & $2299_{-36}^{+35}$ & $1309 \pm 19$ & $878_{-15}^{+17}$ & $470 \pm 7$ & $924 \pm 22$*\\ \vspace{2.5mm} &  &  &  &  &  & $920 \pm 24$\dag \\
         \vspace{1.5mm}
         
         $T_{0,\rm c}$\,[BJD~2450000] & - & $7405.732 \pm 0.002$ & NT & NT & - & NT \\
         \vspace{1.5mm}
         $P_{\rm c}$\,[days] & - & $13.3399 \pm 0.0007$  & $15.679 \pm 0.006$ & $147.5 \pm 3.3$ & - & $\ddagger$ \\
         \vspace{1.5mm}
         $M_{\rm c}\,{\rm or}\,M_{\rm c}\sin i$\,[$M_{\oplus}$] & - & $7.32_{-2.38}^{+2.49}$ & $11.3 \pm 1.1$ & $\ddagger$ & - & $\ddagger$ \\
         \vspace{1.5mm}
         $R_{\rm c}$\,[$R_{\oplus}$] & - & $2.80_{-0.08}^{+0.10}$ & NT & NT & - & NT \\
        \vspace{1.5mm}
         $\rho_{\rm c}$\,[$\rho_{\oplus}$] & - & $0.33_{-0.11}^{+0.12}$ & NT & NT & - & NT \\
         \vspace{1.5mm}
         $T_{\rm eq, c}$ [K] (A$_B$=0) & - & $804_{-12}^{+13}$ & $915 \pm 13$ & $446 \pm 14$ & - & $\ddagger$ \\
         \hline
         \textit{Source} & \citet{Nava2022} & \citet{Guenther2024} & \citet{Mortier2020} & \citet{Nava2022} & \citet{Mortier2018} & \citet{Cloutier2021}* \\ & & & & & & \citet{Hirano2021}\dag\\
         \hline
    \end{tabular}
    }
    \label{tab:stellar_params}
\end{table*}

\section{Observations}
\label{sec:obs}

\subsection{Photometry}
\label{sec:photometry}

\subsubsection{\textit{K2}}
\label{sec:K2}

The \textit{K2} mission \citep{Howell2014} succeeded the original \textit{Kepler} mission \citep{Borucki2010}, utilising the same spacecraft, telescope, and photometer. From 2014 to 2018, \textit{K2} conducted a series of observations across twenty different fields positioned along the ecliptic plane. Each observational period, or `Campaign', lasted approximately 80 days and yielded time-series photometry for bright stars within fields of $\sim$\,100 square degrees.

K2-79 and K2-111 were observed during Campaign 4 (2015 February 10~-- April 20, BJD 2457228.82--2457299.68) of NASA's \textit{K2} mission in long cadence mode, with integration times of 29.4 minutes. K2-106 and K2-222 were similarly observed in long cadence mode during Campaign 8 (2016 January 04~-- March 23, BJD 2457393.74--2457470.75). K2-263 was observed on three occasions: during Campaign 5 (2015 April 27~-- July 10, BJD 2457139.63--2457214.43) in long cadence mode only, and during Campaigns 16 (from 2017 December 07~-- 2018 February 25, BJD 2458095.49--2458175.02) and 18 (from 2018 May 12~-- July 02, BJD 2458251.57--248302.38) in both long and short cadence modes, with the short cadence data having integration times of 1 minute.

We obtained the data from the Mikulski Archive for Space Telescopes (MAST). The \textit{K2} spacecraft operated with only two of the four original reaction wheels, causing the spacecraft to drift over time, with intermittent thruster firings to keep desired targets in the telescope's field of view. This led to a degradation of the light curves, as stars drifted across regions of the detector with non-uniform pixel response, combined with additional instrumental effects such as pointing jitter and temporal variations in the telescopes point-spread function. The light curves were extracted according to procedures developed by \citet{Vanderburg2014}, and the effects of the short timescale spacecraft drift were corrected in a simultaneous fit of the transit shape, systematics, and low-frequency variations following \citet{Vanderburg2016}. The errors of each data point were set equal to the standard deviation of the out-of-transit points in the flattened light curve. In the remaining analyses, we use only the flattened light curves.

A summary of the \textit{K2} observational campaigns for each target is provided in Table \ref{tab:k2_tess_obs_summary}, and the associated Guest Observer (GO) programs are listed in Table \ref{tab:k2_tess_obs_summary_ids}.

\begin{table}
\centering
\caption{Summary of \textit{K2} and \textit{TESS} observations.}
\label{tab:k2_tess_obs_summary}
\begin{tabular}{l l l l}
\hline\hline
Target & Mission(s) & Campaigns\,/\,Sectors & Cadence(s) \\
\hline
K2-79 & \textit{K2} / \textit{TESS} & C4 / S42--44, S70--71 & 29.4\,min / 120\,s \\
K2-106 & \textit{K2} / \textit{TESS} & C8 / S42, S70 & 29.4\,min / 20\,s, 120\,s \\
K2-111 & \textit{K2} / \textit{TESS} & C4 / S70, S71 & 29.4\,min / 120\,s \\
K2-222 & \textit{K2} / \textit{TESS} & C8 / S42, S43, S70 & 29.4\,min / 20\,s, 120\,s \\
K2-263 & \textit{K2} / \textit{TESS} & C5, C16, C18 / & 29.4\,min, 1\,min / \\ & & S44--46, S72 & 120\,s \\
TOI-1634 & \textit{TESS} only & Sectors 18, 86 & 120\,s, 20\,s \\
\hline
\end{tabular}
\end{table}

\subsubsection{\textit{TESS}}
\label{sec:TESS}
The \textit{Transiting Exoplanet Survey Satellite} \citep[\textit{TESS};][]{Ricker2014} is a NASA mission launched in 2018 to perform an all-sky survey for transiting exoplanets. Equipped with four wide-field cameras, \textit{TESS} observes large swaths of the sky in $24\degree \times 96\degree$ sectors, each monitored continuously for approximately 27 days. Over its nominal and extended missions, \textit{TESS} has produced high-cadence (2-minute and 20-second) light curves for selected targets, as well as 30-minute, 10-minute, and 200-second full-frame images for broader stellar populations.

TOI-1634 (TIC 201186294) was observed in Sector 18 (2019 November 03~-- 27, BJD 2458790.66--2458813.75) during year 2 of the \textit{TESS} primary mission. The observations were taken with 120\,s cadence with CCD 4 on camera 1. It was re-observed in Sector 86 (2021 November 21~-- December 18, BJD 2460636.27--2460662.83) during year 7 of the second ecliptic plane survey with 20\,s and 120\,s cadence with CCD 3 camera 1. A total of 20 transits were observed in Sector 18 and 17 transits in Sector 86 with several transits missed during the data transfer events near perigee passage in both sectors.

K2-79 (TIC 435339558) was observed in Sectors 42 (2021 August 20~-- September 16, BJD 2459447.69--2459470.08) (CCD 3 camera 4), 43 (2021 September 16~-- October 12, BJD 2459474.17--25459498.56) (CCD 2 camera 3), and 44 (2021 October 12~-- November 06, BJD 2459500.24--2459524.45) (CCD 1 camera 1), during year 4 of the \textit{TESS} first extended mission \citep{Ricker2021}. It was then re-observed in year 6 during its second ecliptic plane survey in Sectors 70 (2023 September 20~-- October 16, BJD 2460209.00--2460233.03) (CCD 2 camera 4) and 71 (2023 October 16~-- November 11, BJD 2460235.56--2460259.41) (CCD 4 camera 2). Observations in all sectors were taken with 120\,s cadence, with a total of 8 transits observed.

K2-106 (TIC 266015990) was observed with both 20\,s and 120\,s cadence in Sector 42 during year 4 of the first extended mission with CCD 1 on camera 3. It was then re-observed in Sector 70 during year 6 of the second ecliptic plane survey with 120\,s cadence with CCD 2 on camera 2. A total of 66 transits of K2-106\,b were observed; however all transits of K2-106\,c fall in the gaps caused by data transfer events.

K2-111 (TIC 14227229) was observed in Sectors 70 (CCD 4 camera 4) and 71 (CCD 3 camera 2) (as part of the \textit{TESS} mission-selected target list) during year 6 of the second ecliptic plane survey with 120\,s cadence. A total of 8 transits were observed, with a further 2 narrowly missed due to data transfer events.

K2-222 (TIC 257774438) was observed in Sectors 42 (CCD 1 camera 3) and 43 (CCD 2 camera 1) with 120\,s cadence during year 4 of the second \textit{TESS} extended mission, and in Sector 70 (CCD 2 camera 2) with both 20\,s and 120\,s cadence, during year 6 of the second ecliptic plane survey. A total of 4 transits were observed, with two of them observed in the shorter cadence.

K2-263 (TIC 21276520) was observed in Sectors 44 (CCD 3 camera 4), 45 (2021 November 06~-- December 02, BJD 2459526.47--2459550.63) (CCD 2 camera 3), 46 (2021 December 02-- 30, BJD 2459553.02--2459578.71) (CCD 1 camera 1), during year 4 of the second \textit{TESS} extended mission. It was then re-observed during Sector 72 (2023 November 11~-- December 07, BJD 2460262.21--2460285.59) (CCD 1 camera 2) during year 6 of the second ecliptic plane survey. All observations were taken with 120\,s cadence, however due to the long period nature of the orbit ($P\sim$\,50\,d), only one transit of the planet (during Sector 45) was observed across all sectors.

The data were processed in the \textit{TESS} Science Processing Operations Center (SPOC; \citealt{Jenkins2016}) pipeline at NASA Ames Research Center.

For the photometric analysis, we downloaded the \textit{TESS} photometry from MAST and used the Pre-search Data Conditioning Simple Aperture Photometry (PDCSAP; \citealt{Stumpe2012, Stumpe2014, Smith2012}) light curve reduced by the SPOC. We use the quality flags provided by the SPOC pipeline to filter out poor quality data. 

An overview of the \textit{TESS} observational sectors and cadences for each system is also included in Table \ref{tab:k2_tess_obs_summary}, with the relevant General Investigator (GI) program IDs provided in Table \ref{tab:k2_tess_obs_summary_ids}.

\subsubsection{CHEOPS}
\label{sec:CHEOPS}

The \textit{CHEOPS} spacecraft \citep{Benz2021}, launched in 2019, is designed to deliver ultra-high precision photometry to support both the discovery of new exoplanets \citep{Osborn2022, Wilson2022, Luque2023, Gliese122024} and the refinement of their properties \citep{Bonfanti2021, Lacedelli2022, Palethorpe2024}. As part of this study, we obtained two transit visits each for K2-79\,b, K2-106\,b, K2-106\,c, K2-111\,b, K2-222\,b, K2-263\,b, and TOI-1634\,b through the \textit{CHEOPS} AO-2 Guest Observers program ID:07 (PI: Mortier). In addition, TOI-1634 was observed using Director's Discretionary Time (DDT) (ID:15, PI: Mortier), with the aim of capturing a possible transit of TOI-1634\,c, as suggested by previous RV analyses. This DDT observation spanned $\sim106$ hours, with a $\sim25$ hour gap after the first 65 hours due to a scheduling conflict with a higher-priority target. A summary of these observations is provided in Table \ref{tab:cheops_obs}.

The data were processed using the \textit{CHEOPS} Data Reduction Pipeline (DRP v14.1.2; \citealt{Hoyer2020}) that conducts frame calibration, instrumental and environmental correction, and aperture photometry using pre-defined radii ($R$ = 22.5\arcsec [RINF], 25.0\arcsec [DEFAULT], 30.0\arcsec [RSUP], and OPTIMAL radius) as well as a noise-optimised radius [ROPT]. The size of the OPTIMAL radius is determined on a per-visit basis to account for the differing brightnesses of different targets and is determined by minimising the noise-to-signal ratio \citep{Hoyer2020}. The DRP produced flux contamination (see \citet{Hoyer2020} and \citet{Wilson2022} for computation and usage) that was subtracted from the light curves. We retrieved the data and corresponding instrumental basis vectors and assessed the quality using the \texttt{pycheops} Python package \citep{Maxted2022}. For each visit, we follow the recommendation of the DRP and decorrelate with the parameters suggested by this package, which can be found in Table \ref{tab:cheops_obs}. The detrending vectors are then passed as free parameters when fitting the transits with a normal distribution centred on 0 ($\mathcal{N} (0,1)$) using the \texttt{lm\_fit} function. To detrend from stray light due to nearby objects reflecting in the telescope (or `glint') we fit a spline against the roll angle of the telescope. This is done with \texttt{pycheops} using the \texttt{add\_glint} function, which is fitted using uniform distribution between 0 and 2 ($\mathcal{U}(0,2)$). Outliers were also trimmed from the light curves, with points that were 4$\sigma$ away or further from the median value removed. We concluded that using the OPTIMAL radius was the best option as it gave the highest signal-to-noise. 

In addition to the DRP + \texttt{pycheops} workflow described above, we also processed all \textit{CHEOPS} visits using the independent \texttt{PIPE} reduction software \citep{Brandeker2024}, which performs PSF-based photometry, dark-frame modelling, background-star subtraction, and PCA-based decorrelation of instrumental systematics. This alternative reduction produced light curves that were fully consistent with the DRP outputs, with no statistically significant improvement in precision or transit depth recovery for any of the targets. Given the small amplitudes of the transits studied here and the relatively modest brightness of our targets, both pipelines gave comparable results. For consistency across the dataset we therefore adopt the DRP + \texttt{pycheops} light curves for all subsequent modelling.

\subsubsection{MuSCAT2}
\label{sec:MuSCAT2}

We acquired five transits of TOI-1634.01 from \citet{Hirano2021}, taken on 2020 February 07 (BJD~2458887.35), 2020 February 10 (BJD~2458890.34), 2020 February 11 (BJD~2458891.35), 2021 February 14 (BJD~2459259.34), and 2021 February 16 (BJD~2459261.34) using the multiband imager MuSCAT2 \citep{Narita2019} mounted on the 1.52\,m Telescopio Carlos Sánchez (TCS) telescope at Teide Observatory in Tenerife, Spain. MuSCAT2 is a sibling of MuSCAT but has four channels for the $g, r, i$ and $z_s$ bands. The field of view of MuSCAT2 is 7.4$'$ $\times$ 7.4$'$ with a pixel scale of 0.44$\arcsec$ per pixel. The observations were conducted with an exposure time of 3--60\,s depending on the band and sky condition. Aperture radii of 8--12 pixels were adopted depending on the band and night, which means that the companion star at 2.5$\arcsec$ away is included in the photometric apertures in all bands. 

The obtained data were reduced using the same methodology described in \citet{Wilson2025}. This approach models systematics arising from changes in the PSF shape using a principal component analysis (PCA)-based framework. For each MuSCAT2 transit, PCA on the relevant instrumental parameters was performed, and the optimal number of principal components, $\theta$, was determined using leave-one-out cross-validation (LOOCV) to minimise overfitting. The fluxes were then linearly regressed against the principal components, masking the in-transit data to avoid biasing the detrending process. The resulting linear models were used to decorrelate the MuSCAT2 light curves. To propagate uncertainties from the detrending, samples were drawn from the posterior distributions of the regression coefficients and the resulting model uncertainties were added in quadrature to the photometric errors.

\subsubsection{OAA, LCOGT, RCO}

Several of the ground-based light curves of TOI-1634.01 analysed here were obtained as part of the \textit{TESS} Follow-up Observing Program Sub Group 1 (TFOP SG1; \citealt{TFOPSG1}) by \citet{Cloutier2021} and \citet{Hirano2021}, which coordinates ground-based photometric monitoring of \textit{TESS} planet candidates to confirm on-target transits. In this work, we re-analysed the raw versions of these light curves, which were provided directly by \citet{Cloutier2021} and \citet{Hirano2021}, in order to ensure consistent detrending and modelling across all ground-based datasets.

We acquired two full transits of TOI-1634.01 from \citet{Cloutier2021}, taken on 2020 February 13 (BJD~2458893.31) and 21 (BJD~2458901.27) using the main 0.4\,m instrument ensemble at Observatori Astron{\`o}nomic Albanya{\`a} (OAA) with stable observation conditions in the valley. Differential photometry was performed in a 36$'$ $\times$ 36$'$ star field centred on TOI-1634 using the \textit{Sloan-i'} filter with 10.0$\arcsec$ photometric aperture (in 7.4$\arcsec$ FWHM conditions) using the \texttt{AstroImageJ} pipeline. The sequences consisted of 88 and 148 frames of 120\,s and 100\,s exposure times, respectively. A small number of outlying points during transit due to instrumental anomalies ($>10\sigma$) were removed before the transit fit.

We acquired a full transit of TOI-1634.01 from \citet{Cloutier2021} taken on 2020 September 30 (BJD~2459122.84) in the Pan-STARRS $z_s$ band from the Las Cumbres Observatory Global Telescope \citep[LCOGT;][]{Brown2013} 1\,m network node at McDonald Observatory. The 4096 $\times$ 4096 LCOGT SINISTRO cameras have an image scale of 0.39$\arcsec$ per pixel, resulting in a 26$'$ $\times$ 26$'$ field of view. The images were calibrated by the standard LCOGT \texttt{BANZAI} pipeline \citep{McCully2018}, and photometric data were extracted with \texttt{AstroImageJ} \citep{Collins2017}. The TOI-1634.01 observation used 40-second exposures and a photometric aperture radius of 2.5$\arcsec$ to extract the differential photometry.

A full transit observation on 2020 February 20 (BJD 2458900.27) of TOI-1634.01 was obtained using the RCO 40\,cm telescope located at the Grand-Pra Observatory, Switzerland, from \citet{Cloutier2021}. The full transit was observed in the \textit{Sloan-i'} passband with an exposure time of 90\,s. The light curve of TOI-1634.01 was produced using the \texttt{AstroImageJ} pipeline with 8.0$\arcsec$ aperture and by detrending against airmass and FWHM.

As with the MuSCAT2 data, the raw light curves were reprocessed using the same PCA-based detrending approach described in Section \ref{sec:MuSCAT2}, ensuring consistent treatment of systematics across all ground-based light curves.

\subsection{Spectroscopy}
\label{sec:spectroscopy}

In order to characterise the planetary systems identified in our sample, we obtained a series of spectroscopic observations using several high-resolution spectrographs. These data were used to derive precise RVs and to assess stellar activity. Table \ref{tab:spec_obs_summary} summarises the spectroscopic observations, listing for each target the instruments employed, the number of spectra acquired, and the observing periods. Further details on the data obtained with each telescope are provided in the following subsections.

\begin{table}
\centering
\caption{Summary of spectroscopic observations. Dates are given in DD--MM--YY format.}
\label{tab:spec_obs_summary}
\begin{tabular}{l l l l}
\hline\hline
Target & Telescope & $\rm N_{\rm obs}$ & Observing span \\
\hline
K2-79 & HARPS-N & 80 & 04/11/15 -- 02/11/24\\
 \vspace{1.5mm}
 & HIRES & 63 & 31/10/15 -- 09/11/19\\
K2-106 & HARPS-N & 45 & 14/08/14 -- 26/12/19\\
 & HIRES & 74 & 12/08/16 -- 18/08/17\\
 & ESPRESSO & 23 & 08/08/19 -- 16/11/19\\
 & HARPS & 20 & 25/10/16 -- 27/11/16\\
 & PFS & 13 & 14/08/16 -- 14/01/17\\
 & FIES & 6 & 05/10/16 -- 25/11/16\\
 \vspace{1.5mm}
    & HDS & 3 & 12/10/16 -- 14/11/16\\
K2-111 & HARPS-N & 116 & 26/10/15 -- 24/02/19\\
 & HIRES & 55 & 26/07/16 -- 09/11/19 \\
 & ESPRESSO & 41 & 20/10/18 -- 03/03/20 \\
\vspace{1.5mm}
 & FIES & 6 & 16/11/15 -- 20/11/15 \\
K2-222 & HARPS-N & 75 & 24/07/16 -- 19/01/25 \\
 \vspace{1.5mm}
 & HIRES & 55 & 04/02/17 -- 05/02/18 \\[1.5mm]
 K2-263 & HARPS-N & 96 & 23/12/15 -- 16/03/20 \\[1.5mm]
TOI-1634 & HARPS-N & 32 & 07/07/20 -- 04/03/21 \\
 & IRD & 48 & 27/09/20 -- 02/03/21 \\
\hline
\end{tabular}
\end{table}

\subsubsection{HARPS-N}
\label{sec:HARPS-N}

HARPS-N is a high-precision spectrograph mounted on the Telescopio Nazionale Galileo at the Observatorio del Roque de los Muchachos in La Palma, Spain \citep{Cosentino_2012}. The spectrograph covers wavelengths in the range 383--690\,nm, with average resolving power R = 115,000. 

As part of the HARPS-N GTO program, we analysed a total of 443 spectra across six targets, combining new and archival observations. Unless otherwise noted, all spectra were reduced using the HARPS-N Data Reduction Software (DRS version 3.0.1), with RVs and activity indicators extracted accordingly.

For K2-79, we analysed 80 spectra between 2015 November 04~-- 2024 November 02 (BJD 2457730.53--2460616.64). Of these, 79 were taken over 4 seasons and published by \citet{Nava2022}, with one additional observation acquired in 2024. All exposures were 30 minutes, yielding a mean signal-to-noise ratio (S/N) of $\sim$30 in order 50.

For K2-106, we analysed 45 spectra between 2014 August 14~-- 2019 December 26 (BJD 2457614.68--2458844.36). Twelve spectra, acquired between 2016 October 30~-- 2017 January 28 (BJD 2457699.49--2457770.37), taken by \citet{Guenther2017}, were reduced with DRS version 3.7, and had a mean S/N of 21.79 in order 50. The remaining spectra, all with 30-minute exposures, achieved a higher mean S/N of $\sim$27 in the same order.

For K2-111, we analysed 116 spectra between 2015 October 26~-- 2019 February 24 (BJD 2457321.72--2458539.44). This includes 104 from \citet{Mortier2020}, while 12 additional public observations from \citet{Fridlund2017} were reduced using DRS version 3.7 and had a mean S/N of 39.6 in order 50. All observations used 30-minute exposure times, with the remaining spectra achieving a mean S/N of $\sim$47 in order 50.

For K2-222, we analysed 74 spectra between 2016 August 14~-- 2025 January 19 (BJD 2457614.70--2460695.36). One observation taken on 2017 September 03 (BJD~2457999.74) was identified as anomalous and excluded. A total of 63 spectra were previously reported by \citet{Nava2022}, spanning three seasons from 2016 August 14 to 2019 December 23 (BJD 2457614.70--2458841.37). Exposure times ranged between 15--30 minutes, yielding a mean S/N of $\sim$74 in order 50. All observations were acquired as part of the HARPS-N GTO program.

For K2-263, we analysed 96 spectra between 2015 December 23~-- 2020 March 16 (BJD 2457379.63--2458924.55). Of these, 67 were published in \citet{Mortier2018} between 2015 December 23~-- 2018 January 31 (BJD 2457379.63--2458120.66). All observations used 30-minute exposures and achieved a mean S/N of $\sim$35 in order 50.

For TOI-1634, we analysed 32 spectra from \citet{Cloutier2021}, taken between 2020 August 07~-- 2021 March 04 (BJD 2459068.72--2459278.39), also as part of the HARPS-N GTO program. All observations had 30-minute exposures and a mean S/N of $\sim$107. The RVs were extracted via template-matching using the \texttt{TERRA} pipeline \citep{Anglada2012}, which is optimised for M-dwarf spectra and provides improved RV precision over traditional cross-correlation techniques.

Additionally, we quantify the chromospheric activity level of each host star using the Ca II H\&K S-index from the HARPS-N spectra. The S-index values were converted to log R$'_{\rm HK}$ following the procedure of \citet{Noyes1984a} for all targets, bar TOI-1634 for which \citet{Astuillo2017} was used, adopting the stellar $B-V$ colours listed in Table \ref{tab:stellar_params}. The resulting mean log R$'_{\rm HK}$ and their dispersions are reported in Table \ref{tab:spec_activity}. Stellar magnetic activity contributes to RV variability and can therefore affect the precision with which planetary masses are determined. The measured activity levels for our targets indicate low/moderate activity consistent with the observed RV scatter, and are taken into account in our GP modelling where appropriate (see Section \ref{sec:modelling} for individual target discussions).

All RVs with their errors are available in the supplementary material.

\begin{table}
\centering
\caption{Chromospheric activity indicators for the host stars derived from the HARPS-N spectra. The table lists the mean Ca II H\&K activity index, expressed as $\langle \log \rm R'_{\rm HK} \rangle$, the standard deviation of the activity time series $\sigma_{\log \rm R'_{\rm HK}}$, and the average error $\langle \sigma_{\log \rm R'_{\rm HK}} \rangle_{\rm meas}$.}
\label{tab:spec_activity}
\begin{tabular}{l c c c}
\hline\hline
Target & $\langle \log \rm R'_{\rm HK} \rangle$ & $\sigma_{\log \rm R'_{\rm HK}}$ & $\langle \sigma_{\log \rm R'_{\rm HK}} \rangle_{\rm meas}$ \\
\hline
K2-79 & -5.08 & 0.07 & 0.02 \\
K2-111 & -4.96 & 0.46 & 0.008 \\
K2-222 & -4.81 & 0.56 & 0.003 \\
K2-263 & -5.01 & 0.09 & 0.02 \\
TOI-1634 & -5.25 & 0.11 & 0.02 \\
\hline
\end{tabular}
\end{table}

\subsubsection{ESPRESSO}
\label{sec:ESPRESSO}

The fibre-fed Echelle Spectrograph for Rocky Exoplanets and Stable Spectroscopic Observations \citep[ESPRESSO;][]{Pepe2014, Pepe2021} is installed in the combined Coudé facility of the Very Large Telescope (VLT) at the Paranal Observatory, Chile. It delivers high-precision RVs with a resolving power of R $\sim140 000$ in its high-resolution mode, covering a wavelength range of 378--789\,nm.

We obtained 23 ESPRESSO spectra of K2-106, previously published by \citet{Guenther2024}, taken between 2019 August 08~-- 2019 November 16 (BJD 2458703.75--2458788.76). All calibration frames were taken using the standard procedures of this instrument. There were typical exposure times of 900\,s, with a mean S/N of $\sim$20 in order 50.

For K2-111, we obtained 41 spectra from \citet{Mortier2020}, taken between 2018 October 30~-- 2020 March 03 (BJD 2458421.74--2458911.50). Typical exposure times were 900\,s per observation, resulting in S/N values of 45--80 at 550\,nm, with a few lower S/N values. A technical intervention on the instrument in June 2019 introduced an RV zero-point shift. As such, we treated the 16 pre-intervention and 25 post-intervention spectra as independent data sets with a fitted offset. 

All spectra were reduced and extracted using the dedicated ESPRESSO pipeline.

\subsubsection{FIES}
\label{sec:FIES}
The FIbre-fed Echelle Spectrograph \citep[FIES;][]{Frandsen1999, Telting2014} is mounted on the 2.56\,m Nordic Optical Telescope (NOT) at the Roque de los Muchachos Observatory, La Palma, Spain.

We obtained six FIES spectra of K2-106 from \citet{Guenther2017}, acquired between 2016 October 05~-- 2016 November 25 (BJD 2457666.65--2457717.51). Each observation used an exposure time of 2700\,s, yielding a S/N of the extracted spectra of about 35 per pixel at 550\,nm.

For K2-111, six FIES spectra were obtained from \citet{Fridlund2017}, taken in 2015 November (BJD 2457342.50--2457347.47). Exposure times ranged from 2400--3600\,s, with resulting S/N values of 40--60 per pixel at 550\,nm.

The data for both targets were originally reduced by \citet{Guenther2017} and \citet{Fridlund2017} using standard \texttt{IRAF} and \texttt{IDL} routines. RVs were extracted via S/N-weighted, multi-order cross-correlations, using the highest S/N stellar spectrum as the template.

\subsubsection{HARPS}
\label{sec: HARPS}

We retrieved 20 spectra of K2-106 publicly available from \citet{Guenther2017}, taken between 2016 October 25~-- 2016 November 27 (BJD 2457686.68--2457719.60) with typical exposure times of 1800\,s, with the High Accuracy Radial velocity Planet Searcher Spectrograph \citep[HARPS;][]{Mayor2003} on the 3.6\,m ESO telescope at La Silla. The spectra were reduced and extracted using the dedicated HARPS pipeline, with RVs determined using a cross-correlation method with a numerical mask that corresponds to a G2 star, and the RV measurements obtained by fitting a Gaussian function to the average cross-correlation function (CCF).

\subsubsection{HDS}
\label{sec:HDS}

We obtained three spectra of K2-106 with the High Dispersion Spectrograph \citep[HDS;][]{Noguchi2002}, publicly available from \citet{Guenther2017}. The spectra were obtained from 2016 October 12~-- 2016 October 14 (BJD 2457673.98--2457676.02), with a spectral resolution of R $\sim$ 85,000 and a typical S/N of 70--80 per pixel close to the sodium D lines. This instrument uses an I$_2$ absorption cell for precise wavelength calibration, with RVs extracted using a high S/N template taken by the same instrument without the I$_2$-cell.

\subsubsection{HIRES}
\label{sec:HIRES}

The High Resolution Echelle Spectrometer \citep[HIRES;][]{Vogt1994}, mounted on the Keck I telescope at the W. M. Keck Observatory, provides spectra spanning 264--799\,nm with a typical resolution for the following targets of R=60,000.

For K2-79, we obtained 63 HIRES spectra from \citet{Howard2025}, from 2015 October 31 to 2019 November 09 (BJD 2457326.90--2458796.94) using an exposure meter setting of 50,000 counts.

For K2-106, we obtained 74 spectra between 2016 August 12~-- 2017 August 18 (BJD 2457612.93--2457984.08), including 35 public observations from \citet{Sinukoff2017} and four additional spectra from \citet{Howard2025}. Exposures continued until S/N $\approx$\,125 were reached, with typical exposure times of $\sim$\,25 minutes, and an exposure meter setting of 80,000 counts.

For K2-111, we obtained 55 spectra between 2016 July 26~-- 2019 November 09 (BJD 2457596.10--2458796.95), all from \citet{Howard2025}, with an exposure meter setting of 125,000.

For K2-222, we obtained 55 spectra between 2017 February 04~-- 2018 February 05 (BJD 2457788.73--2458154.72), also from \citet{Howard2025}, with an exposure meter setting of 250,000.

The standard California Planet Search (CPS; \citealp{Howard2010}) Doppler pipeline was used for data reduction and RV measurements.

\subsubsection{IRD}
\label{sec:IRD}

We obtained 48 near-infrared spectra observations of TOI-1634 from \citet{Hirano2021}, acquired using the InfraRed Doppler spectrograph \citep[IRD;][]{Tamura2012, Kotani2018} on the 8.2\,m Subaru Telescope, conducted between 2020 September 27~-- 2021 March 02 (BJD 2459120.04--2459275.74). Exposure times ranged from 720--1200\,s depending on observing conditions. Raw data were reduced using IRAF \citep{Tody1993} and custom routines \citep{Kuzuhara2018, Hirano2020}, with resulting RVs having typical S/N of 60--95 per pixel at 1000\,nm.

\subsubsection{PFS}
\label{sec:PFS}

We obtained 13 spectra of K2-106 with the Carnegie Planet Finder Spectrograph \citep[PFS;][]{Crane2006, Crane2008, Crane2010} publicly available from \citet{Guenther2017}. PFS is an échelle spectrograph on the 6.5\,m Magellan/Clay Telescope at Las Campanas Observatory in Chile. Observations were carried out between 2016 August 14~-- 2017 January 14 (BJD 2457614.82--2457767.55). Exposure times ranged from 20--40 minutes, yielding a S/N of 50--140 per pixel. The spectra were acquired at a resolving power of R $\sim$ 76,000 over the wavelength range that includes the iodine absorption lines. The relative RVs were extracted from the spectrum using the techniques described by \citet{Butler1996}.

\section{Transit \& RV modelling}
\label{sec:modelling}

To calculate the best-fit planetary parameters, we performed joint fits to the transit and RV datasets described in Section \ref{sec:obs}, using the \texttt{PyORBIT}\footnote{\url{https://github.com/LucaMalavolta/PyORBIT}, v10.5.5.} code \citep{malavoltaetal16, Malavolta2018} modelling framework. \texttt{PyORBIT} employs the \texttt{BATMAN} package \citep{kreidberg2015} to model transit light curves and supports multiple algorithms for parameter estimation, including nested sampling and Markov Chain Monte Carlo (MCMC) methods. In this study, we adopted the \texttt{emcee} \citep{emcee} MCMC sampler. For the transit modelling, we assumed  a quadratic limb-darkening law and specified instrument-specific exposure times and supersampling factors: 1800.0\,s for \textit{K2} long cadence, 60.0\,s for \textit{K2} short cadence, 120.0\,s for the \textit{TESS} long cadence observations, 20.0\,s for the \textit{TESS} short cadence observations, 60.0\,s for the \textit{CHEOPS} observations, 90.0\,s for the RCO observations, 40.0\,s for the LCOGT observations, 120.0\,s for the OAA long observations, 100.0\,s for the OAA short observations, 46.0\,s for the MuSCAT2 $g$-band observations, 21.0\,s for the MuSCAT2 $i$- and $r$-band observations, and 11.0\,s for the MuSCAT2 $z_s$-band observations as inputs to the light curve model.

Quadratic limb-darkening coefficients were assigned Gaussian priors based on interpolated values from published model grids, using Table 9 from \citet{Claret2018} for the \textit{K2} light curves, Table 25 from \citet{Claret2017} for the \textit{TESS} light curves, and Table 8 from \citet{Claret2021} for the \textit{CHEOPS} light curves. For the ground-based observations, the ExoCTK Limb-Darkening Calculator\footnote{\url{https://exoctk.stsci.edu/limb_darkening}} \citep{ExoCTK} was used, adopting the Phoenix ACES model grid and a 1 grism setting, based on the stellar parameters listed in Table \ref{tab:stellar_params}. Gaussian priors were also placed on the stellar mass, radius, and density using the literature values listed in Table \ref{tab:stellar_params}. Although these parameters are not independent, only the stellar density directly enters the transit model, as it is the quantity constrained by the transit geometry. The stellar mass and radius priors were included solely to propagate uncertainties consistently into the derived planetary properties, and were not used to impose additional constraints on the transit fit itself.

While the orbital and transit analyses in this work are performed using a uniform modelling framework, the adopted stellar parameters are taken from the relevant discovery or characterisation studies for each system (see Table \ref{tab:stellar_params}). For the majority of targets, these stellar properties were determined in analyses led by the HARPS-N collaboration and are therefore largely homogenous in terms of methodology and underlying data. An exception is K2-106, whose stellar parameters were derived in earlier studies using different approaches and, in some cases, different \textit{Gaia} data releases. We acknowledge that this introduces a limited degree of heterogeneity in the stellar parameters across the sample. However, the stellar properties are generally well constrained, and their uncertainties are propagated into the derived planetary parameters through the joint modelling. We therefore do not expect this residual heterogeneity to have a significant impact on the results presented here.

Planetary orbits were assumed to be Keplerian, with eccentricity parametrised using the $\sqrt{e} \cos \omega$ and $\sqrt{e}\sin\omega$ formalism from \citet{Eastman2013} to minimise bias in the low-eccentricity regime. Uniform priors were adopted on inclination, eccentricity, and RV semi-amplitude, with physically motivated bounds of $K \in [0.1, 100]\,\mathrm{m\,s^{-1}}$, $i \in [80^{\circ}, 90^{\circ}]$, and $e \in [0, 0.9]$, where the eccentricity prior is placed directly on $e$ rather than on the transformed parameters $\sqrt{e}\cos\omega$ or $\sqrt{e}\sin\omega$. For the orbital period ($P$) and transit centre time ($T_0$), we adopted informed Uniform priors based on values reported in the literature (see Table \ref{tab:stellar_params}) and extrapolated forward to the epochs of the more recent \textit{CHEOPS} observations. For example, for K2-79\,b, we used the period of $P = 10.99475 \pm 0.00047$ days and transit centre time $T_0 = 7103.2276 \pm 0.0017$ (BJD-2450000) previously reported by \citet{Nava2022} to define uniform priors of $P \in [10.994, 10.996]$ and $T_0 \in [9917.0, 9918.0]$. This approach ensured that the fits remained anchored in prior knowledge while being flexible enough to accommodate updated constraints from newer datasets. 

All available photometric and RV datasets were combined in a single joint fit using \texttt{PyORBIT}. Each photometric dataset (defined by instrument and cadence) was assigned an independent normalisation factor with broad uniform priors ($0.8$--$1.2$) to account for relative flux offsets. For multi-planet systems, the transit model included all relevant planets simultaneously, ensuring proper treatment of overlapping transits, while RV-only planets were included without a corresponding transit component. Per-instrument RV offsets and jitter terms were fitted where required, allowing each dataset to contribute coherently to the global solution while accounting for instrumental and baseline differences.

The parameter space was explored using the MCMC sampler, with a population size set to four times the number of free parameters. Each walker was run for 50,000 steps, with the first 1,000 steps discarded as burn-in. The resulting chains were thinned by a factor of 100 to reduce autocorrelation and saved at every 1,000 steps. Final parameter values were taken as the medians of the marginalised posterior distributions, with 1$\sigma$ uncertainties corresponding to the 16th and 84th percentiles, and are shown in Table \ref{tab:planetresults}.

\begin{table*}
    \centering
    \caption{Planetary properties calculated from the combined transit and RV fits. `NT' indicates a non-transiting planet.}
    \resizebox{\linewidth}{!}{
    \begin{tabular}{llcccccc}
         \hline\hline 
         Parameter & Description & \textbf{K2-79} & \textbf{K2-106} & \textbf{K2-111} & \textbf{K2-222} & \textbf{K2-263} & \textbf{TOI-1634} \\ 
         \hline
         \multicolumn{4}{l}{\textit{Orbital parameters}} \\
         \vspace{1.5mm}
         $T_{\rm 0, b}$\,[BJD~2450000] & Transit centre & $9917.9583_{-0.0040}^{+0.0033}$ & $9473.5712_{-0.0019}^{+0.0020}$ & $9519.1348_{-0.0054}^{+0.0038}$ & $9491.95470_{-0.00091}^{+0.00083}$ & $9534.0662_{-0.0021}^{+0.0020}$ & $9584.97039 \pm 0.00022$ \\
         \vspace{1.5mm}
         $P_{\rm b}$\,[days] & Orbital period & $10.995039_{-0.000016}^{+0.000013}$ & $0.57130782_{-0.00000057}^{+0.00000056}$ & $5.3518963_{-0.000012}^{+0.0000089}$ & $15.3889274_{-0.000011}^{+0.0000099}$ & $50.819172_{-0.000068}^{+0.000066}$ & $0.98934551_{-0.00000037}^{+0.00000036}$ \\
         \vspace{1.5mm}
         $T_{14,b}$\,[hrs] & Transit duration & $4.5655_{-0.5082}^{+0.4295}$ & $1.5990_{-0.0261}^{+0.0241}$ & $3.2362_{-0.2494}^{+0.2379}$ & $4.382_{-0.4838}^{+0.4061}$ & $3.7496_{-0.5247}^{+0.4305}$ & $0.9747_{-0.0472}^{+0.0401}$ \\
         \vspace{1.5mm}
         $e_{\rm b}$ & Eccentricity & $0.032_{-0.025}^{+0.057}$ & $0$ & $0.134_{-0.071}^{+0.093}$ & $0.021_{-0.017}^{+0.051}$ & $0.080_{-0.057}^{+0.089}$ & 0 \\
         \vspace{1.5mm}
         $i_{\rm b}$\,[deg] & Inclination & $88.42_{-0.48}^{+0.58}$ & $88.5_{-1.6}^{+1.1}$ & $86.34_{-0.22}^{+0.24}$ & $88.91_{-0.32}^{+0.36}$ & $89.271_{-0.035}^{+0.045}$ & $86.66_{-0.47}^{+0.29}$ \\
         \vspace{1.5mm}
         $b_{\rm b}$ & Impact parameter & $0.4589_{-0.1760}^{+0.1414}$ & $0.0766_{-0.555}^{+0.014}$ & $0.6022_{-0.0795}^{+0.0933}$ & $0.4512_{-0.1573}^{+0.1359}$ & $0.8315_{-0.0945}^{+0.0587}$ & $0.4307_{-0.0413}^{+0.0614}$ \\
         \vspace{1.5mm}
         $a_{\rm b}$\,[AU] & Semi-major axis & $0.0987_{-0.0016}^{+0.0015}$ & $0.01320\pm0.00014$ & $0.05650_{-0.00045}^{+0.00044}$ & $0.1186_{-0.0021}^{+0.0021}$ & $0.2573_{-0.0030}^{+0.0029}$ & $0.01544_{-0.00015}^{+0.00014}$ \\
         \vspace{1.5mm}
         $a_{\rm b}/R_*$ & Semi-major axis ratio & $16.9_{-1.1}^{+1.0}$ & $2.858_{-0.050}^{+0.047}$ & $9.73_{-0.23}^{+0.22}$ & $24.0_{-1.5}^{+1.3}$ & $65.06_{-0.93}^{+0.90}$ & $7.054_{-0.17}^{+0.072}$\\
         \vspace{1.5mm}
         $R_{\rm b}/R_*$ & Radius ratio & $0.0300_{-0.0018}^{+0.0019}$ & $0.0148 \pm 0.0003$ & $0.0135\pm0.0006$ & $0.0195\pm0.0012$ & $0.0271_{-0.0013}^{+0.0013}$ & $0.0358\pm0.0016$ \\
         \vspace{2.5mm}
         $K_{\rm b}$\,[$\mathrm{m\,s^{-1}}$] & RV semi-amplitude & $1.62_{-0.73}^{+0.73}$ & $6.05 \pm 0.50$ & $2.23 \pm 0.28$ & $2.01_{-0.40}^{+0.39}$ & $2.91_{-0.46}^{+0.45}$ & $5.0_{-1.1}^{+1.0}$ \\
         \vspace{1.5mm}
         $T_{\rm 0,c}$\,[BJD~2450000] & & - & $9473.5249_{-0.0030}^{+0.0028}$ & NT & NT & - & - \\
         \vspace{1.5mm}
         $P_{\rm c}$\,[days] & & - & $13.340603_{-0.000021}^{+0.000021}$ & $15.6748_{-0.0061}^{+0.0062}$ & $141.22_{-0.96}^{+2.9}$ & - & -\\
         \vspace{1.5mm}
         $T_{14,c}$\,[hrs] & & - & $3.7822_{-0.4607}^{+0.7720}$ & NT & NT & - & - \\
         \vspace{1.5mm}
         $R_{\rm c}/R_*$ & & - & $0.0250\pm0.0007$  & NT & NT & - & - \\
         \vspace{1.5mm}
         $b_{\rm c}$ & & - & $0.2603_{-0.1797}^{+0.1671}$ & NT & NT & - & - \\
         \vspace{1.5mm}
         $e_{\rm c}$ & & - & $0.180_{-0.048}^{+0.038}$ & $0.069_{-0.049}^{+0.072}$ & $0$ & - & -\\
         \vspace{1.5mm}
         $i_{\rm c}$\,[deg] & & - & $89.34_{-0.40}^{+0.44}$ & NT & NT & - & - \\
         \vspace{1.5mm}
         $a_{\rm c}$\,[AU] & & - & $0.1078_{-0.0012}^{+0.0011}$ & $0.11565_{-0.00092}^{+0.00092}$ & $0.521_{-0.010}^{+0.010}$ & - & - \\
         \vspace{1.5mm}
         $a_{\rm c}/R_*$ & & - & $23.35_{-0.41}^{+0.38}$ & $19.91_{-0.48}^{+0.44}$ & $105.3_{-6.6}^{+5.9}$ & - & -\\
         \vspace{1.5mm}
         $K_{\rm c}$\,[$\,\mathrm{m\,s^{-1}}$] & & - & $2.27_{-0.62}^{+0.60}$ & $3.19 \pm 0.30$ & $1.57_{-0.46}^{+0.42}$ & - & - \\
         \hline
         \multicolumn{4}{l}{\textit{Planetary parameters}} \\
         \vspace{1.5mm}
         $M_{\rm b}$\,[$M_{\oplus}$] & Planet mass & $5.9 \pm 2.6$ & $7.54_{-0.64}^{+0.65}$ & $5.35_{-0.69}^{+0.68}$ & $7.5 \pm 1.5$ & $15.4_{-2.5}^{+2.4}$ & $4.9_{-1.0}^{+1.0}$ \\
         \vspace{1.5mm}
         $R_{\rm b}$\,[$R_{\oplus}$] & Planet radius & $4.15_{-0.18}^{+0.19}$ & $1.606_{-0.033}^{+0.034}$ & $1.837_{-0.071}^{+0.076}$ & $2.275 \pm 0.10$ & $2.513_{-0.095}^{+0.11}$ & $1.759_{-0.057}^{+0.058}$ \\
         \vspace{1.5mm}
         $\rho_{\rm b}$\,[$\rho_{\oplus}$] & Planet density & $0.0865_{-0.0402}^{+0.0423}$ & $1.9379_{-0.2087}^{+0.2149}$ & $0.9187_{-0.1620}^{+0.1700} $ & $0.6749_{-0.1564}^{+0.1733}$ & $1.0248_{-0.1990}^{+0.2055}$ & $0.9580_{-0.2290}^{+0.2298}$ \\
         \vspace{2.5mm}
         $T_{\rm eq, b}$ [K] (A$_B$=0) &  Equilibrium temperature & $1033_{-31}^{+30}$ & $2324 \pm 35$ & $1293 \pm 18$ & $803_{-25}^{+24}$ & $482_{-8}^{+7}$ & $900 \pm 6$\\
         \vspace{1.5mm}
         $M_{\rm c}$ or $M_{\rm c}\sin i$\,[$M_{\oplus}$] & Planet mass or minimum mass & - & $7.9 \pm 2.1$ & $11.1_{-1.0}^{+1.1}$ & $12.3_{-3.6}^{+3.3}$ & - & - \\
         \vspace{1.5mm}
         $R_{\rm c}$\,[$R_{\oplus}$] & & - & $2.712 \pm 0.071$ & NT & NT & - & - \\
         \vspace{1.5mm}
         $\rho_{\rm c}$\,[$\rho_{\oplus}$] & & - & $0.4237_{-0.1223}^{+0.1210}$ & NT & NT & - & - \\
         \vspace{1.5mm}
         $T_{\rm eq, c}$ [K] (A$_B$=0) &  & - & $838_{-13}^{+12}$ & $923 \pm 13$ & $396 \pm 12$ & - & -\\
         \hline
         
    \end{tabular}
    }
    \label{tab:planetresults}
\end{table*}

\subsection{K2-79}
\label{sec:K2-79}

\begin{figure}
    \centering
    \includegraphics[width=\linewidth]{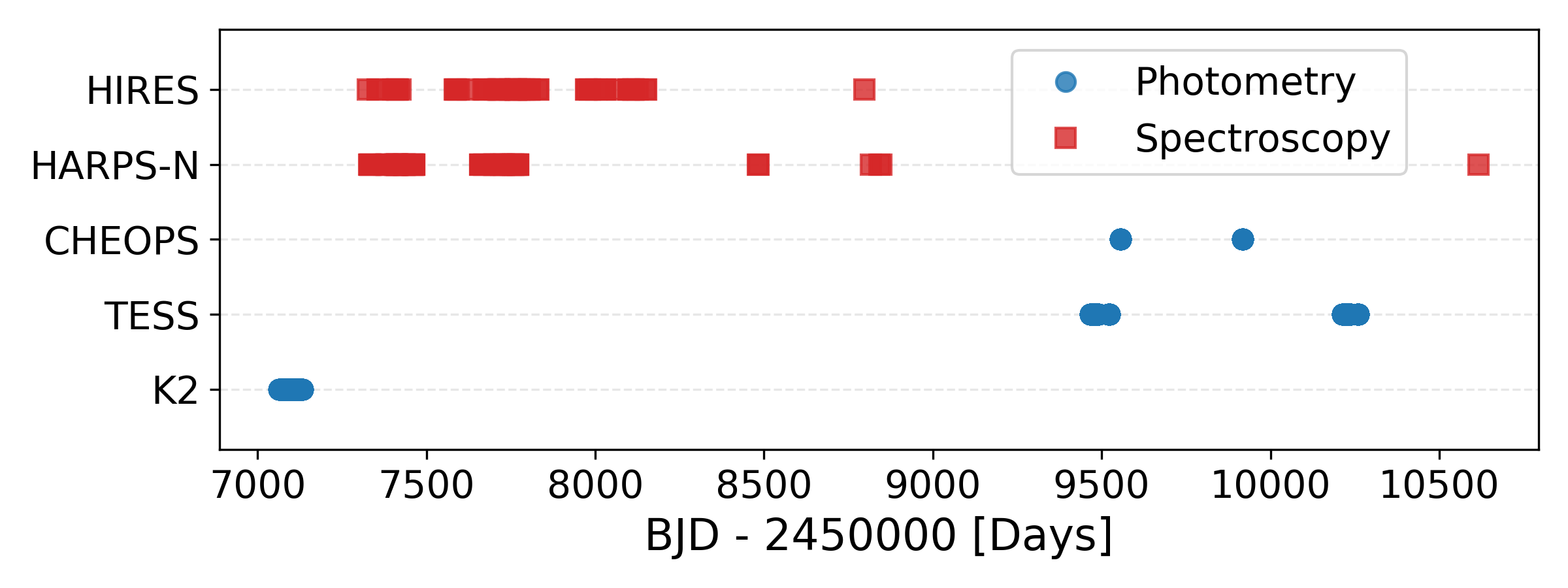}
    \caption{Timeline of photometric and spectroscopic observations of K2-79.}
    \label{fig:K2-79_timeline}
\end{figure}

K2-79\,b was previously characterised by \citet{Nava2022} and more recently by \citet{Bonomo2023} and \citet{Howard2025}, with varying conclusions about its mass and orbital properties. This analysis incorporates a broader dataset, including both HARPS-N and HIRES RVs, along with photometry from \textit{K2}, \textit{TESS}, and \textit{CHEOPS}, as seen in Figure \ref{fig:K2-79_timeline}.

We performed GP regression tests on the RV data within the \texttt{PyORBIT} framework, using a quasi-periodic kernel to account for potential stellar activity-induced variability \citep{rajpaul_gaussian_2015}. The GP hyperparameters included an amplitude term, a periodic timescale linked to stellar rotation, and a decay timescale representing the finite lifetime of active regions. Broad, physically motivated priors were adopted, centred on the stellar rotation period estimated by \citet{Nava2022}. The MCMC chains converged well, but the GP amplitude was found to be low, and the marginal likelihood comparison (as measured by the Bayesian Information Criterion, BIC; \citep{Schwarz1978}, where lower values indicate a preferred model) ($\Delta\mathrm{BIC} \approx 8$) indicated no significant improvement over the white-noise model with per-instrument jitter terms. The quasi-periodic timescale remained poorly constrained due to the sparse temporal sampling and the lack of nightly cadence in the HARPS-N and HIRES datasets. These results mirror those of \citet{Nava2022}, who reported that GP fits for K2-79 tended to overfit the noise and were not well supported by the data. Additionally the combined HARPS-N + HIRES RV time series shows no detectable long-term trend or cycle-like variability over the $\approx9$\,yr span (see Figure \ref{fig:K2-79_transits} third row). Thus, no additional GP component is required to model magnetic-cycle variations, and the white noise + jitter model remains preferred. We therefore adopted the simpler model without a GP component for the final joint fit, as it provides equivalent or better performance with fewer free parameters.

We derive an orbital period of $P_{\rm b}=10.995039_{-0.000016}^{+0.000013}$ days and a transit centre time of $T_{\rm 0, b} = 9917.9583_{-0.0040}^{+0.0033}$ (BJD~2450000), significantly improving upon the formal ephemeris precision reported in the earlier studies. Our mass estimate of $M_{\rm b}=5.9\pm 2.6\,M_{\oplus}$ lies between the higher values reported by \citet{Nava2022} ($11.8 \pm 3.6\,M_{\oplus}$) and \citet{Bonomo2023} ($9.2 \pm 2.4\,M_{\oplus}$), and the much lower less constrained estimate from \citet{Howard2025} ($3.8_{-4.2}^{+4.3}\,M_{\oplus}$). While formally consistent within $1.3\sigma$ with \citet{Nava2022} and $1\sigma$ of \citet{Bonomo2023}, our result favours a moderately lower mass. To assess whether the difference relative to \citet{Nava2022} and \citet{Bonomo2023} might be driven by our inclusion of the HIRES RVs, we repeated the analysis using only the HARPS-N RVs. The HARPS-N only fit results in $M_{\rm b} = 8.0_{-3.0}^{+2.8}\,M_{\oplus}$, fully consistent with both our full dataset solution and the results of \citet{Nava2022} and \citet{Bonomo2023} within the uncertainties. Thus, this discrepancy likely reflects the inclusion of a more extensive RV dataset, joint photometry from multiple space missions, and a more flexible treatment of instrument specific jitter and stellar variability, particularly across observing seasons --- effects that may not have been fully captured in the global models of earlier analyses. We recover a low-eccentricity orbit whose eccentricity posterior is peaked at zero ($e \approx 0.03$), consistent with a circular solution and broadly in agreement with \citet{Nava2022} and \citet{Howard2025}. The planetary radius is well constrained at $R_{\rm b}=4.15_{-0.18}^{+0.19}\,R_{\oplus}$, which is consistent with both previous works. 

Although our HARPS-N only test confirms that the HIRES data are not biasing the semi-amplitude for K2-79\,b, the system nonetheless illustrates the challenges posed by low-amplitude RV signals analysed with heterogeneous extraction pipelines. In principle, modern template-matching pipelines, such as \texttt{S-BART} \citep{Silva2022}, or more advanced spectrum-level corrections like those applied in the recent analysis of Kepler-10 \citep{Bonomo2025}, may help to improve internal consistency across instruments and mitigate subtle systematics that classical CCF pipelines do not fully capture. However, a meaningful application of these techniques would require a homogenous reprocessing of all available spectra which lies beyond the scope of this present study. As in the Kepler-10 case, future work combining additional high-precision RVs with uniform, state-of-the-art extraction may ultimately be necessary to fully resolve the remaining uncertainties in the mass of K2-79\,b.

Figure \ref{fig:K2-79_transits} shows the phase-folded light curves for the \textit{K2} long cadence, \textit{TESS} long cadence, and \textit{CHEOPS} data along with the best fitting transit model, as well as the orbital solution and RV residuals phased on the period of K2-79\,b for the combined fit with the HARPS-N and HIRES RVs. The final joint transit-RV solution reproduces all available datasets well, with residuals consistent with white noise and no evident structure (Figure \ref{fig:K2-79_transits}). The root-mean-square (RMS) scatter of the RV residuals ($\sim4.8\,\mathrm{m\,s^{-1}}$) matches the typical measurement uncertainties, and the residuals from the \textit{K2} and \textit{TESS} transit fits are consistent with their expected photon noise limits. The \textit{CHEOPS} light curve fit is visibly noisier, reflecting both the small transit depth of the planet and the fact that the \textit{CHEOPS} precision for this relatively faint target does not reach the level achieved by the original \textit{K2} data. Despite the increased scatter, the \textit{CHEOPS} transits occur at the predicted epochs and their depths are consistent with those measured from \textit{K2} and \textit{TESS}, supporting the robustness of the global solution. The \textit{TESS} transit displays a slightly asymmetric ingress and egress that could superficially resemble a small timing offset, but this feature is more plausibly explained by residual systematics or local instrument trends. Importantly, the \textit{CHEOPS} observations occurred between the two \textit{TESS} observing sectors (see Figure \ref{fig:K2-79_transits}), $\sim$\,6.5 years after the initial \textit{K2} campaign, and captured the transits at the epochs predicted by the original global ephemeris. This temporal ordering confirms that the period and transit timing derived from the joint fit are accurate, and that no measurable timing offset is present across the different instruments.

\subsection{K2-106}
\label{sec:K2-106b}

\begin{figure}
    \centering
    \includegraphics[width=\linewidth]{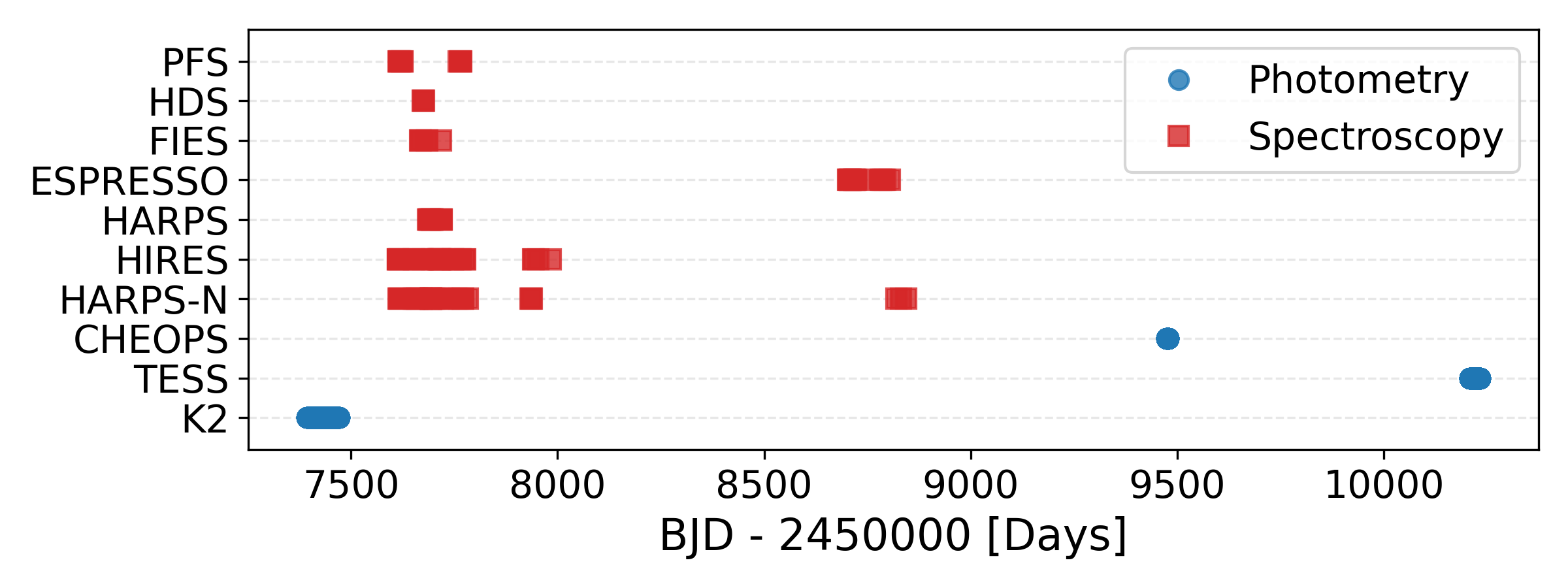}
    \caption{Timeline of photometric and spectroscopic observations of K2-106.}
    \label{fig:K2-106_timeline}
\end{figure}

K2-106 has been the subject of several prior characterisation efforts, including photometric analysis of the original \textit{K2} light curve \citep{Vanderburg2014}, validation of the planetary system \citep{Adams2017}, and detailed RV and transit modelling in multiple follow-up studies \citep{Sinukoff2017, Guenther2017, Bonomo2023, Guenther2024, Howard2025}. For K2-106, we performed a joint fit incorporating data from ESPRESSO, HARPS, HARPS-N, HIRES, PFS, HDS, and FIES, as well as light curves from \textit{K2}, \textit{TESS} (both long and short cadence), and four \textit{CHEOPS} transits (two per planet) (see Figure \ref{fig:K2-106_timeline}).

As for K2-79, we also tested the inclusion of a GP component to model possible correlated stellar activity in the RVs. Given the star's low chromospheric activity index ($\log R'_{\mathrm{HK}} = -5.04 \pm 0.19$; \citealp{Guenther2017}) and slow projected rotation ($v \sin i = 2.8 \pm 0.35 \,\mathrm{km\,s^{-1}}$; \citealp{Guenther2017}), the GP amplitude was expected to be small. The fitted amplitude was negligible, and the marginal likelihood comparison ($\Delta$BIC $\approx$ 8) indicated no significant improvement relative to the simpler white-noise model with per-instrument jitter terms. This outcome agrees with the conclusions of \citet{Guenther2017, Guenther2024}, who found that K2-106 is magnetically quiet and that activity-induced RV variations are below the measurement precision. Additionally inspection of the multi-instrument RV time series (Figure \ref{fig:K2-106b_transits} third row) shows no evidence for a long-term RV trend. We therefore adopted the non-GP solution for the final two planet joint fit.

Our fit confirms a circular orbit for K2-106\,b and a modestly eccentric orbit for K2-106\,c ($e_{\rm c} = 0.180_{-0.048}^{+0.038}$), consistent with previous findings by \citet{Guenther2024}. We derive updated masses and radii of $M_{\rm b}=7.54_{-0.64}^{+0.65}\,M_{\oplus}$, $R_{\rm b}=1.606_{-0.033}^{+0.034}\,R_{\oplus}$, and $M_{\rm c}=7.9\pm2.1\,M_{\oplus}$, $R_{\rm c}=2.712\pm0.071\,R_{\oplus}$, all consistent with parameters reported in \citet{Guenther2024}. The derived RV semi-amplitudes and inclinations agree well with previous results within uncertainties, and the fit provides significantly improved transit ephemerides.

Figure \ref{fig:K2-106b_transits} shows the phase-folded light curves for the \textit{K2} long cadence, \textit{TESS} long cadence, \textit{TESS} short cadence, and \textit{CHEOPS} data along with the best fitting transit model, as well as the orbital solution and RV residuals phased on the period of K2-106\,b for the combined fit with the HARPS-N, HIRES, ESPRESSO, HDS, HARPS, and PFS datasets. Figure \ref{fig:K2-106c_transits} shows the phase-folded light curves for the \textit{K2} long cadence and \textit{CHEOPS} data along with the best fitting transit model, orbital solution, and RV residuals phased on the period of K2-106\,c using the same combined RV dataset. The final two-planet joint model provides a good overall description of both the transit and RV datasets, with residuals consistent with white noise and no evidence for correlated structure (Figures \ref{fig:K2-106b_transits}, \ref{fig:K2-106c_transits}). The RMS scatter of the combined RV residuals ($\simeq 5\,\mathrm{m\,s^{-1}}$) is comparable to the median internal uncertainties, confirming that the per-instrument jitter terms adequately capture the remaining noise. The \textit{K2} and \textit{TESS} transit fits for planet b are precise and mutually consistent, while the \textit{CHEOPS} light curves for both planets exhibit noticeably higher scatter. This is expected given the small transit depths ($\approx$\,500--700\,ppm) and the fact that the host star's apparent magnitude places it near the limit where \textit{CHEOPS} achieves photon-noise-dominated precision. Despite the increased scatter, the \textit{CHEOPS} light curves clearly capture the predicted transits, and their measured depths are consistent with those observed by \textit{K2} and \textit{TESS}, supporting the robustness of the joint fit.

\subsection{K2-111}
\label{sec:K2-111}

\begin{figure}
    \centering
    \includegraphics[width=\linewidth]{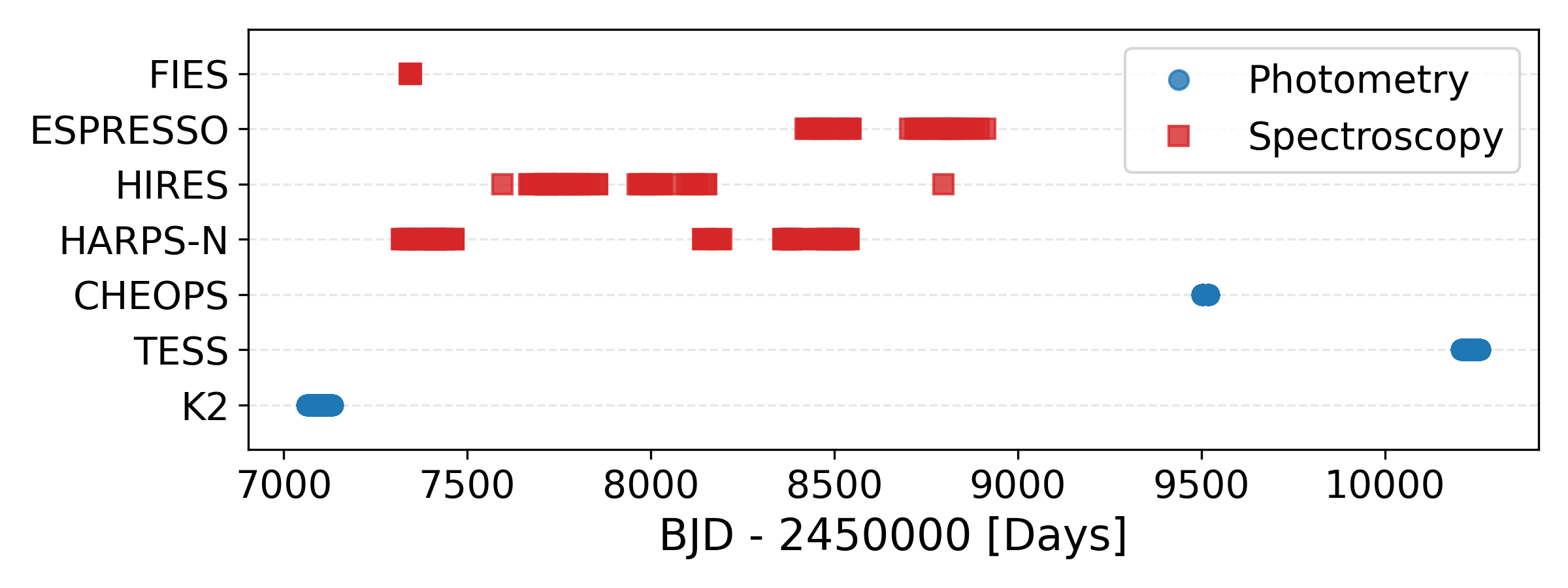}
    \caption{Timeline of photometric and spectroscopic observations of K2-111.}
    \label{fig:K2-111_timeline}
\end{figure}

K2-111 is a compact multi-planet system previously validated by \citet{Fridlund2017} and further characterised by \citet{Mortier2020}, \citet{Bonomo2023}, and \citet{Howard2025}. Our analysis expands substantially on these earlier works by combining HARPS-N, HIRES, ESPRESSO, and FIES RV measurements with photometry from \textit{K2}, \textit{TESS}, and two \textit{CHEOPS} visits, as shown by Figure \ref{fig:K2-111_timeline}. 

We determine an updated orbital period of $P_{\rm b}=5.3518963_{-0.000012}^{+0.0000089}$ days and a transit centre time of $T_{\rm 0, b} = 9519.1348_{-0.0054}^{+0.0038}$ (BJD~2450000), improving significantly over the \textit{K2}-only ephemerides of the previous studies. The planet's radius, $R_{\rm b}=1.837_{-0.071}^{+0.076}\,R_{\oplus}$ is consistent with earlier estimates from \citet{Mortier2020}. 
We recover a mass of $M_{\rm b}=5.35_{-0.69}^{+0.68}\,M_{\oplus}$, consistent with both previous estimates, and find a moderate eccentricity of $e_{\rm_b} = 0.134_{-0.071}^{+0.093}$, in agreement with the non-zero solution of \citet{Mortier2020}. Our joint RV model also includes the outer, non-transiting companion K2-111\,c, for which we measure a period of $P_{\rm c}=15.6748_{-0.0061}^{+0.0062}$ days, a minimum mass of $M_{\rm c}\sin i=11.1_{-1.0}^{+1.1}\,M_{\oplus}$, and a low eccentricity of $e_{\rm c}=0.069_{0.049}^{+0.072}$.

Figure \ref{fig:K2-111_transits} shows the phase-folded light curves for the \textit{K2} long cadence, \textit{TESS} long cadence, and \textit{CHEOPS} data along with the best fitting transit model. The orbital solutions and RV residuals are also shown phased on the periods of K2-111\,b and K2-111\,c from the two-planet combined fit using the HARPS-N, HIRES, ESPRESSO, and FIES RV datasets. The final two-planet solution provides an excellent fit to both the transit and RV data (Figure \ref{fig:K2-111_transits}). The RMS scatter of the combined RV residuals ($\approx\,4.5\,\mathrm{m\,s^{-1}}$) is comparable to the median measurement uncertainties, indicating that the per-instrument jitter terms adequately capture the remaining noise. For the inner transiting planet, K2-111\,b, the \textit{K2} and \textit{TESS} light curves give consistent transit depths and durations, while the \textit{CHEOPS} transits appear noisier owing to the planet's shallow depth and the host star's relative faintness. The RVs for the outer, non-transiting planet K2-111\,c are well modelled by a low-eccentricity Keplerian curve, with no residual structure suggesting additional companions or activity-related variability. Inspection of the multi-instrument RV time series (Figure \ref{fig:K2-111_transits} third row) likewise shows no evidence for a long-term trend or magnetic-cycle driven variations across the $\sim4$\,yr baseline. Overall, the joint two-planet white-noise model captures the available data well without requiring a correlated-noise component.

\subsection{K2-222}
\label{sec:K2-222}

\begin{figure}
    \centering
    \includegraphics[width=\linewidth]{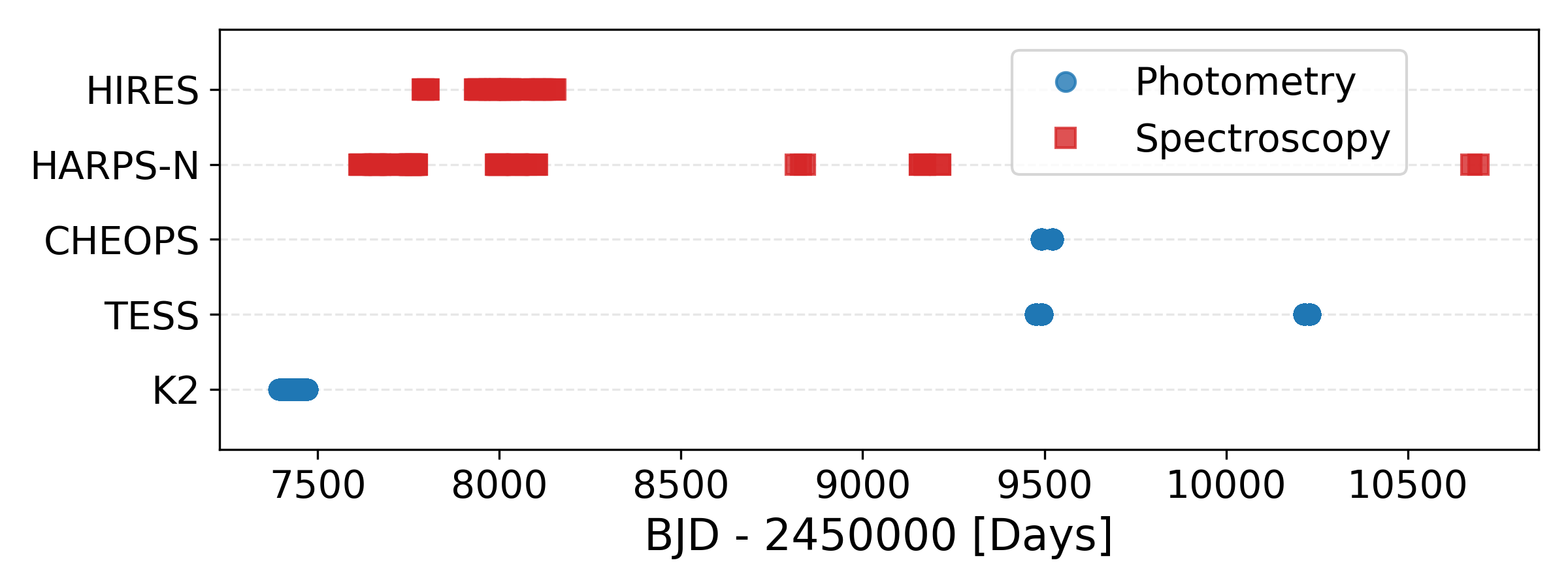}
    \caption{Timeline of photometric and spectroscopic observations of K2-222.}
    \label{fig:K2-222_timeline}
\end{figure}

K2-222\,b was originally discovered and validated using \textit{K2} photometry and HARPS-N RVs by \citet{Nava2022}. A second, longer-period RV signal was tentatively identified in their dataset, interpreted as a non-transiting outer companion, although they did not adopt the two-planet fit as their final solution due to the significant phase gaps in the coverage of the period of planet c.
\citet{Bonomo2023} re-analysed the system using a HARPS-N and \textit{K2} dataset and similarly adopted a single-planet solution, finding no statistically significant evidence for the outer companion. A subsequent one-planet analysis by \citet{Howard2025}, using \textit{K2} photometry and an independent RV dataset from HIRES and APF, did not recover the outer companion and adopted a single-planet model. In this study, we use an updated and expanded set of data, including two \textit{CHEOPS} visits, \textit{TESS} long and short cadence observations, and additional HARPS-N and HIRES RVs (see Figure \ref{fig:K2-222_timeline}).

We tested both single- and two-planet models. Our preferred solution adopts a two-planet model: K2-222\,b on a low-eccentricity orbit who eccentricity posterior is peaked at zero ($e_{\rm b} = 0.021_{-0.017}^{+0.051}$), consistent with being circular, and K2-222\,c as a non-transiting outer planet on a circular orbit. For K2-222\,b, we derive $P_{\rm b} = 15.3889274_{-0.000011}^{+0.0000099}$ days, $M_{\rm b} = 7.5 \pm 1.5\,M_{\oplus}$, and $R_{\rm b} = 2.275 \pm 0.10\,R_{\oplus}$. The outer companion is detected at $P_{\rm c} = 141.22_{-0.96}^{+2.9}$ days with a mass of $M_{\rm c}\sin i = 12.3_{-3.6}^{+3.3}\,M_{\oplus}$. The single planet fit, following a similar approach to \citet{Howard2025}, produced slightly higher estimates for $M_{\rm b} = 8.1 \pm 1.5\,M_{\oplus}$ and $K_{\rm b} = 2.18_{-0.41}^{+0.40}\,\mathrm{m\,s^{-1}}$, though these values are consistent with the two-Keplerian solution within 1$\sigma$. The elevated mass may reflect partial absorption of the longer-period RV signal into the single-Keplerian model. Comparing the BIC, the single-Keplerian model had a log-median BIC1$_{\rm pl}$ of $-268\,384$ whilst the two-Keplerian had a log-median BIC2$_{\rm pl}$ of $-268\,380$. From this the single-Keplerian model is preferred by $\Delta$BIC = BIC2$_{\rm pl}$ $-$ BIC1$_{\rm pl} = 4$, which corresponds to an approximate (natural-log) Bayes factor of $\Delta\ln \mathcal{B} \approx -\frac{1}{2} \Delta \mathrm{BIC} \approx -2$ in favour of the one-planet model. This constitutes only positive/weak evidence (not decisive; \citealp{KassRaftery1995, Trotta_2008}). Despite the marginal BIC preference, the two-planet solution provides a more physically interpretable description of the RVs: it removes the long-period structure from the residuals and avoids eccentricity inflation in the inner planet, whose eccentricity posterior is peaked at zero ($e_{\rm b} = 0.021_{-0.017}^{+0.051}$, consistent with circular), while returning parameters for b that are consistent with the one-planet fit within 1$\sigma$. We therefore adopt the two-planet non-GP model as the preferred solution, and report both fits for completeness. It remains consistent with the transit data and more fully captures the RV structure. Finally, to verify that the long-period signal attributed to K2-222\,c could not instead arise from an instrumental or astrophysical trend, we fitted an alternative model consisting of a single Keplerian for K2-222\,b plus a third-order polynomial term to the RVs. This polynomial captures any low-frequency curvature without invoking a second planet. The resulting maximum \textit{a posteriori} (MAP) BIC of $-268\,369$ was significantly less favourable than the two-Keplerian value ($-268\,383$), corresponding to $\Delta$BIC\,$\approx$\,14 in favour of the planetary model. The clear preference for the Keplerian description confirms that the 141-day modulation is best explained by a coherent, periodic signal rather than a smooth instrumental or stellar trend. This conclusion is further supported by the multi-year RV time series (Figure \ref{fig:K2-222_transits} third row), which shows no evidence for a long-term instrumental or stellar activity trend over the $\sim7$\,yr baseline. Our results for planet b are in good agreement with \citet{Nava2022} ($M_{\rm b} = 6.0 \pm 1.9\,M_{\oplus}$), though we find a smaller eccentricity in comparison to theirs ($e_{\rm b} = 0.23_{-0.12}^{+0.11}$). This discrepancy may reflect the tendency for RV fits to favour non-zero eccentricities in the presence of unmodelled noise or long-period signals, which are more fully accounted for in the two-planet fit. The derived mass and radius are also consistent within $1\sigma$ with the single-planet solution of \citet{Bonomo2023}.

Figure \ref{fig:K2-222_transits} shows the phase-folded light curves from the \textit{K2} long cadence, \textit{TESS} long and short cadence, and \textit{CHEOPS} observations along with the best fitting transit model. The corresponding orbital solutions and RV residuals, phased on the periods of K2-222\,b and K2-222\,c, are shown for the two-planet combined fit using the HARPS-N and HIRES RVs. The adopted two-planet non-GP model reproduces both the transit and RV datasets well. The RMS of the RV residuals ($\approx 3.5\,\mathrm{m\,s^{-1}}$) is consistent with the median instrumental uncertainties for both HARPS-N and HIRES, and no correlated structure remains once the second planet is included. The \textit{K2} and \textit{TESS} transits of K2-222\,b show consistent depths and durations, and the \textit{CHEOPS} light curves exhibit similar features, albeit with higher noise. The outer, non-transiting companion of K2-222\,c is well modelled by a low-amplitude, near-circular Keplerian signal with no indication of additional long-term trends. 

\subsection{K2-263}
\label{sec:K2-263}

\begin{figure}
    \centering
    \includegraphics[width=\linewidth]{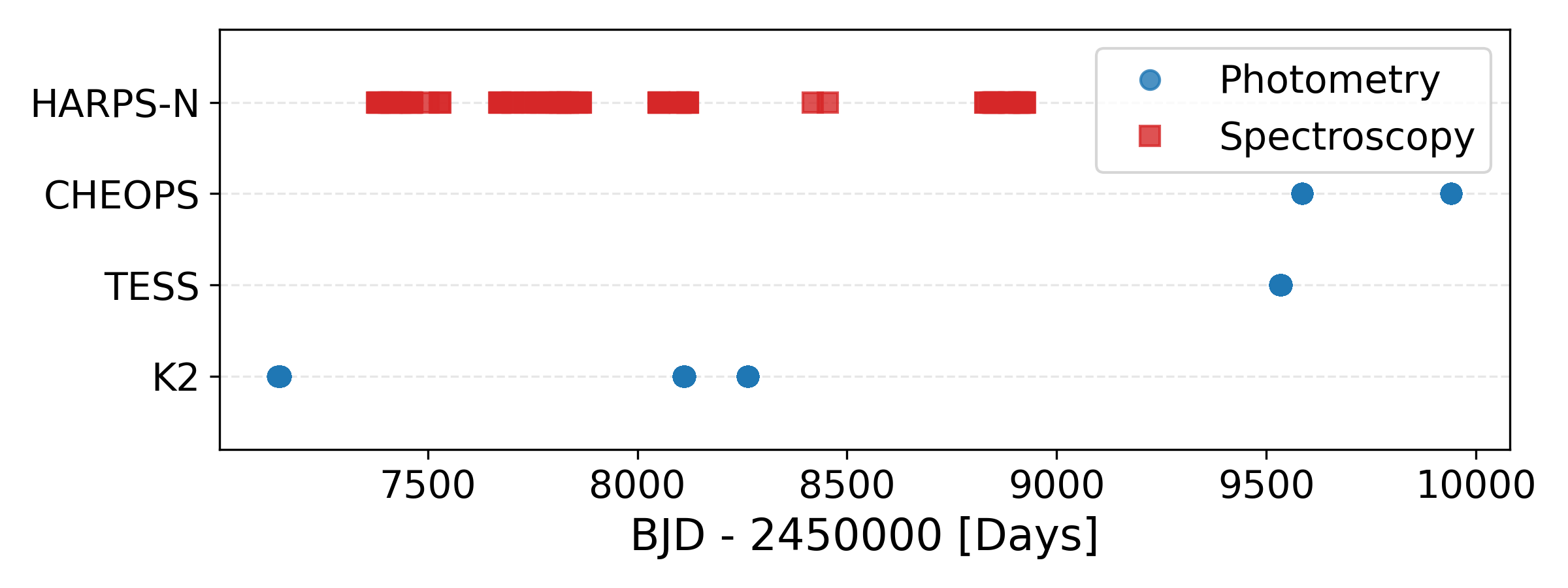}
    \caption{Timeline of photometric and spectroscopic observations of K2-263.}
    \label{fig:K2-263_timeline}
\end{figure}

K2-263\,b was first identified as a planet candidate by \citet{Mayo2016} and subsequently characterised by \citet{Mortier2018} and \citet{Bonomo2023}, based on \textit{K2} photometry and HARPS-N RVs. They reported a mass of $14.8 \pm 3.1\,M_{\oplus}$ and found no significant improvement when including a GP noise model in their analysis to mitigate stellar activity. In this study, we incorporate a significantly expanded dataset --- 29 further HARPS-N RVs, long- and short-cadence \textit{K2} photometry, four \textit{TESS} sectors, and two \textit{CHEOPS} transits, as shown in Figure \ref{fig:K2-263_timeline}.

We explored a suite of circular and eccentric models, both with and without a GP regression. In line with \citet{Mortier2018}, the inclusion of a GP did not materially change the inferred planetary parameters or improve the fit, and its higher BIC disfavours its additional complexity. Inspection of the HARPS-N RV time series (Figure \ref{fig:K2-263_transits} third row) likewise shows no evidence for a long-term instrumental or stellar trend across the $\sim4$\,yr baseline, supporting the adequacy of of the white-noise model without a GP component. Among the non-GP models, the eccentric solution is preferred over the circular one with a $\Delta$BIC corresponding to $\Delta \ln \mathcal{B} \approx 9$ \citep{KassRaftery1995, Trotta_2008}. Although the preference is modest, the eccentric model gives an eccentricity posterior that is peaked at a non-zero value and shows no significant differences in any other fitted parameters. We therefore adopt the eccentric, non-GP model as the final solution, deriving $P_{\rm b} =  50.819172_{-0.000068}^{+0.000066}$ days, $T_{\rm 0,b} = 9534.0662_{-0.0021}^{+0.0020}$ (BJD~2450000), $M_{\rm b} = 15.4_{-2.5}^{+2.4}\,M_{\oplus}$, and $R_{\rm b} = 2.513_{-0.095}^{+0.11}\,R_{\oplus}$, all consistent with but more precise than the values of \citet{Mortier2018}. The orbit is therefore mildly eccentric ($e_{\rm b} = 0.080_{-0.057}^{+0.089}$) and the inclination remains tightly constrained at $i_{\rm b} = 89.271_{-0.035}^{+0.045}$ degrees.

\subsection{TOI-1634}
\label{sec:TOI-1634}

TOI-1634 is a nearby M dwarf hosting at least one transiting planet (TOI-1634\,b), with a possible second, longer period planet (TOI-1634\,c) detected in RVs. The system has been previously characterised by multiple teams with notably different derived masses for TOI-1634\,b: \citet{Cloutier2021} reported $M_{\rm b} = 4.91_{-0.70}^{+0.68}\,M_{\oplus}$ based on HARPS-N data, while \citet{Hirano2021} found $M_{\rm b} = 10.14 \pm 0.95\,M_{\oplus}$ using IRD. A subsequent reanalysis by \citet{Luque2022} combining both datasets yielded an intermediate mass of $M_{\rm b} = 7.57 \pm 0.71\,M_{\oplus}$. Given these discrepancies, we conducted an extensive suite of model fits to explore the effects of dataset choice and planet multiplicity. Specifically, we tested both single- and two-Keplerian circular models using HARPS-N data alone, IRD data alone, and a combined HARPS-N + IRD dataset. In each two-Keplerian case, we incorporated the tentative long-period signal attributed to TOI-1634\,c. In total, we evaluated six fits: circular one- and two-Keplerian models for each of the three RV dataset combinations. Across these fits, we observed clear disagreement between the HARPS-N and IRD datasets. This is evident in the HARPS-N + IRD RV time series (Figure \ref{fig:TOI-1634_RVs} second row), where the IRD measurements exhibit both larger scatter and epoch-dependent offsets. The short $\sim7$ month baseline also provides no meaningful leverage on multi-year stellar or instrumental trends, limiting our ability to distinguish a smooth drift from a long-period Keplerian signal. The HARPS-N only single-Keplerian model yields a mass of $M_{\rm b} = 4.98_{-0.80}^{+0.79}\,M_{\oplus}$, closely matching \citet{Cloutier2021}, whereas the IRD only fit returns $M_{\rm b} = 11.3 \pm 1.0\,M_{\oplus}$, in line with \citet{Hirano2021}, whilst joint HARPS-N + IRD models produced intermediate masses ($\sim7.4\,M_{\oplus}$), in agreement with \citet{Luque2022}. However, the IRD dataset exhibits large scatter and unexplained systematic offsets that likely impacted the fits it was included in. 

\begin{figure}
    \centering
    \includegraphics[width=\linewidth]{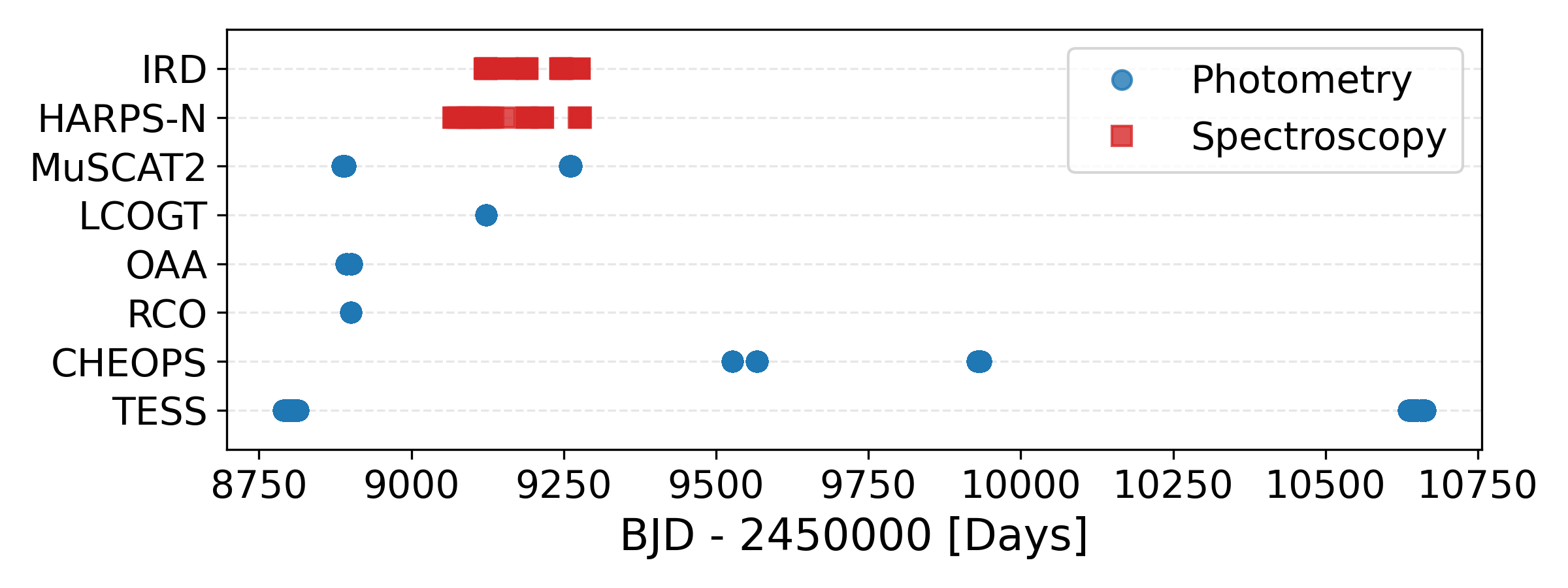}
    \caption{Timeline of photometric and spectroscopic observations of TOI-1634.}
    \label{fig:TOI-1634_timeline}
\end{figure}

We compiled a comprehensive photometric dataset for TOI-1634\,b, combining observations from a range of instruments. These include \textit{TESS} long and short cadence light curves, two transits observed with \textit{CHEOPS}, as well as ground-based photometry from MuSCAT2 in the $g$-, $r$-, $i$-, and $z_s$-bands, LCOGT, RCO, and OAA (at both 100\,s and 120\,s cadences) (see Figure \ref{fig:TOI-1634_timeline}). Although the long-baseline \textit{CHEOPS} DDT observations did not attain its aim of detecting transits of TOI-1634\,c, likely due to the large ephemeris uncertainty and an unfortunate timing gap in the observations, numerous transits of TOI-1634\,b (due to its ultra-short-period (USP)) were captured, which helped to refine its ephemeris.

We also tested a GP regression model using the HARPS-N, \textit{TESS}, and \textit{CHEOPS} datasets to assess the potential impact of stellar activity on the RVs. The MCMC chains converged well, but the GP amplitude was found to be low and the hyperparameters remained unconstrained, consistent with the findings of \citet{Cloutier2021}, who reported that a GP fit tended to absorb the planetary signal rather than model correlated stellar noise. The marginal likelihood comparison, quantified using the BIC, gave a difference of $\Delta\mathrm{BIC} \approx 3800$ in favour of the simpler, non-GP two-Keplerian model (log-median $\mathrm{BIC}_{\rm GP} = -274\,004$). This substantial difference, combined with the low stellar activity inferred from the absence of Ca~II~H\&K or photometric variability by \citet{Cloutier2021}, and the slow rotation and lack of flares in the \textit{TESS} data reported by \citet{Hirano2021}, supports the conclusion that a GP component is unnecessary.

We find that the HARPS-N RVs require an additional low-frequency component beyond the signal of TOI-1634\,b. When parametrised as a second Keplerian, this component corresponds to a period of order $\sim100$\,d, but with a low semi-amplitude that is not significantly different from zero. To test whether the long-period RV signal attributed to a possible TOI-1634\,c could instead arise from a smooth instrumental or astrophysical trend, we fitted an additional model consisting of a single Keplerian for TOI-1634\,b plus a third-order polynomial term in time. The resulting MAP BIC of $-970\,216.4$ was statistically indistinguishable from the two-Keplerian value ($-970\,215.8$), corresponding to $\Delta$BIC\,$\approx$\,$0.6$. The negligible difference indicates that both models fit the data equally well, providing no significant evidence for a second Keplerian signal. This suggests that the apparent long-period modulation can be explained by a smooth trend, and that the inclusion of TOI-1634\,c should be regarded as tentative pending further RV monitoring. A Bayesian general Lomb-Scargle (BGLS; \citealt{Boisse2011, mortieretal15}) periodogram of the HARPS-N RVs of TOI-1634 after subtraction of the signal of TOI-1634\,b shows a broad low-amplitude excess of power around $\sim100$\,d, but no sharply defined or statistically significant peak. This behaviour is consistent with a smooth low-frequency trend rather than a cohere periodic signal. We note that the mildly elevated \textit{Gaia} EDR3 RUWE ($\sim 1.23$; \citealt{Lindegren2021}) for TOI-1634 does not provide independent evidence for an additional companion as nearby M dwarfs frequently exhibit higher RUWE values for reasons unrelated to orbiting planets \citep{Sozzetti2023}. In the final single-planet model, TOI-1634\,b has $P_{\rm b} = 0.98934551_{-0.00000037}^{+0.00000036}$ days, $M_{\rm b} = 4.9 \pm 1.0\,M_{\oplus}$, and $R_{\rm b} = 1.759_{-0.057}^{+0.058}\,R_{\oplus}$. A Keplerian parametrisation of this low-frequency variability results in a characteristic period of $P \approx 102$\,d and semi-amplitude $K \approx 0.4\,\mathrm{m\,s^{-1}}$, but this should not be interpreted as a confirmed planetary detection. Independent transit-timing analyses of TOI-1634\,b \citep{Naponiello2025} similarly find no significant TTVs, consistent with the expectation that a companion on a wide, non-resonant orbit such as TOI-1634\,c would no induce detectable short-term timing variations.

Figure \ref{fig:TOI-1634_RVs} shows the orbital solutions and RV residuals of TOI-1634\,b and TOI-1634\,c, phased on their respective periods, from the combined two-planet fit using only the HARPS-N RVs. Figure \ref{fig:TOI-1634_transits} shows the phase-folded light curves for the \textit{TESS} long and short cadence, \textit{CHEOPS}, both long and short cadence OAA observations, LCOGT, RCO, and multi-band MuSCAT2 ($g$-, $r$-, $i$-, and $z_s$-bands) data along with the best fitting transit model. The transit fits across all instruments show excellent consistency in both depth and shape, with residuals consistent with photon noise, demonstrating systematics are well controlled. The HARPS-N RVs are well described by the adopted single-planet model, with a smooth low-frequency term, with low-amplitude residuals and no evidence for additional short-period structure. Overall, the combined photometric and spectroscopic fits provide a consistent and physically plausible description of the system. 

\section{Discussion}
\label{sec:discussion}

\subsection{Interior structure and M--R relations}
\label{sec:int_struc}
\begin{figure}
    \centering
    \includegraphics[width=\linewidth]{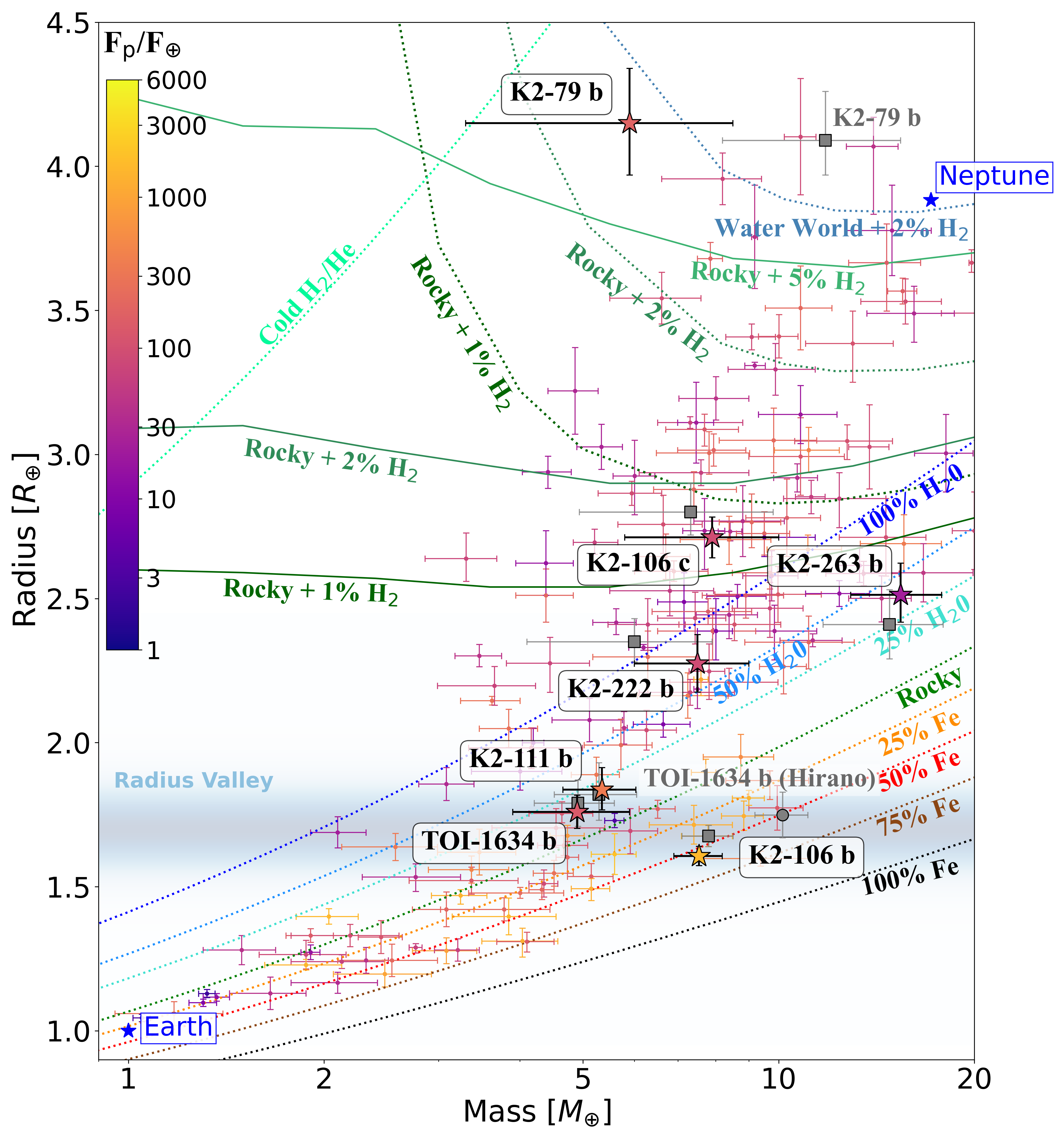}
    \caption{M--R diagram. Planet results from this study are shown as coloured stars, while archival literature values for the same planets (Table \ref{tab:stellar_params}) appear as grey symbols with error bars; squares for most planets and a circle for the alternative TOI-1634\,b value from \citet{Hirano2021}. Other confirmed planets with mass and radius precisions better than 20\% and 5\%, taken from \texttt{TepCat} \citep{TepCat}, are also plotted. Dotted lines show compositions from \citet{zeng_growth_2019} (1000\,K), and solid lines from \citet{Lopez_2014} (1 Gyr, solar metallicity, 1000\,$F_{\oplus}$). Points are colour-coded by incident flux. Earth, Neptune, and the approximate radius valley \citep{Fulton_2017} are indicated. Archival points are not individually labelled; exceptions are K2-79\,b and one of the TOI-1634\,b measurement, where the closest grey point corresponds to the \citet{Cloutier2021} solution.}
    \label{fig:M-R_plot}
\end{figure}

We place the mass and radius measurements of the transiting planets into context and compare them to composition models in Figure \ref{fig:M-R_plot}, which shows a mass-radius (M-R) plot for small planets ($R < 4.5$ $R_{\oplus}$ and $M <$ 20 $M_{\oplus}$). Confirmed planets, with a mass and radius estimated to a precision better than 20\% and 5\%, respectively, taken from \texttt{TepCat} \citep{TepCat} are also shown. Furthermore, the coloured lines in the M-R plot show different compositions, taken from
\citet{zeng_growth_2019}\footnote{Models are available online at \url{https://lweb.cfa.harvard.edu/~lzeng/planetmodels.html}} (dashed lines) and \citet{Lopez_2014} (solid lines). We include both models for completeness, but note that the H/He relations from \citet{zeng_growth_2019} are not appropriate for planets actively losing their atmospheres, as shown by \citet{Rogers2023}. While the 1000\,K \citet{zeng_growth_2019} curves remain useful for illustrating the broader sample, for K2-106\,c we adopt the \citet{Lopez_2014} models as the default, since they assume a fixed system age and therefore better capture planets undergoing mass loss.

As a next step, we used the planetary properties derived from the combined transit and RV fits to model the internal structure of the transiting planets. This was done using the \texttt{plaNETic} framework\footnote{\url{https://github.com/joannegger/plaNETic}} \citep{Egger+2024}, which uses a neural network trained on the \texttt{BICEPS} forward model \citep{Haldemann+2024} as a fast surrogate model in combination with a full grid accept-reject sampling scheme. Each planet is assumed to consist of a volatile layer (H, He, H$_2$O), a mantle layer (SiO$_2$, MgO, FeO) and an inner core (Fe, FeS). 

There is a high degeneracy when inferring the internal structure of exoplanets based on their mean density alone. The resulting posterior distributions will therefore to some extent depend on the chosen priors. We run here six different models for each planet, combining different prior options for the planetary Si/Mg/Fe ratios with prior options for the envelope compositions. More specifically, explore three assumptions for the planetary Si/Mg/Fe ratios: (1) that they match those of the host star exactly \citep[following e.g.][]{Thiabaud+2015}; (2) that the planets are iron-enriched relative to their host star, using the empirical relation of \citet{Adibekyan+2021}; and (3) that the planetary Si/Mg/Fe ratios are allowed to vary freely, independent of the stellar abundances. In addition, we consider two possibilities for the volatile envelope: (A) a water-enriched composition, consistent with formation beyond the ice line, and (B) a primarily H/He-dominated envelope, consistent with formation inside the ice line. Further details on these priors can be found in \cite{Egger+2024}. The resulting constraints for the transiting planets studied here are summarised in the following.

\subsubsection{K2-79 b}
\label{sec:K2-79b}


K2-79\,b, with a measured mass of $M_{\rm b} = 5.9 \pm 2.6\,M_{\oplus}$ and radius of $R_{\rm b} = 4.15_{-0.18}^{+0.19}\,R_{\oplus}$, lies well above the bulk of the small-planet population in Figure \ref{fig:M-R_plot}, occupying the low-density sub-Saturn regime. Its resulting bulk density of $\rho_{\rm b} \approx 0.45 \mathrm{g\,cm^{-3}}$ indicates the presence of a substantial volatile-rich envelope. While \citet{Nava2022} reported a somewhat higher mass and slightly smaller radius, the updated parameters presented here place K2-79\,b even father from rocky or water-world compositions, reinforcing its classification as a highly inflated low-mass planet.

Our interior modelling confirms that a wide range of compositions remain viable. Under water-rich priors (prior~A), we infer envelope mass fractions spanning from a few per cent up to $\sim$\,60\%, consistent with envelopes containing a broad range of water mass fractions. For the water-poor prior, the models favour H/He-dominated atmospheres with envelope mass fraction of $6^{+2}_{-3}$\%, in agreement with expectations from formation and evolution scenarios for low-density sub-Saturns. Given the planet's high irradiation ($F_{\rm b} \approx 180\,F_{\oplus}$) and low surface gravity ($g_{\rm b} \approx 0.34\, g_{\oplus}$), such an H/He envelope would be susceptible to long-term escape unless supported by heavier volatiles, whereas a water-rich composition naturally gives a more resilient atmosphere of Gyr timescales.

Transmission spectroscopy with the \textit{James Webb Space Telescope} \citep[\textit{JWST};][]{JWST} or the \textit{Atmospheric Remote-sensing Infrared Exoplanet Large-survey} \citep[\textit{Ariel};][]{Tinetti2018} could help discriminate between these possibilities. A hydrogen-rich atmosphere is expected to produce spectral modulations of $\Delta d \sim$230--240\,ppm over five scale heights, while a more water-enriched enveople would reduce the signal to $\sim$\,130\,ppm, as also noted by \citet{Nava2022}.
 
\subsubsection{K2-106 b \& c}
\label{sec:K2-106}


K2-106\,b, with a mass of $M_{\rm b} = 7.54_{-0.64}^{+0.65}\,M_{\oplus}$ and radius $R_{\rm b} = 1.606_{-0.033}^{+0.034}\,R_{\oplus}$, lies well inside the rocky region of the M--R diagram (Figure \ref{fig:M-R_plot}). Its bulk density of $\rho_{\rm b} \approx 9.3\,\mathrm{g\,cm^{-3}}$ places it among the densest known USP planets, overlapping with the compositional curves for rocky or iron-rich interiors. Although its position on the M--R diagram overlaps the rocky and iron-rich composition curves, this does not uniquely imply an Earth-like core-mantle fraction: such a structure is only permitted in our interior modelling if the mantle is enriched in iron relative to solar-composition silicates. Degeneracies in interior structure models \citep{Hakim2018, Suissa2018} mean that a range of refractory compositions are possible, particularly given the extreme surface conditions expected for such a highly irradiated ($F_{\rm b} \approx 4620\,F_{\oplus}$ and short-period ($P_{\rm b} = 0.57130782_{-0.00000057}^{+0.00000056}$ days) planet.

Our interior modelling further supports the interpretation of K2-106\,b as a largely atmosphere-free core. In the water-poor case, reproducing the observed density requires envelope mass fractions below $10^{-6}$, far too small for a low-mean-molecular-weight (MMW) atmosphere to survive long-term photoevaporation. This strongly indicates that the planet is effectively a bare core. In the water-rich case, fixing the planetary Si/Mg/Fe ratios to stellar abundances fails to reproduce the observed density, consistent with \citet{RodriguezMartinez+2023}, who argue that K2-106\,b cannot simultaneously host a steam atmosphere and reflect the refractory abundances of its host star. Allowing the interior elemental ratios to vary does permit thin steam envelopes with mass fractions of $0.3^{+0.7}_{-0.2}$\% and $0.4^{+0.8}_{-0.3}$\% for priors A2 and A3, respectively. However, even these very small volatile layers may struggle to remain stable against hydrodynamic escape over Gyr timescales given the planet's extreme irradiation. A more detailed atmospheric-loss calculation is beyond the scope of this work, but the plausibility of such long-lived steam envelopes should be approached with caution.

Despite its high density, K2-106\,b could still host a transient or hybrid atmosphere generated by surface outgassing or a magma ocean \citep{Ito2015, Bower2022}. Such atmospheres would be dominated by heavy species (e.g. CO, SiO) and may be detectable through infrared spectroscopy or thermal phase-curve measurements with \textit{JWST} or future high-resolution instruments such as the \textit{CRyogenic InfraRed Echelle Spectrograph} Upgrade project \citep[\textit{CRIRES+};][]{crires}.

K2-106\,c has a similar mass to K2-106\,b but a substantially larger radius ($R_{\rm b} = 2.712 \pm 0.071\,R_{\oplus}$), placing it above the radius valley and implying the presence of a significant volatile envelope. In the water-poor case, we infer tightly constrained H/He-dominated envelope mass fractions of $1.5\pm0.3$\%, $1.9\pm0.5$\%, and $1.8^{+0.7}_{-0.6}$\% for priors B1, B2, and B3, respectively. In the water-rich scenario, a broad range of compositions remains possible. Its location on M--R diagram is compatible with either a sub-Neptune-like H/He envelope of $\sim$1-2\% by mass (consistent with \citet{Lopez_2014}) or a water-rich composition \citep{zeng_growth_2019}.

K2-106\,c receives a moderate flux ($F_{\rm b} \approx 70\,F_\oplus$), close to the empirical threshold where atmospheric escape becomes efficient over Gyr timescales \citep{Lopez_2013, Owen2017, Fulton2018}. The contrasting atmospheric outcomes of planets b and c --- nearly identical in mass yet markedly different in radius --- are naturally explained by atmospheric loss mechanisms such as XUV-driven escape or core-powered mass loss \citep{Lopez_2014, Ginzburg_2018}, similar to the interpretation proposed for other compact multi-planet systems spanning the radius valley, such as GJ 9827 \citep{Rice2019}.

Together, the K2-106 planets provide a valuable test case for models of atmospheric evolution: K2-106\,b likely represents an exposed rocky core, while K2-106\,c has retained a volatile envelope. Future observations with \textit{JWST} or high-resolution spectroscopy will be crucial for distinguishing between H/He- and H$_2$O-rich scenarios for K2-106\,c and for probing any residual atmosphere or surface vapour on K2-106\,b.

\subsubsection{K2-111 b}
\label{sec:K2-111b}

K2-111\,b resides just above the radius valley and falls between the 50\% H$_2$O and rocky composition curves in Figure \ref{fig:M-R_plot}. Its bulk density of $\rho_{\rm b} \approx 4.7\,\mathrm{g\,cm^{-3}}$ is lower than that of a iron-free rocky composition, indicating that K2-111\,b cannot be explained as a bare silicate core and must retain at least modest volatile envelope. At the same time, its high incident flux ($F_{\rm b}\approx 490\,F_\oplus$) and likely old age imply that any long-lived atmosphere cannot be dominated by low-MMW species, which would be efficiently removed by photoevaporation. This places the planet in a particularly informative region of the M--R diagram: too dense to be a stripped rocky core, yet too irradiated to host a primordial H/He envelope. 

Our interior modelling reflects this tension. For water-poor priors, we obtain extremely small envelope mass fractions of $10^{-6}-10^{-5}$, which would be rapidly lost to hydrodynamic escape and are therefore not viable long-term solutions. In contrast, the water-rich priors allow steam-dominated atmospheres with mass fractions of $10^{+8}_{-6}$\%, $13^{+9}_{-7}$\%, and $16^{+10}_{-9}$\% for priors A1, A2, and A3. These heavier, higher-MMW envelopes are far more resistant to atmospheric escape, and their presence would naturally explain the observed density.

This interpretation is further supported by the chemical properties of the host star. K2-111 is iron-poor but $\alpha$-enhanced ([$\alpha$/Fe] = 0.27; \citealp{Mortier2020}), which can favour lower planetary core densities for a given mass \citep{Santos2017}. A steam-dominated envelope atop a composition influenced by the star's non-solar refractory ratios therefore provides a consistent explanation for the planet's bulk properties. The requirement for heavier volatiles closely parallels the case of TOI-238\,b, where an evaporation analysis showed that a steam atmosphere can indeed remain stable over Gyr timescales despite high irradiation \citep{Egger2025}. A similar analysis for K2-111\,b would likely confirm that a water-rich or mixed volatile envelope is able to survive long-term atmospheric escape, whereas a H/He-dominated envelope would not.

Taken together, these results suggest that K2-111\,b is best described as a rocky planet that retains a significant steam-rich envelope shaped both by its host star's chemistry and by strong atmospheric escape over its lifetime.

\subsubsection{K2-222 b}
\label{sec:K2-222b}

K2-222\,b lies above the radius valley in Figure \ref{fig:M-R_plot}, with a bulk density of $\rho_{\rm b} \approx 3.5\,\mathrm{g\,cm^{-3}}$, which is significantly lower than that of a purely rocky composition, indicating the presence of a volatile-rich envelope. Using the revised orbital parameters, we derive an incident flux of $F_{\rm b} \approx 67\,F_\oplus$, lower than the $\sim 95\,F_\oplus$ estimated by \citet{Nava2022}. The discrepancy stems from differing luminosities: their value is model-derived, whereas ours is scaled from the measured stellar radius and effective temperature. The updated flux places K2-222\,b closer to the regime where long-term atmospheric retention is plausible, though not guaranteed.

Our interior modelling allows both water-rich and water-poor solutions. Under the water-rich priors (prior~A), we infer high-MMW envelopes with mass fractions $35^{+11}_{-15}$\% and water mass fractions between 94--100\%. These solutions correspond to planets with substantial steam-rich or mixed volatile atmospheres. In contrast, the water-poor priors result in very low-mass H/He-dominated envelopes with mass fractions of $0.2^{+0.2}_{-0.1}$\%, $0.3^{+0.3}_{-0.2}$\%, $0.3^{+0.4}_{-0.2}$\% for priors B1--B3. Given K2-222\,b's high flux and moderate surface gravity ($g_{\rm b} \sim 1.44\,g_\oplus$), such extremely small H/He layers lie near the threshold of long-term stability and may be vulnerable to photoevaporation.

This places K2-222\,b in a boundary region similar to K2-111\,b: too large to be a bare rocky core, but potentially unable to retain a low-MMW atmosphere over Gyr timescales. If the tenuous H/He envelopes allowed by the water-poor priors are unstable, the planet would require a higher-MMW (e.g. steam-rich) envelope to survive to the present day. Such water-rich compositions also remain fully consistent with its mass and radius, as show by the \citet{zeng_growth_2019} models.

\subsubsection{K2-263 b}
\label{sec:K2-263b}

K2-263\,b is a temperate, intermediate-mass planet whose updated properties reinforce its classification as a cold sub-Neptune with a mixed ice-rock interior with a bulk density of $\rho_{\rm b} \approx 3.5\,\mathrm{g\,cm^{-3}}$. On the M--R diagram, the planet lies between 25\% and 50\% H$_2$O composition curves, consistent with a substantial volatile or ice-rich interior. This is in line with the interpretation of \citet{Mortier2018}, who suggested that K2-263\,b likely contains comparable fractions of rocky and icy material, pointing to formation beyond the snow line followed by inward migration.

Our interior modelling supports this picture and allows both water-rich and water-poor envelope compositions. Under the water-rich priors (A1--A3), K2-263\,b is compatible with steam-dominated envelopes with mass fractions of $25^{+12}_{-11}$\%, $31^{+11}_{-14}$\%, and $26^{+13}_{-12}$\%. Such solutions correspond to high-MMW atmospheres overlaying a substantial ice-rock interior. Meanwhile, the water-poor priors (B1--B3) permit H/He-dominated envelopes with mass fractions of $0.6^{+0.5}_{-0.3}$\%, $0.7^{+0.6}_{-0.5}$\%, and $0.8^{+0.8}_{-0.6}$\%. These thin, hydrogen-rich envelopes also reproduce the observed density due to the planet's comparatively large solid interior.

With a long orbital period, low incident flux ($F_{\rm b} \sim 8.1\,F_{\oplus}$), and an equilibrium temperature of $T_{\rm eq,b} \approx 480$\,K, K2-263\,b occupies the cool end of the sub-Neptune population. It is comparable in mass and radius to Kepler-131\,b \citep{Marcy2014} and HD 106315\,b \citep{Barros2017, Crossfield2017}, but stands out for its significantly lower irradiation and higher mean density. These characteristics make K2-263\,b a valuable analogue for the predicted population of cold, water-rich super-Earths and an important benchmark for models of planet formation, migration, and volatile retention in the low irradiation regime above the radius valley. K2-263\,b therefore joins the emerging population of sub-Neptunes with $T_{\rm eq} < 600$\,K that are likely to possess water-rich compositions around FGK dwarfs (e.g. the cold sub-Neptune candidates discussed in \citet{Bonomo2025}). This group remains sparsely sampled, and K2-263\,b provides a valuable addition that helps anchor the low-irradiation end of the M-R distribution.

\subsubsection{TOI-1634 b}
\label{sec:TOI-1634b}

TOI-1634\,b occupies a compositionally intriguing region of the M--R diagram, lying just above the 100\% silicate (iron-free) bare-core line and near the upper edge of the radius valley. With updated parameters, the planet has a bulk density of $\rho_{\rm b} \approx 4.9\,\mathrm{g\,cm}^{-3}$. This value is lower than that of a pure silicate composition, indicating that TOI-1634\,b cannot be explained as a bare rocky core and must retain at least a small volatile component or a non-Earth-like refractory mixture. Given its intense isolation ($F_{\rm b} \approx 121\,F_\oplus$) and USP-like characteristics, TOI-1634\,b occupies a transitional regime where atmospheric loss is expected to be strong, making it a key system for understanding the mechanisms that shape the radius valley.

Our interior modelling reflects this balance between density and evaporation. For the water-poor priors, we obtain extremely small H/He envelope mass fractions, which would be rapidly removed by photoevaporation and are therefore not viable long-term solutions. In contrast, the water-rich priors result in steam-dominated envelopes with mass fractions of $17^{+11}_{-9}$\%, $12^{+10}_{-7}$\%, and $22^{+11}_{-11}$\% for priors A1--A3. These higher-MMW envelopes are far more stable against hydrodynamic escape and naturally account for the planet's position above the pure-silicate line. As with K2-111\,b, TOI-1634\,b appears to lie in a regime where only heavier volatile envelopes --- no primordial hydrogen --- can survive over Gyr timescales.

These findings help place the planet within the broader context of previous work. \citet{Cloutier2021} suggested that an extended H/He envelope could explain the radius, but argued that such an atmosphere would be highly vulnerable to photoevaporation. \citet{Hirano2021}, using IRD RVs, instead found a density consistent with a purely rocky interior. Overall, TOI-1634\,b represents an important benchmark for understanding USP planets near the radius valley, highlighting the interplay between the composition, irradiation-driven atmospheric escape, and the requirement for higher-MMW volatiles to maintain even a thin envelope over the system's lifetime.

\subsection{Prospects for mission planning}
\label{sec:HWO}

An accurate knowledge of an exoplanet's transit ephemeris is essential for planning future time-critical observations, for example \citet{Dragomir2020} showed that 81\% of \textit{TESS} planet ephemerides will have mid-transit uncertainties exceeding 30\,min after just one year. For high-demand facilities, such as \textit{JWST} and the upcoming \textit{Ariel} and \textit{HWO}, maintaining precise ephemerides is critical to avoid wasted observing time and ensure full transit coverage. For many transiting systems discovered by missions like \textit{Kepler} and \textit{K2}, formal uncertainties in orbital period and transit epoch propagate into substantial timing uncertainties when extrapolated into the mid-2040s, when \textit{HWO} is due to launch \citep{Ikwut2020}. Without ephemeris updates, many promising targets will have transit windows too uncertain to schedule efficiently, without significant and time-critical re-analysis, which is often infeasible close to observation deadlines. In this study, we demonstrate that combining all published transit observations, especially with the addition of just two recent \textit{CHEOPS} transits, for a given planet can significantly reduce these uncertainties, in many cases restoring sub-hour precision at the expected launch of \textit{HWO} (assumed here as $\sim$ BJD 2467981.50, i.e. $\sim$2045).

\begin{table*}
    \centering
    \caption{Summary of improvements in transit ephemerides for all transiting planets in this sample. Columns show the original, \textit{TESS}-updated, and \textit{TESS}+\textit{CHEOPS}-updated period uncertainties ($P_{\rm err}$) and corresponding transit time uncertainties ($T_{0,{\rm err}}$) propagated to 2045 January 01 (BJD~2467981.50). The factors $\sigma P$ and $\sigma T$ quantify the improvement relative to the preceding fit stage, and the final three columns list the ROP (return on precision), defined as seconds of uncertainty reduced per hour of new observing time for \textit{TESS}, \textit{CHEOPS}, and the combined dataset, respectively.}
    \resizebox{\linewidth}{!}{
    \begin{tabular}{llllllllllllll}
         \hline\hline \vspace{1mm}
         Planet & $P_{\rm err}$ [s] & $P_{\rm err}$ [s] & $P_{\rm err}$ [s] & $\sigma P_{TESS}$ & $\sigma P_{CHEOPS}$ & $T_{0_{\rm err}}$ [hr] & $T_{0_{\rm err}}$ [hr] & $T_{0_{\rm err}}$ [hr] & $\sigma T_{TESS}$ & $\sigma T_{CHEOPS}$ & ROP$_{TESS}$ & ROP$_{CHEOPS}$ & ROP$_{\rm total}$\\ \vspace{1.5mm}
          & (Literature) & (+\textit{TESS}) & (+\textit{CHEOPS}) &  &  & (Literature) & (+\textit{TESS}) & (+\textit{CHEOPS}) &  &  & [$\mathrm{s\,hr}^{-1}$] & [$\mathrm{s\,hr}^{-1}$] & [$\mathrm{s\,hr}^{-1}$] \\
         \hline
         \vspace{1.5mm}
         K2-79\,b & 33.70 & 2.89 & 1.25 & 11.64 & 2.31 & 9.26 & 0.62 & 0.27 & 14.87 & 2.31  &  12.00 & 66.42 & 12.40 \\
         \vspace{1.5mm}
         K2-106\,b & 0.48 & 0.05 & 0.05 & 9.65 & 1.07 & 2.45 & 0.21 & 0.20 & 11.69 & 1.07 & 6.21 & 2.06 & 6.14 \\
         \vspace{1.5mm}
         K2-106\,c & 59.62 & 59.62 & 1.81 & 1.00 & 40.00 & 13.13 & 13.13 & 0.33 & 1.00 & 40.03 & 0.0 & 1470.89 & 34.71 \\
         \vspace{1.5mm}
         K2-111\,b & 34.56 & 0.90 & 0.90 & 38.28 & 1.00 & 19.52 & 0.41 & 0.41 & 47.45 & 1.00 & 53.08 & 0.00 & 51.74 \\
         \vspace{1.5mm}
         K2-222\,b & 76.03 & 1.08 & 0.90 & 70.40 & 1.20 & 14.52 & 0.17 & 0.14 & 86.12 & 1.21 & 26.58 & 5.72 & 26.39 \\
         \vspace{1.5mm}
         K2-263\,b & 8.12 & 9.50 & 5.79 & 0.85 & 1.64 & 0.44 & 0.45 & 0.27 & 0.98 & 1.64 & -0.01 & 24.51 & 0.23 \\
         TOI-1634\,b$^{\dagger}$ & 1.30 & 0.04 & 0.03 & 37.50 & 1.10 & 3.34 & 0.08 & 0.07 & 40.94 & 1.10 & 9.06 & 0.20 & 8.27 \\
         \hline
         \multicolumn{14}{@{}p{\dimexpr0.85\linewidth-2\tabcolsep}@{}}{Note: $^{\dagger}$ TOI-1634\,b was initially validated with \textit{TESS} and ground-based photometry rather than \textit{K2}; the second column corresponds to the improvement from an additional \textit{TESS} sector, and the third column includes the final \textit{CHEOPS} observations.}

    \end{tabular}
    }
    \label{tab:transit_improv}
\end{table*}

\begin{figure}
    \centering
    \includegraphics[width=\linewidth]{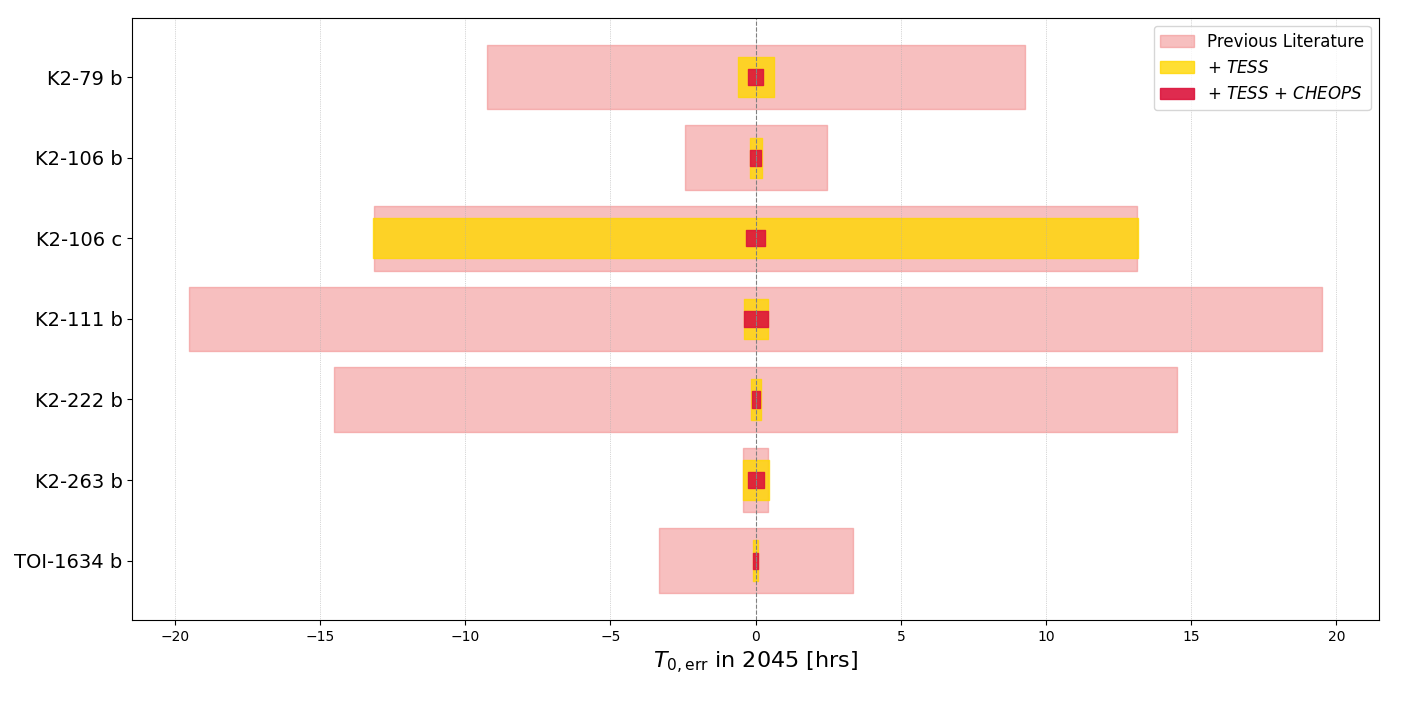}
    \caption{Projected absolute transit time uncertainties ($T_{\mathit{0},\, err}$) at the 2045 epoch for the seven transiting planets analysed in this study. Each bar represents the 1$\sigma$ timing window centred on the nominal transit mid-time. Light red bars correspond to the original transit ephemerides published in the literature (source papers referenced in Table \ref{tab:stellar_params}), yellow bars show the improvement obtained after adding the \textit{TESS} transits, and the dark red bars indicate the final precision after including \textit{CHEOPS} observations.}
    \label{fig:transit_improv}
\end{figure}

We analysed seven transiting planets with archival and new photometry spanning \textit{K2}, \textit{TESS}, \textit{CHEOPS}, and other ground-based sources, producing updated ephemerides via a global fit of all available transit times. For each planet, three sequential fits were performed: (1) using the original discovery data only, (2) adding \textit{TESS} transits, and (3) subsequently adding \textit{CHEOPS} transits. This stepwise approach isolates the specific contribution of each mission to the final precision achieved. The resulting improvements are shown in Figure \ref{fig:transit_improv}, which presents the absolute transit time uncertainties in hours for each planet in $\sim$\,2045, comparing the original predictions (source papers referenced in Table \ref{tab:stellar_params}) with the ones from this study. 

To estimate the projected transit time uncertainty at that epoch, we use:

\begin{equation}
    \sigma_T = \sqrt{\sigma_{T_{0}}^2 + (N \cdot\sigma_P)^2}
    \label{eqn:uncertainty}
\end{equation}

where $\sigma_{T_{0}}$ corresponds to the error on the transit centre time, $\sigma_P$ corresponds to the error on the orbital period, and $N$ is the number of orbits between the reported transit centre time and the 2045 epoch (calculated as $N = (T_{2045} - T_0)/P$). 

To quantify the efficiency of these ephemeris refinements, we also introduce a metric for return on precision (ROP), defined as the number of seconds of future timing uncertainty reduced per hour of new observing time:

\begin{equation}
    \text{ROP}\,[\mathrm{s\,hr}^{-1}] = 3600 \times\frac{T_{0_{\rm err,\, 1}} - T_{0_{\rm err,\,2}}}{t_{\rm obs}}
\end{equation}

where $T_{0_{\rm err,\, 1}}$ and $T_{0_{\rm err,\,2}}$ are the predicted transit time uncertainties before and after the addition of new observations, respectively, and $t_{\rm obs}$ is the total duration in hours of the new observing data included in that stage. We computed three related quantities: (1) ROP$_{TESS}$ --- the improvement achieved by adding \textit{TESS} data relative to the original literature ephemeris, (2) ROP$_{CHEOPS}$ --- the improvement achieved by subsequently adding \textit{CHEOPS} data relative to the \textit{TESS}-only fit, and (3) ROP$_{\rm total}$ --- the cumulative improvement from both missions combined. This formulation allows the relative contributions of each mission to be separated, while still expressing their combined impact on long-term predictability. Both missions extend the temporal baseline and provide additional transits that constrain $P$ and $T_0$. More recent observations from \textit{CHEOPS} are particularly valuable for this reason, while \textit{TESS} often provides the dominant improvement for systems where multiple transits were observed per sector. A full numerical summary of the results is provided in Table \ref{tab:transit_improv}. 

Planets with relatively long orbital periods or limited early baselines benefited the most from these multi-mission ephemeris refinements, as their timing uncertainties grow fastest when extrapolated into the future. For example, K2-222\,b originally had a predicted uncertainty of $\sim$\,14.5 hours in 2045, which improved by a factor of $\sim$\,70$\times$ after including \textit{TESS} data alone, and by a further $\sim$\,1.2$\times$ once the new \textit{CHEOPS} transits were added. This corresponds to an overall ROP of 26.39\,$\mathrm{s\,hr}^{-1}$, meaning that for every hour of new observing time from \textit{TESS} and \textit{CHEOPS} combined, $\sim$\,26 seconds of future scheduling are recovered --- amounting to just over 14 hours saved in total. K2-79\,b showed a similar trend, improving by $\sim$\,12$\times$ after adding \textit{TESS} and then an additional $\sim$\,2.3$\times$ after incorporating the \textit{CHEOPS} transits, highlighting the cumulative impact of combining wide-field survey data with targeted follow-up. For K2-263\,b, only a single \textit{TESS} transit was detected across four sectors due to its 50.8-day period. The limited coverage and lower timing precision led to a slight degradation in the ephemeris relative to the \textit{K2}-only solution. The addition of two high-quality \textit{CHEOPS} transits, however, recovered and modestly improved the overall precision, yielding a new $\sim$\,1.6$\times$ gain. Given the planet's extensive prior \textit{K2} coverage (one long-cadence and two short-cadence campaigns) the projected timing uncertainty by 2045 was already small, so the absolute improvement time remains marginal. This case illustrates how, for well-constrained long-period systems, \textit{CHEOPS} can refine ephemerides where \textit{TESS} contributes little due to sparse transit sampling. K2-106\,b was already published with one \textit{TESS} sector \citep{Guenther2024}, so the addition of a single new \textit{TESS} sector only modestly improved the precision (period uncertainty improved by $\sim$\,9.7$\times$, and the projected 2045 timing uncertainty decreased from $\sim$\,2.45 hours to $\sim$\,0.21 hours). In contrast, K2-106\,c saw no improvement from the \textit{TESS} data, as all transits fell within sector data gaps; the full $\sim$\,40$\times$ enhancement in period precision arises directly from the dedicated \textit{CHEOPS} visits. This clearly demonstrates the importance of dedicated follow-up even for targets nominally covered by wide-field surveys. This stepwise comparison illustrates that while \textit{TESS} generally provides the largest single boost to ephemeris precision, \textit{CHEOPS} plays a critical role in maintaining and restoring sub-hour predictability for targets otherwise missed or poorly sampled by \textit{TESS}.

Without updated ephemerides, these targets would have been unsuitable for transmission spectroscopy, high-contrast imaging, or precision scheduling in the extrapolated future --- such as when \textit{HWO} is expected to be operational. K2-111\,b, whose projected timing uncertainty at the launch of the \textit{HWO} era was $\sim$\,20 hours based on its original \textit{K2}-only ephemeris, would have posed a significant challenge for precise scheduling. By incorporating all available data, this was reduced to just $\sim$\,0.41 hours (a $\sim$\,47$\sigma$ improvement). This corresponds to an ROP$_{\rm total}$ of 52\,$\mathrm{s\,hr}^{-1}$, meaning that each hour of follow-up observation recovered roughly 52 seconds of future scheduling uncertainty. Although the gain per hour is modest due to the large cumulative observing due to \textit{TESS}, the overall 19-hour reduction in projected uncertainty makes the planet far more viable for time-critical follow-up during the \textit{HWO} era. For TOI-1634\,b, which was initially validated using \textit{TESS} and ground-based photometry, adding one additional \textit{TESS} sector modestly reduced the projected 2045 timing uncertainty, and the \textit{CHEOPS} observations further improved it by $\sim$\,12\%. This demonstrates that even recently discovered systems benefit from regular ephemeris maintenance to maintain sub-hour predictability over multi-decade baselines, particularly as \textit{TESS} is expected to cease Northern Hemisphere observations in the near future. As shown in Figure \ref{fig:transit_improv}, the improvements in orbital period uncertainty ($\sigma_P$) directly translate into long-term gains, since timing error increases linearly with time due to $N \sigma_P$ accumulation across orbits. Overall, the updated period uncertainties in this study are typically reduced by roughly an order of magnitude or more compared to previously published values, with the largest gains occurring for systems with sparse initial coverage or long orbital periods, and smaller but still measurable improvements for planets with long baselines or previously precise ephemerides.

In addition to this, several of the systems we examined also host non-transiting outer planets identified via RV measurements: K2-111\,c and K2-222\,c. These companions orbit at periods approximately 3--10$\times$ longer than their transiting counterparts and may induce subtle dynamical perturbations, such as TTVs or long-term precession. In this analysis, no significant TTVs were detected, and the observed \textit{CHEOPS} transits occur at the epochs predicted by the extrapolated ephemerides. This indicates that any dynamical effects are either negligible or not detectable with the current data. More broadly, systems with known but non-transiting companions present an additional motivation for maintaining accurate ephemerides. In such cases, the transiting planet may be the only directly observable tracer of the system's architecture. Ensuring that its future transits are precisely predictable not only enables atmospheric characterisation but also preserves the possibility of detecting long-term dynamical perturbations, which may be caused by additional, non-transiting planets in the system. These effects, often undetectable in sparse datasets, can accumulate over time and introduce deviations from linear ephemerides. Thus, ephemeris precision plays a foundational role in enabling follow-up science in multi-planet systems, whether or not all planets are seen in transit.

\section{Conclusions}
\label{sec:conclusion}

These results underscore the importance of routine ephemeris updates, especially for high-priority targets with long orbital periods or sparse follow-up coverage. As the exoplanet community prepares for missions with increasingly demanding scheduling constraints, such as \textit{HWO} and \textit{Ariel}, the timely maintenance of accurate ephemerides will be essential to ensure that observing opportunities are not lost. This work demonstrates a straightforward and scalable approach to doing so, using publicly available photometry and accessible fitting tools to combine archival and recent transit timings. 

When considering the relative contributions of \textit{TESS} and \textit{CHEOPS}, both missions play distinct but complementary roles in maintaining precise transit ephemerides. For most systems, \textit{TESS} delivered the largest improvement by extending the temporal baseline and adding multiple new transits, typically reducing the projected 2045 timing uncertainties by an order of magnitude or more relative to the original \textit{K2} fits. The subsequent \textit{CHEOPS} observations --- 151 hours across six of the AO-2 targets (excluding TOI-1634\,b as the DDT skews the sample) --- provided a further cumulative reduction of about 61 hours in predicted timing uncertainty. These AO-2 systems exhibit a mean ROP of $\sim$\,21\,$\mathrm{s\,hr^{-1}}$, corresponding to roughly 21 seconds of future scheduling precision recovered per hour. While this per-hour figure may appear modest, these small gains compound significantly over multi-decade baselines, translating into many hours of improved transit predictability by the 2040s. Together, \textit{TESS} and \textit{CHEOPS} demonstrate how combining broad-survey coverage with targeted, high-cadence follow-up can restore sub-hour predictability for established systems, safeguarding them for future atmospheric and dynamical studies with future facilities.

Beyond refined ephemerides, this analysis provides significantly improved planetary parameters for seven small transiting planets, resulting in more precise masses, radii, and bulk densities. These worlds span a wide compositional range: from the dense, rocky or iron-rich interior of K2-106\,b to the moderately dense, steam-rich or higher-MMW envelopes inferred for TOI-1634\,b and K2-111\,b, and the volatile-rich or mixed rock-ice compositions of K2-263\,b and K2-79\,b. Their diversity emphasises the complexity of small planet formation and the importance of well-constrained M--R relationships for testing models of atmospheric loss and interior structure. Several of these systems also host non-transiting companions inferred from long-term RV trends --- such as K2-111\,c and K2-222\,c --- which may play an important role in shaping the dynamical evolution and composition of their inner planets. Continued RV monitoring will be crucial for fully resolving these outer companions and characterising their impact on system architecture.

Overall, this work illustrated the lasting value of coordinated photometric and spectroscopic follow-up. The combination of \textit{CHEOPS} transits, HARPS-N RVs, and archival \textit{K2}, \textit{TESS}, and ground-based data has produced ephemerides precise enough to enable atmospheric and interior studies well into the 2040s. As future characterisation missions target cooler and smaller exoplanets, reliable ephemerides and accurate physical parameters will be essential foundations. Sustained investment in precision follow-up --- particularly for legacy systems discovered a decade or more ago --- remains essential to building a statistically robust population of small exoplanets and maximising the scientific return of upcoming missions.

\section*{Acknowledgements}

We thank Aleksandar Atanasov for carrying out initial analyses on these systems.

This paper includes data collected by the \textit{TESS} mission, which are publicly available from the Mikulski Archive for Space Telescopes (MAST). Funding for the \textit{TESS} mission is provided by NASA’s Science Mission directorate. We acknowledge the use of public \textit{TESS} Alert data from pipelines at the \textit{TESS} Science Office and at the \textit{TESS} Science Processing Operations Centre. Resources supporting this work were provided by the NASA High-End Computing (HEC) Program through the NASA Advanced Supercomputing (NAS) Division at Ames Research Centre for the production of the SPOC data products. 

This work makes use of data from the European Space Agency (ESA) \textit{CHEOPS} mission, acquired through the \textit{CHEOPS} AO-2 Guest Observers programme ID:07 (PI: Mortier) and DDT programme ID:15 (PI: Mortier). \textit{CHEOPS} is an ESA mission in partnership with Switzerland with important contributions to the payload and the ground segment from Austria, Belgium, France, Germany, Hungary, Italy, Portugal, Spain, Sweden and the United Kingdom. We thank support from the \textit{CHEOPS} GO Programme and Science Operations Centre for help in the preparation and analysis of the \textit{CHEOPS} observations. This research has made use of the Exoplanet Follow-up Observation Program (ExoFOP) website, which is operated by the California Institute of Technology, under contract with the National Aeronautics and Space Administration under the Exoplanet Exploration Program.

This paper includes data collected by the \textit{K2} mission. Funding for the \textit{K2} mission is provided by the NASA Science Mission directorate. Some of the data presented in this paper were
obtained from the Mikulski Archive for Space Telescopes (MAST). STScI is operated by the Association of Universities for Research in Astronomy, Inc., under NASA contract NAS5-26555. Support for MAST for non-HST data is provided by the NASA Office of Space Science via grant NNX13AC07G and by other grants and contracts.

This research made use of the open-source Python package exoctk, the Exoplanet Characterization Toolkit (Bourque et al, 2021).

This work is based on observations made with the Italian Telescopio Nazionale Galileo (TNG) operated on the island of La Palma by the Fundaci\'on Galileo Galilei of the INAF (Istituto Nazionale di Astrofisica) at the Spanish Observatorio del Roque de los Muchachos of the Instituto de Astrofisica de Canarias. The HARPS-N project was funded by the Prodex Program of the Swiss Space Office (SSO), the Harvard University Origin of Life Initiative (HUOLI), the Scottish Universities Physics Alliance (SUPA), the University of Geneva, the Smithsonian Astrophysical Observatory (SAO), the Italian National Astrophysical Institute (INAF), University of St. Andrews, Queen’s University Belfast, and University of Edinburgh.

This article is based on observations made with the MuSCAT2 instrument, developed by ABC, at Telescopio Carlos Sánchez, operated on the island of Tenerife by the IAC in the Spanish Observatorio del Teide.

This publication makes use of The Data \& Analysis Center for Exoplanets (DACE), which is a facility based at the University of Geneva (CH) dedicated to extrasolar planets data visualisation, exchange and analysis. DACE is a platform of the Swiss National Centre of Competence in Research (NCCR) PlanetS, federating the Swiss expertise in Exoplanet research. The DACE platform is available at \url{https://dace.unige.ch}.

L.P. acknowledges funding from the Royal Society Career Development Fellowship, grant number CDF\textbackslash
R1\textbackslash251054.

A.M. acknowledges funding from a UKRI Future Leader Fellowship, grant number MR/X033244/1 and a UK Science and Technology Facilities Council (STFC) small grant ST/Y002334/1. 

This work has been carried out within the framework of the NCCR PlanetS supported by the Swiss National Science Foundation under grants 51NF40\_182901 and 51NF40\_205606. JAE acknowledges support through the European Space Agency (ESA) Research Fellowship Programme in Space Science.

T.G.W. acknowledges support from the University of Warwick and UKSA.

A.S.B. acknowledges financial contribution from the INAF Large Grant 2023 ``EXODEMO".

A.C.C. acknowledges support from STFC consolidated grant number ST/V000861/1, UKRI/ERC Synergy Grant EP/Z000181/1 (REVEAL), and UKSA grant number ST/B001077/1. STFC consolidated grant number ST/V000861/1 UKRI/ERC Synergy Grant EP/Z000181/1 (REVEAL).

Y.N.E.E. acknowledges support from a Science and Technology Facilities Council (STFC) studentship, grant number ST/Y509693/1.

L.M. acknoledges  financial contribution from the INAF Large Grant 2023 ``EXODEMO''.

M.S. acknowledges financial support from the Belgian Federal Science Policy Office (BELSPO) in the framework of the PRODEX Programme of the European Space Agency (ESA) under contract number C4000140754.

V.V.E. has been supported by UK’s Science \& Technology Facilities Council through the STFC grants ST/W001136/1 and ST/S000216/1.

C.A.W. would like to acknowledge support from the UK Science and Technology Facilities Council (STFC, grant number ST/X00094X/1).
\section*{Data Availability}

This paper includes raw data collected by the \textit{K2} mission, which are publicly available from MAST (\url{https://archive.stsci.edu/missions-and-data/k2}).
Raw data collected by the \textit{TESS} mission are also publicly available from MAST (\url{https://archive.stsci.edu/tess}). Raw data collected by \textit{CHEOPS} can be found using the file keys in Table \ref{tab:cheops_obs} at \url{https://cheops-archive.astro.unige.ch/archive_browser/}. Observations made with HARPS-N on the Telescopio Nazionale Galileo 3.6\,m telescope are available at the CDS (\url{https://cds.unistra.fr}).



\bibliographystyle{mnras}
\bibliography{main} 




\appendix

\section{Photometric Observations Summary}

\begin{table*}
\centering
\caption{Summary of \textit{K2} and \textit{TESS} Observations.}
\label{tab:k2_tess_obs_summary_ids}
\begin{tabular}{l l l }
\hline\hline
Target & Campaign / Sector & Program IDs \\
\hline
K2-79 & C4 & GO4060 LC, GO4029 LC, GO4033 LC, GO4007 LC \\ 
K2-79 & Sectors 42--44  & G04205, G04242, G04223 \\ 
K2-79 & Sectors 70--71 & G06165, G06058 \\
K2-106 & C8 & GO8068 LC, GO8032 LC, GO8077 LC \\
K2-106 & Sector 42 & G04205, G04242, G04223 \\
K2-106 & Sector 70 & G06165, G06058 \\
K2-111 & C4 & GO4060 LC, GO4033 LC, GO4007 LC \\
K2-111 & Sectors 70-71 & Mission-selected target (no GI ID)\\
K2-222 & C8 & GO8068 LC, GO8051 LC, GO8032 LC, GO8077 LC \\
K2-222 & Sectors 42-43 & G04205, G04059, G04242, G04191, G04223 \\ 
K2-222 & Sector 70 & G06165, G06200, G06058 \\
K2-263 & C5 & GO5007 LC, GO5029 LC, GO5033 LC, GO5104 LC, GO5106 LC, and GO5060 LC \\
K2-263 & C16 & GO16009 LC \& SC, GO16011 LC, GO16015 LC \& SC, GO16020 LC, GO16021 LC, GO16101 LC \& SC \\
K2-263 & C18 & GO18027 LC \& SC, GO18048 LC, GO18049 LC, GO18065 LC \& SC, GO18901 LC \\
K2-263 & Sector 44 & G04231, G04205, G04242 \\
K2-263 & Sectors 45-46, 72 & G06058 \\
TOI-1634 & Sector 18 & G022198 \\
TOI-1634 & Sector 86 & G07157, G07066\\
\hline
\end{tabular}
\end{table*}

\begin{landscape}
    \begin{table}
        \centering
        \renewcommand\multirowsetup{\raggedright}
      \caption{Log of \textit{CHEOPS} observations. The column T\textsubscript{exp} gives the exposure time in terms of the integration time per image multiplied by the number of images stacked on-board prior to download. N\textsubscript{obs} is the number of frames. Effic. is the proportion of the time in which unobstructed observations of the target occurred. R\textsubscript{ap} is the aperture radius used for the photometric extraction. RMS is the standard deviation of the residuals from the best fit. The variables in the final column are as follows: time, t; spacecraft roll angle, $\phi$, PSF centroid position, (x, y); smear correction, \texttt{smear}; aperture contamination, \texttt{contam}; image background level, \texttt{bg}.}
        \resizebox{\linewidth}{!}{
        \begin{tabular}{lllllllllll}
             \hline\hline 
             Target & Start date & Duration & T\textsubscript{exp} & N\textsubscript{obs} & Effic. & File key & APER & R\textsubscript{ap} & RMS & Decorrelation \\ 
             & (UTC) & (h) & (s) & & (\%) & & & (pixels) & (ppm) & \\
             \hline 
             K2-106\,b & 2021-09-17T00:49:50 & 11.61 & 1 x 60.0 & 560 & 80.3 & CH\_PR220007\_TG000101\_V0300 & DEFAULT & 25 & 18297 & sin($\phi$), cos($\phi$), \texttt{bg, contam, smear}\\ 
             K2-106\,b & 2021-09-20T10:47:56 & 11.61 & 1 x 60.0 & 560 & 80.3 & CH\_PR220007\_TG000102\_V0300 & DEFAULT & 25 & 17837 & sin($\phi$), cos($\phi$), \texttt{bg, contam, smear}\\ 
             K2-106\,c & 2021-09-15T19:31:53 & 15.72 & 1 x 60.0 & 755 & 79.9 & CH\_PR220007\_TG000201\_V0300 & DEFAULT & 25 & 10514 & sin($\phi$), cos($\phi$), \texttt{bg, contam, smear}\\ 
             K2-106\,c & 2021-09-29T04:02:52 & 15.72 & 1 x 60.0 & 638 & 67.6 & CH\_PR220007\_TG000202\_V0300 & DEFAULT & 25 & 7758 & sin($\phi$), cos($\phi$), \texttt{bg, contam, smear} \\ 
             K2-222\,b & 2021-10-04T05:02:00 & 8.84 & 1 x 60.0 & 366 & 68.9 & CH\_PR220007\_TG000301\_V0300 & DEFAULT & 25 & 567 & No decorrelation needed.\\ 
             K2-222\,b & 2021-11-03T23:42:00 & 9.22 & 1 x 60.0 & 375 & 67.7 & CH\_PR220007\_TG000302\_V0300 & DEFAULT & 25 & 1094 & sin($2\phi$), cos($2\phi$)\\ 
             K2-79 b & 2021-12-06T08:12:52 & 8.39 & 1 x 60.0 & 312 & 61.9 & CH\_PR220007\_TG000401\_V0300 & DEFAULT & 25 & 34339 & sin($\phi$), cos($\phi$), \texttt{bg, contam, smear}, t$^2$, x$^2$, y$^2$, \\ & & & & & & & & & & sin(2$\phi$), cos(2$\phi$)\\ 
             TOI-1634\,b & 2021-11-07T21:20:51 & 10.69 & 1 x 60.0 & 391 & 60.9 & CH\_PR220007\_TG000501\_V0300 & DEFAULT & 25 & 1928 & sin($\phi$), cos($\phi$), \texttt{bg, contam, smear} \\ 
             TOI-1634\,b & 2021-12-18T11:04:54 & 11.36 & 1 x 60.0 & 363 & 53.2 & CH\_PR220007\_TG000502\_V0300 & DEFAULT & 25 & 18585 & t, x, y, sin($\phi$), cos($\phi$), \texttt{bg}\\ 
             K2-111\,b & 2021-10-15T05:39:50 & 16.66 & 1 x 60.0 & 547 & 54.7 & CH\_PR220007\_TG000601\_V0300 & DEFAULT & 25 & 2979 & sin($\phi$), cos($\phi$), \texttt{bg, contam, smear}, t$^2$, x$^2$, y$^2$, \\ & & & & & & & & & & sin(2$\phi$), cos(2$\phi$)\\
             K2-111\,b & 2021-10-31T07:04:52 & 16.84 & 1 x 60.0 & 602 & 59.5 & CH\_PR220007\_TG000602\_V0300 & DEFAULT & 25 & 13916 & y, sin($\phi$), cos($\phi$), \texttt{bg, contam, smear} \\ 
             K2-263\,b & 2022-01-05T04:36:52 & 12.41 & 1 x 60.0 & 432 & 58 & CH\_PR220007\_TG000901\_V0300 & DEFAULT & 25 & 10306 & No decorrelation needed. \\ 
             K2-263\,b & 2022-12-26T19:47:00 & 13.26 & 1 x 60.0 & 431 & 54.1 & CH\_PR220007\_TG001501\_V0300 & DEFAULT & 25 & 1246 & t, x, y, sin($\phi$), cos($\phi$), \texttt{bg}\\ 
             K2-79\,b & 2022-12-04T03:56:00 & 10.74 & 1 x 60.0 & 411 & 63.7 & CH\_PR220007\_TG001601\_V0300 & DEFAULT & 25 & 13256 & t, x, y, sin($\phi$), cos($\phi$), \texttt{bg} \\
             TOI-1634 & 2022-12-15T01:18:54 & 65.08 & 1 x 60.0 & 2202 & 56.4 & CH\_PR430015\_TG000101\_V0300 & DEFAULT & 25 & 2223 & t, x, y, sin($\phi$), cos($\phi$), \texttt{bg, contam, smear}, t$^2$, \\ (DDT) & & & & & & & & & & x$^2$, y$^2$, sin(2$\phi$), cos(2$\phi$) \\
             TOI-1634 & 2022-12-18T19:13:54 & 40.87 & 1 x 60.0 & 1331 & 54.3 & CH\_PR430015\_TG000301\_V0300 & DEFAULT & 25 & 2210 & sin($\phi$), cos($\phi$), \texttt{contam, smear}, t$^2$, x$^2$, y$^2$, \\  (DDT) & & & & & & & & & & sin(2$\phi$), cos(2$\phi$)\\
             \hline  
        \end{tabular}
        }
        \label{tab:cheops_obs}
    \end{table}
\end{landscape}

\section{Joint Fits Figures}

\begin{figure*}
    \centering
    \includegraphics[width=0.33\linewidth]{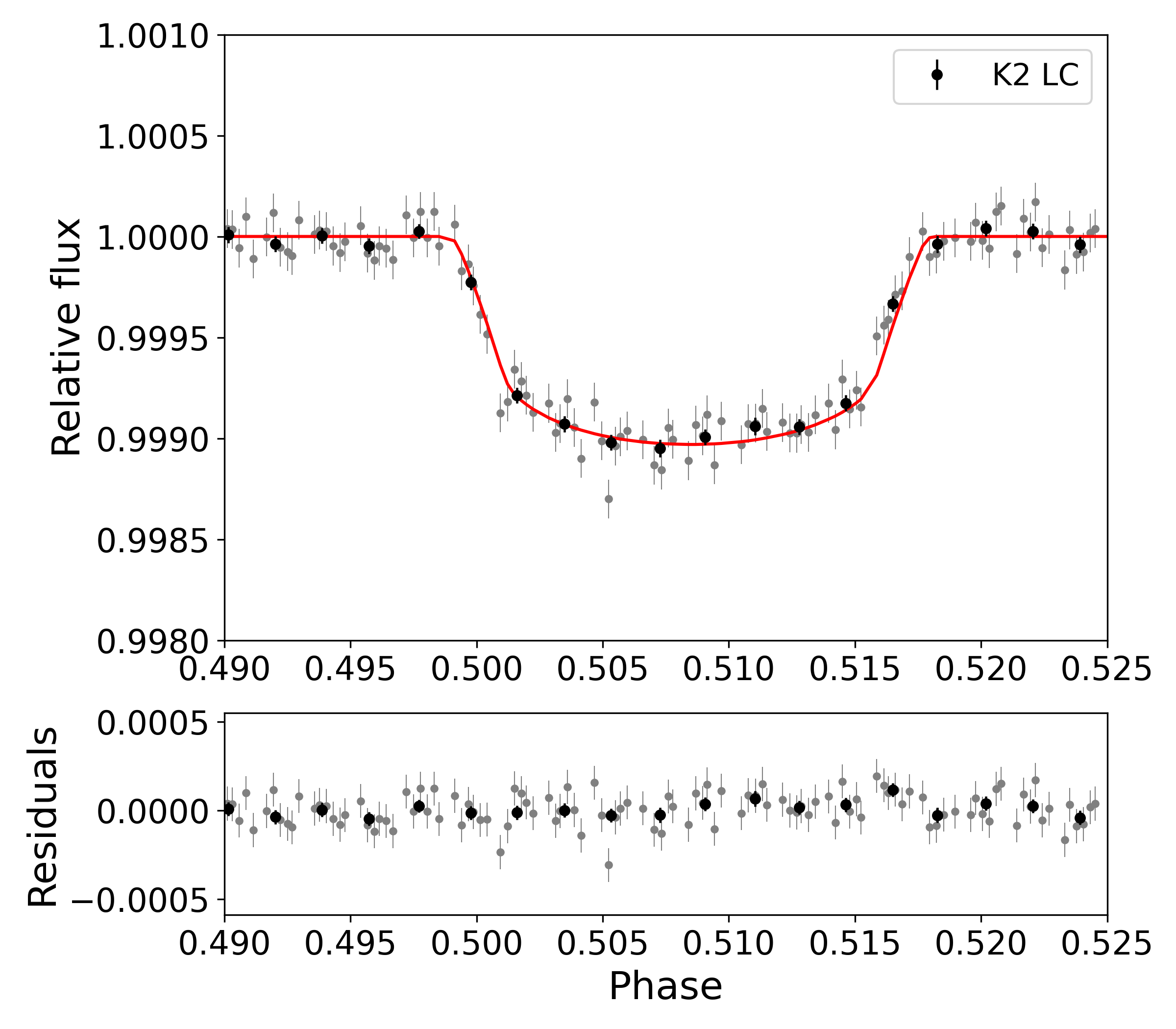}
    \includegraphics[width=0.33\linewidth]{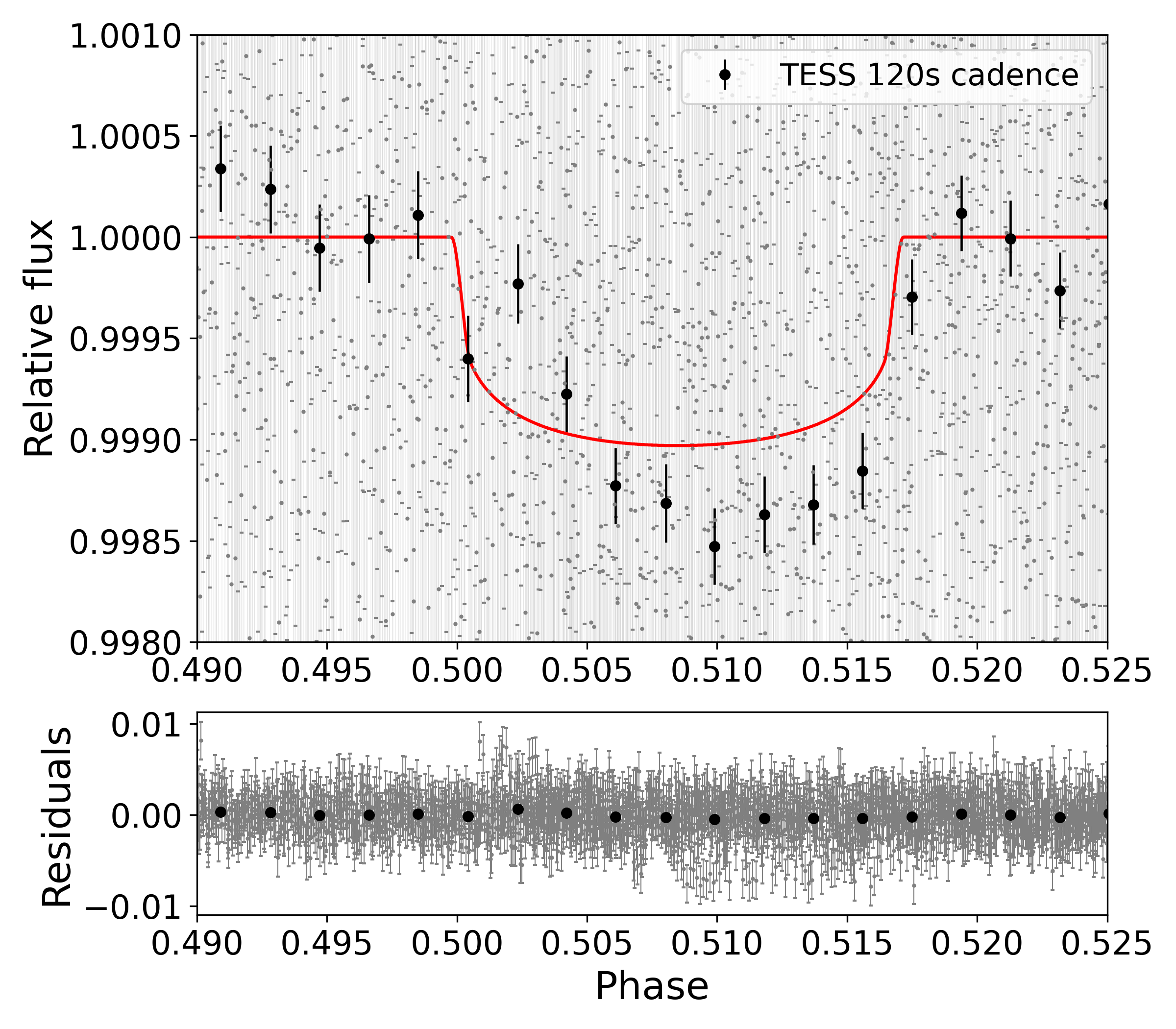}
    \includegraphics[width=0.33\linewidth]{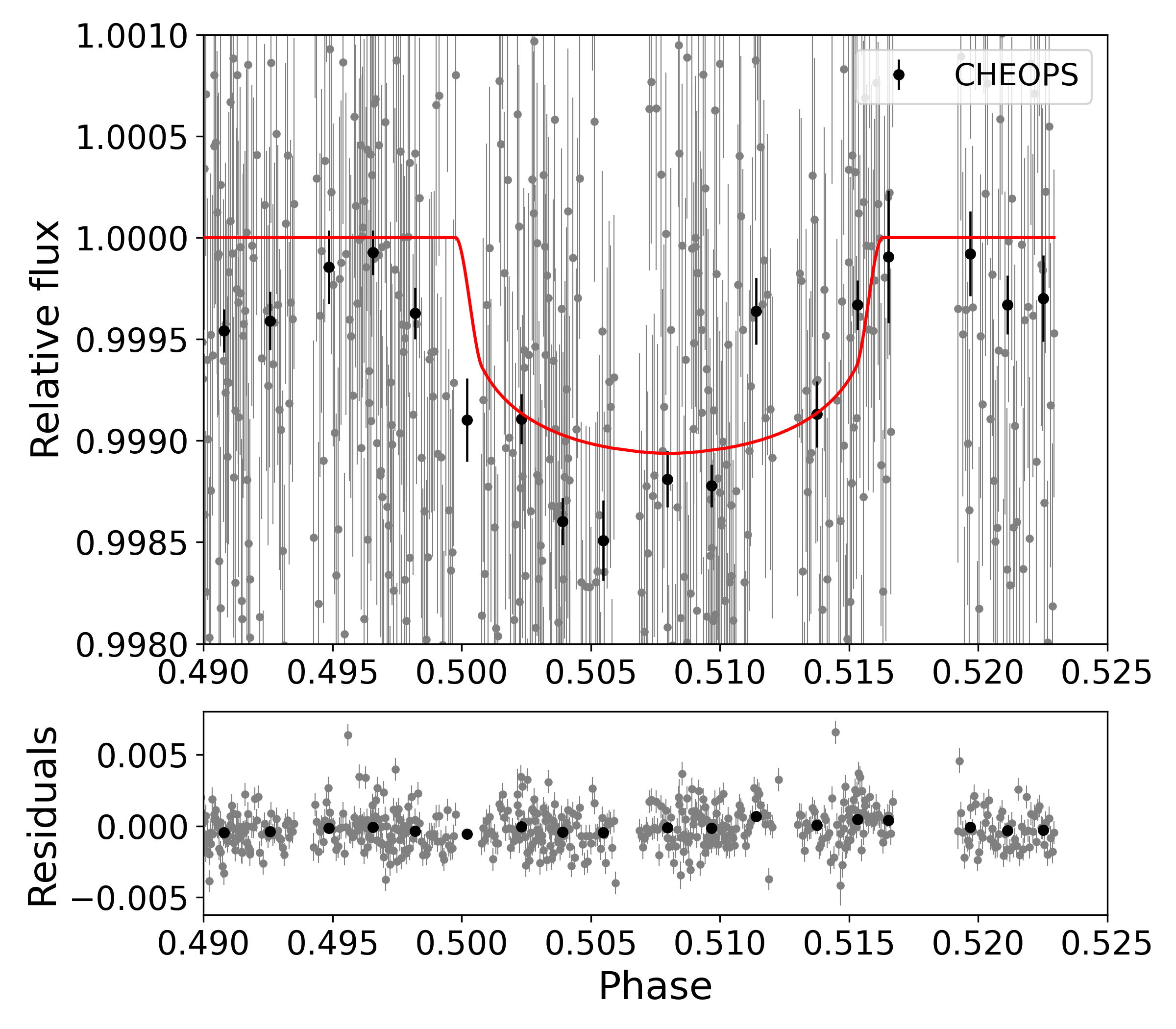}\\
    \includegraphics[width=0.33\linewidth]{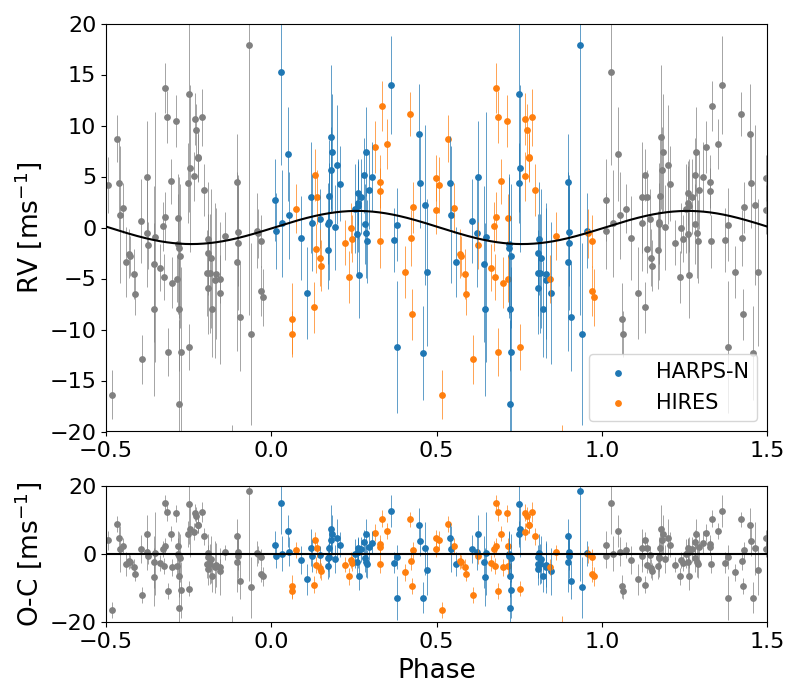}\\
    \includegraphics[width=0.5\linewidth]{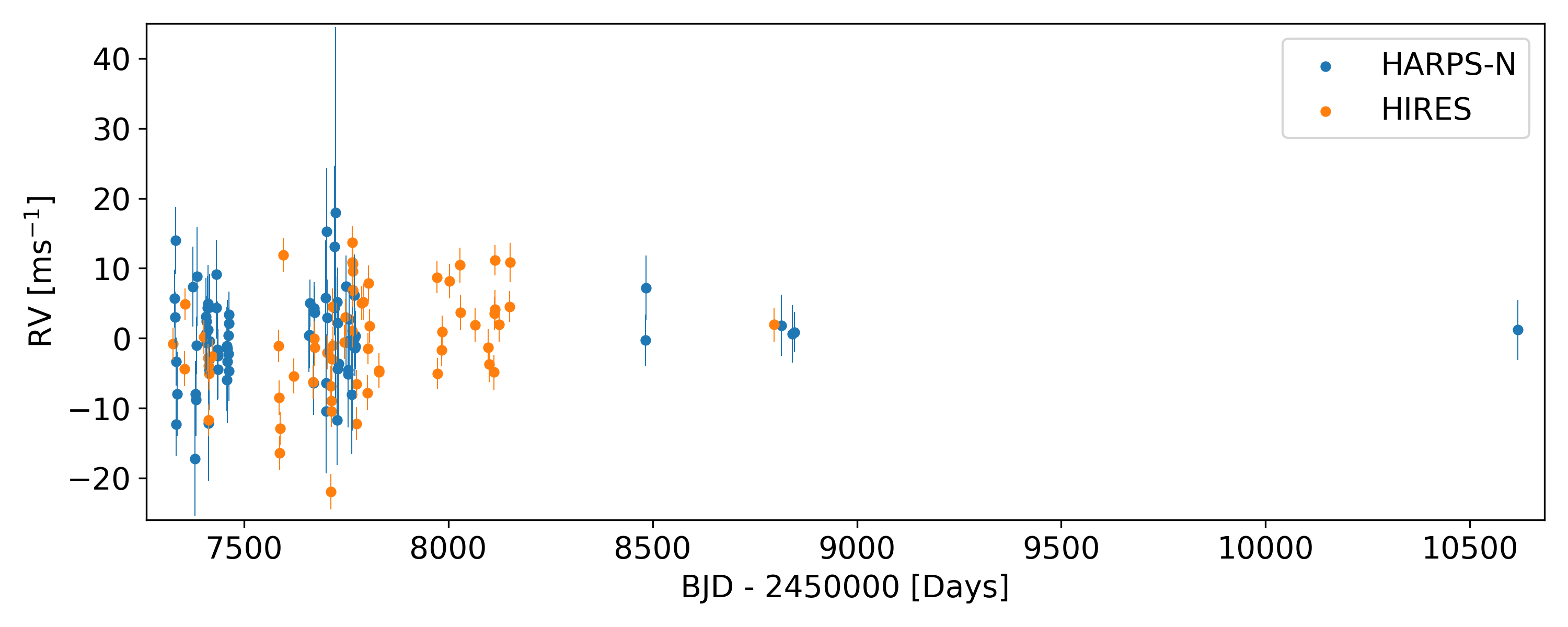}
    \caption{Combined fit results for K2-79\,b. Transit fit to the light curve data from \textit{K2} long cadence (\textbf{first row left}), \textit{TESS} long cadence (\textbf{first row centre}), and \textit{CHEOPS} (\textbf{first row right}). The cadence (29.4\,min for \textit{K2} long cadence, 120\,s for \textit{TESS} long cadence, and 60\,s for \textit{CHEOPS}) fluxes are plotted in grey, the fluxes binned every 30\,min are over-plotted in black, and the fitted transit is shown by the red solid line. The RV fit to the HARPS-N \& HIRES RVs is shown in the \textbf{second row} for K2-79\,b. Coloured circular points show the phase folded RVs, grey points show these same values in subsequent phases, and the black line shows the best fit Keplerian model. The \textbf{third row} shows the RVs over the full $\sim9$\,yr baseline, illustrating the absence of any significant long term trend.}
    \label{fig:K2-79_transits}
\end{figure*}

\begin{figure*}
    \centering
    \includegraphics[width=0.33\linewidth]{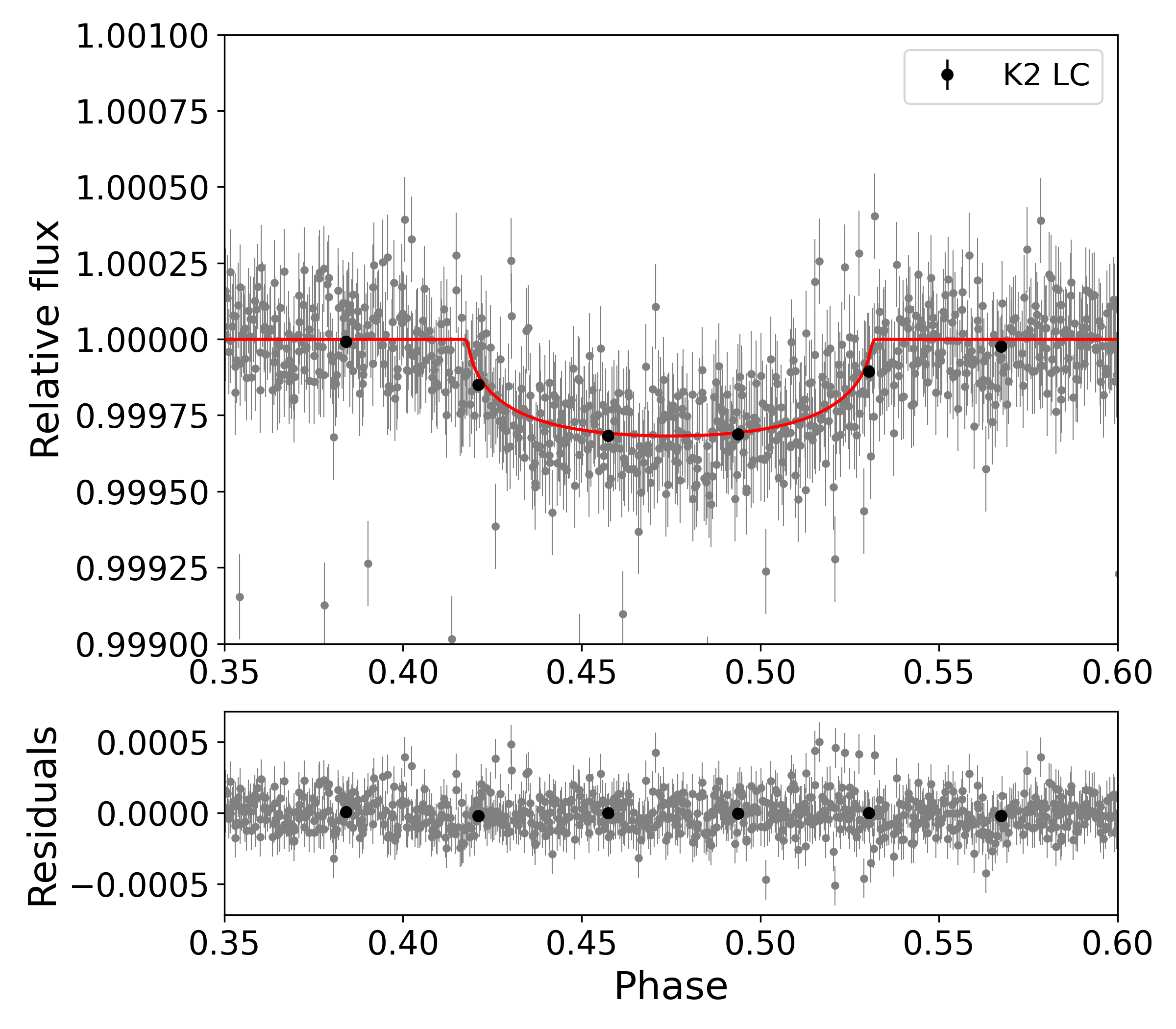}
    \includegraphics[width=0.33\linewidth]{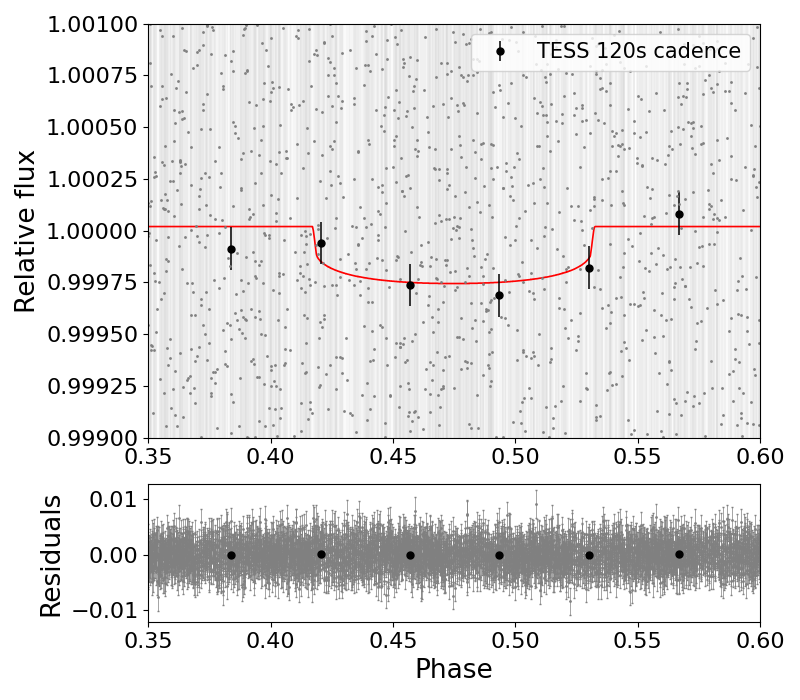}
    \includegraphics[width=0.33\linewidth]{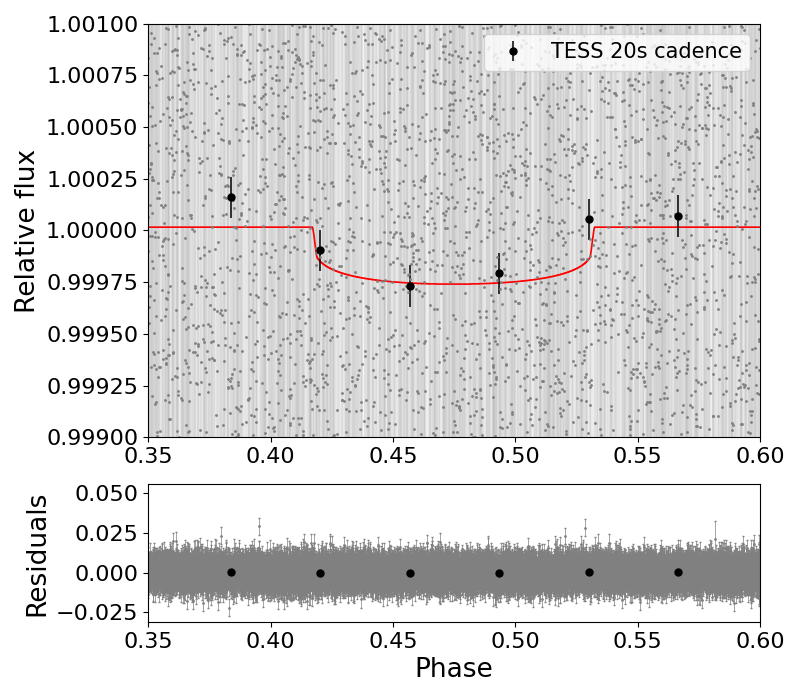}\\
    \includegraphics[width=0.33\linewidth]{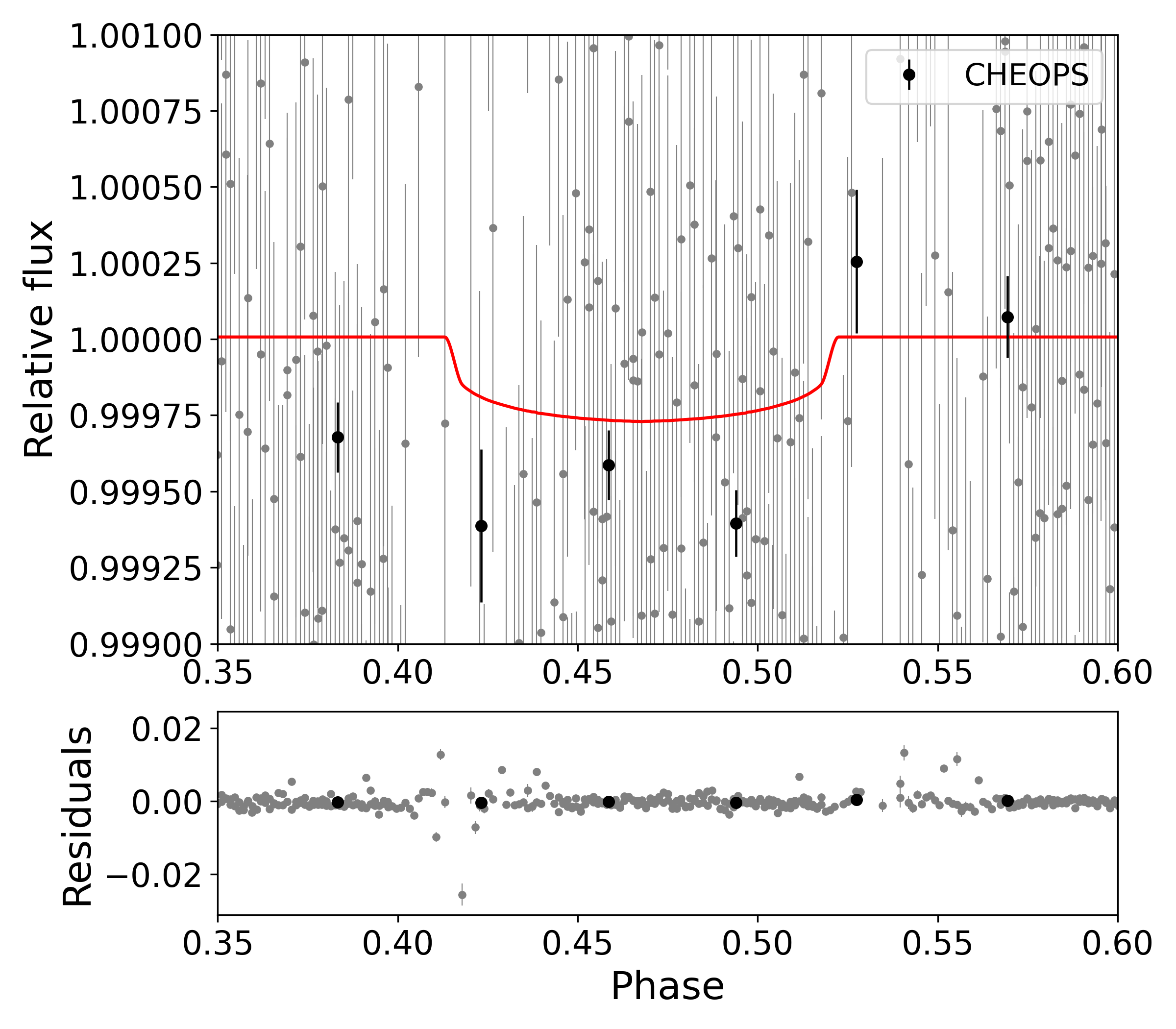}
    \includegraphics[width=0.33\linewidth]{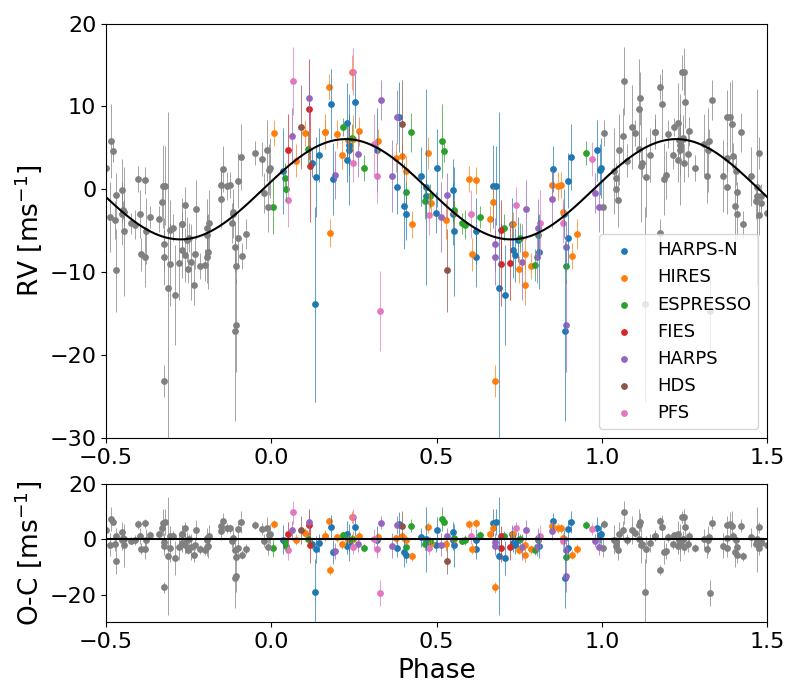}\\
    \includegraphics[width=0.5\linewidth]{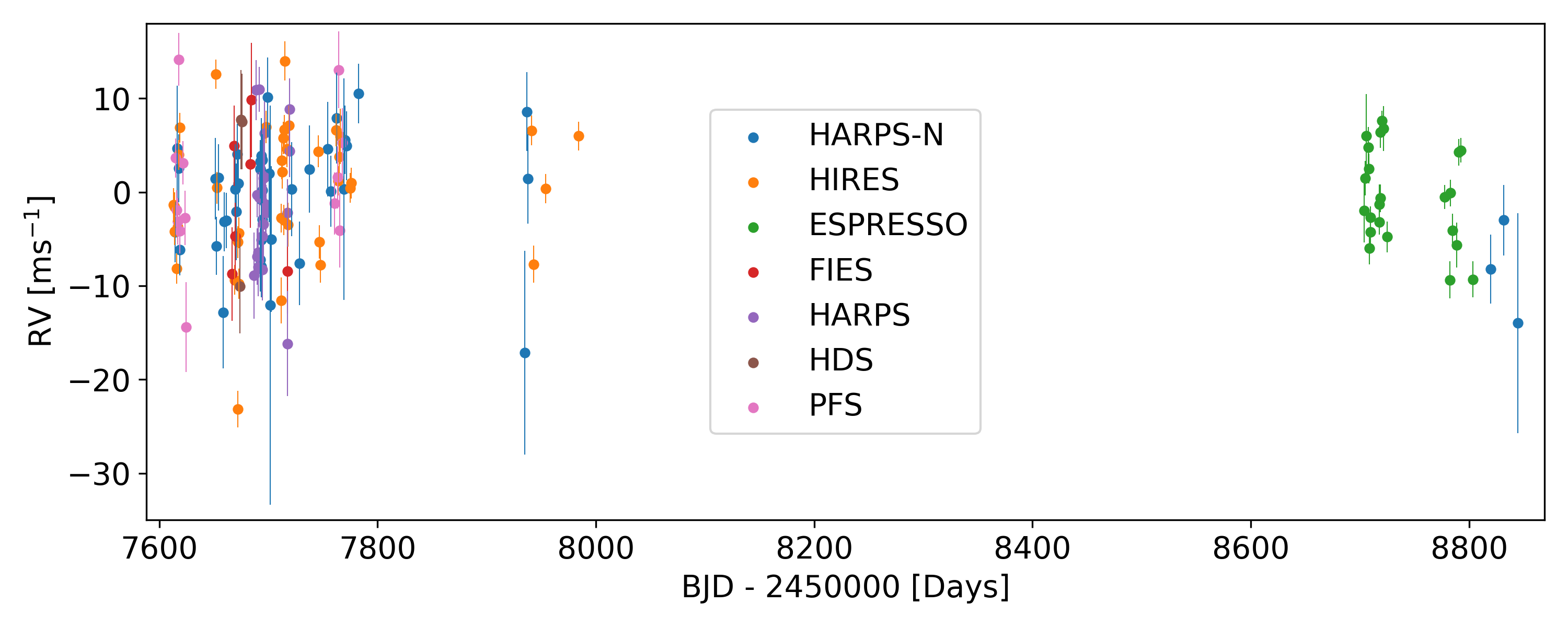}
    \caption{Combined fit results for K2-106\,b. Transit fit to the light curve data from \textit{K2} long cadence (\textbf{first row left}), \textit{TESS} long cadence (\textbf{first row centre}), \textit{TESS} short cadence (\textbf{first row right}), and \textit{CHEOPS} (\textbf{second row left}). The cadence (29.4\,min for \textit{K2} long cadence, 120\,s for \textit{TESS} long cadence, 20\,s for \textit{TESS} short cadence, and 60\,s for \textit{CHEOPS}) fluxes are plotted in grey, the fluxes binned every 30\,min are over-plotted in black, and the fitted transit is shown by the red solid line. The RV fit to the HARPS-N, HIRES, ESPRESSO, FIES, HARPS, HDS, and PFS RVs are shown in the \textbf{second row right} for K2-106\,b. Coloured circular points show the phase folded RVs, grey points show these same values in subsequent phases, and the black line shows the best fit Keplerian model. The \textbf{third row} shows the RVs over the full $\sim4$\,yr baseline.}
    \label{fig:K2-106b_transits}
\end{figure*}

\begin{figure*}
    \centering
    \includegraphics[width=0.33\linewidth]{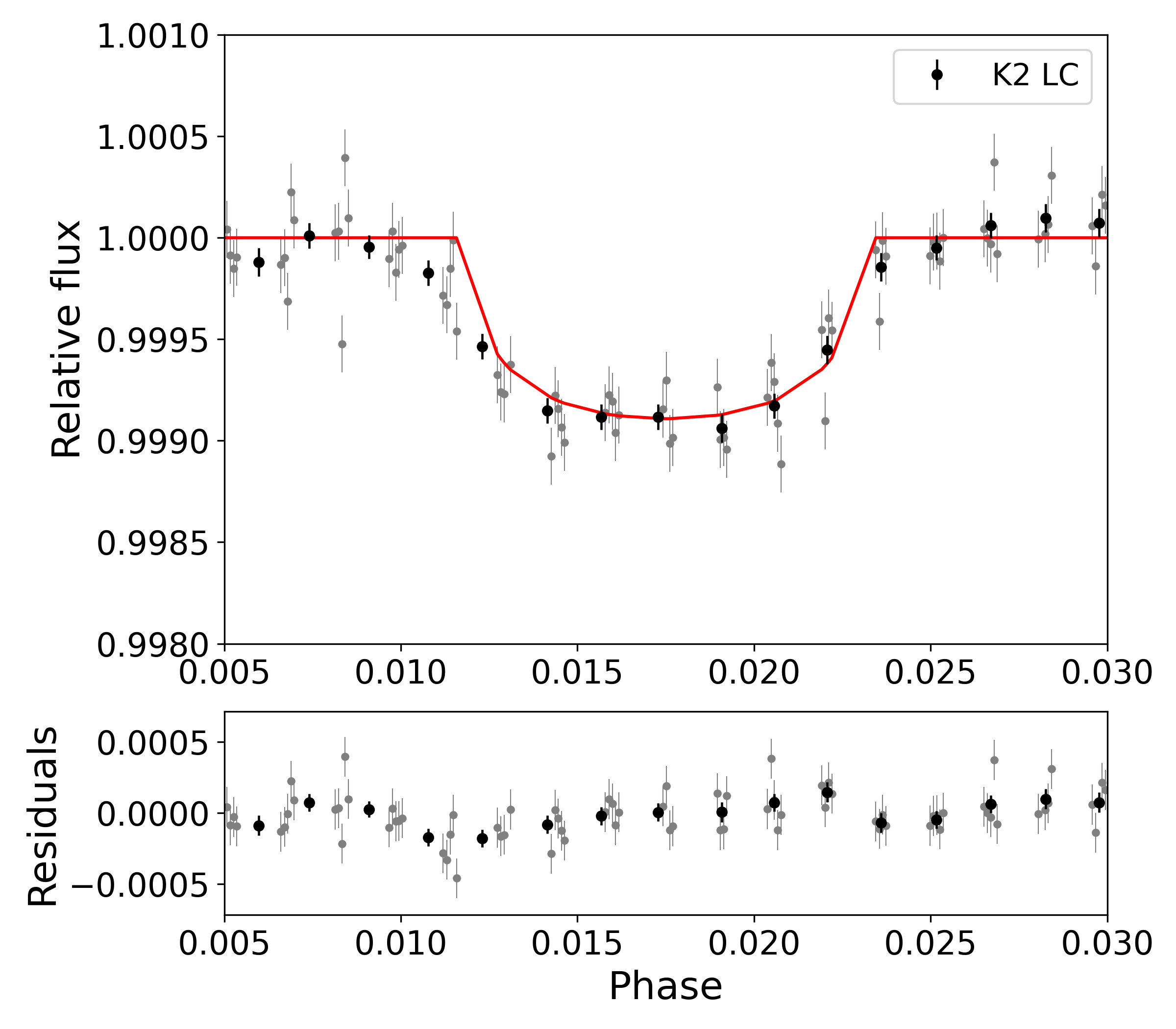}
    \includegraphics[width=0.33\linewidth]{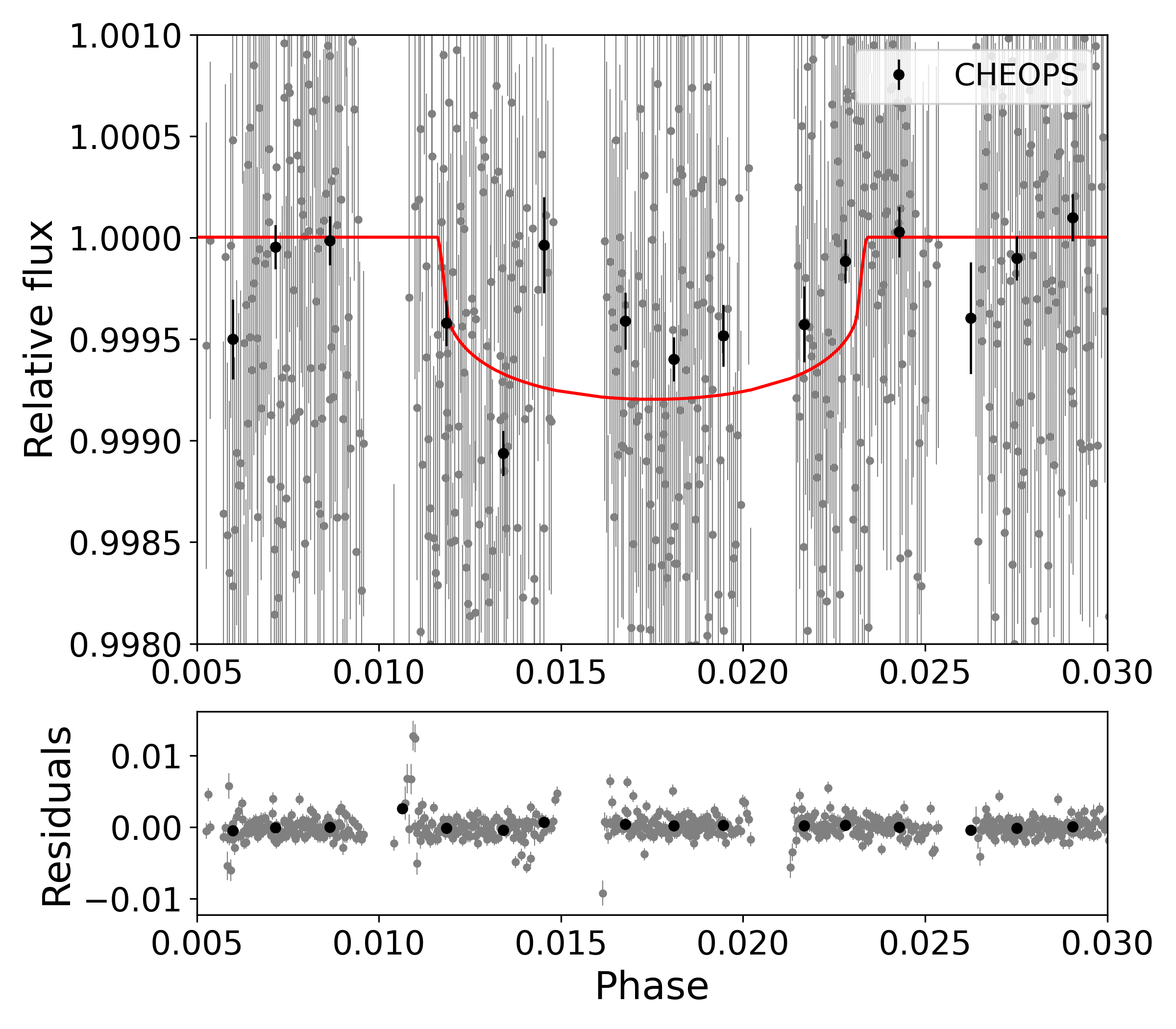}
    \includegraphics[width=0.33\linewidth]{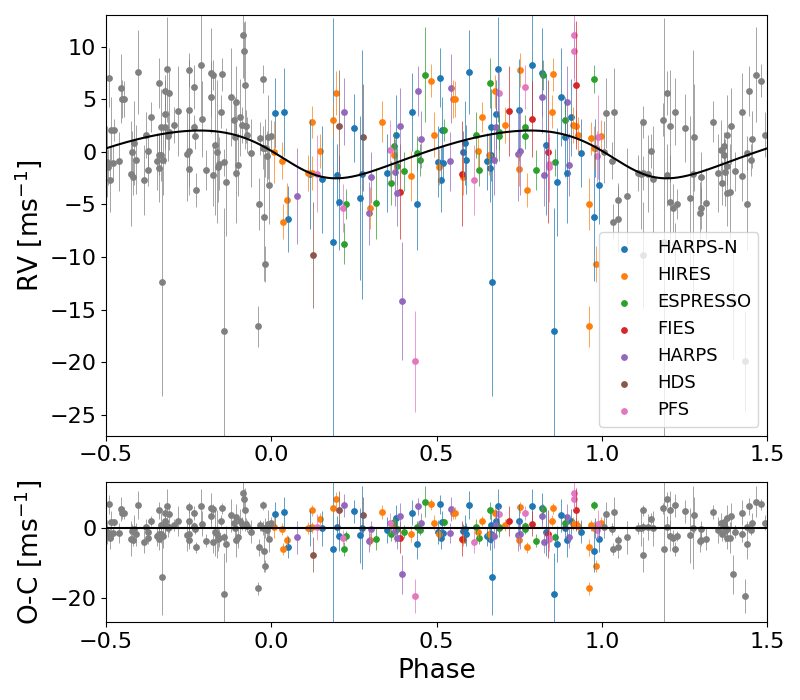}
    \caption{Combined fit results for K2-106\,c. Transit fit to the light curve data from \textit{K2} long cadence (\textbf{left}) and \textit{CHEOPS} (\textbf{centre}). The cadence (29.4\,min for \textit{K2} long cadence and 60\,s for \textit{CHEOPS}) fluxes are plotted in grey, the fluxes binned every 30\,min are over-plotted in black, and the fitted transit is shown by the red solid line. The RV fit to the HARPS-N, HIRES, ESPRESSO, FIES, HARPS, HDS, and PFS RVs are shown in the \textbf{right} for K2-106\,c. Coloured circular points show the phase folded RVs, grey points show these same values in subsequent phases, and the black line shows the best fit Keplerian model.}
    \label{fig:K2-106c_transits}
\end{figure*}

\begin{figure*}
    \centering
    \includegraphics[width=0.33\linewidth]{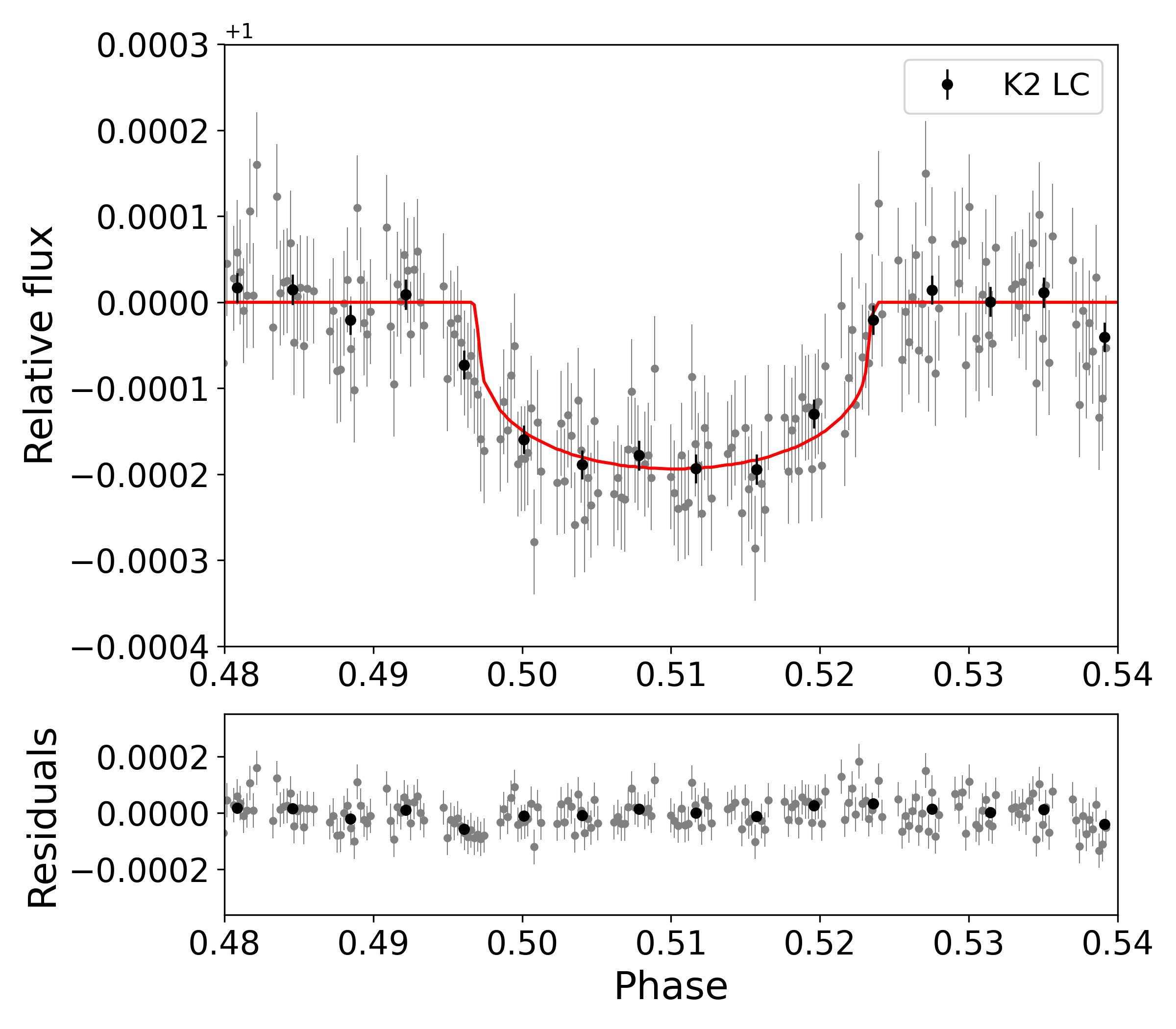}
    \includegraphics[width=0.33\linewidth]{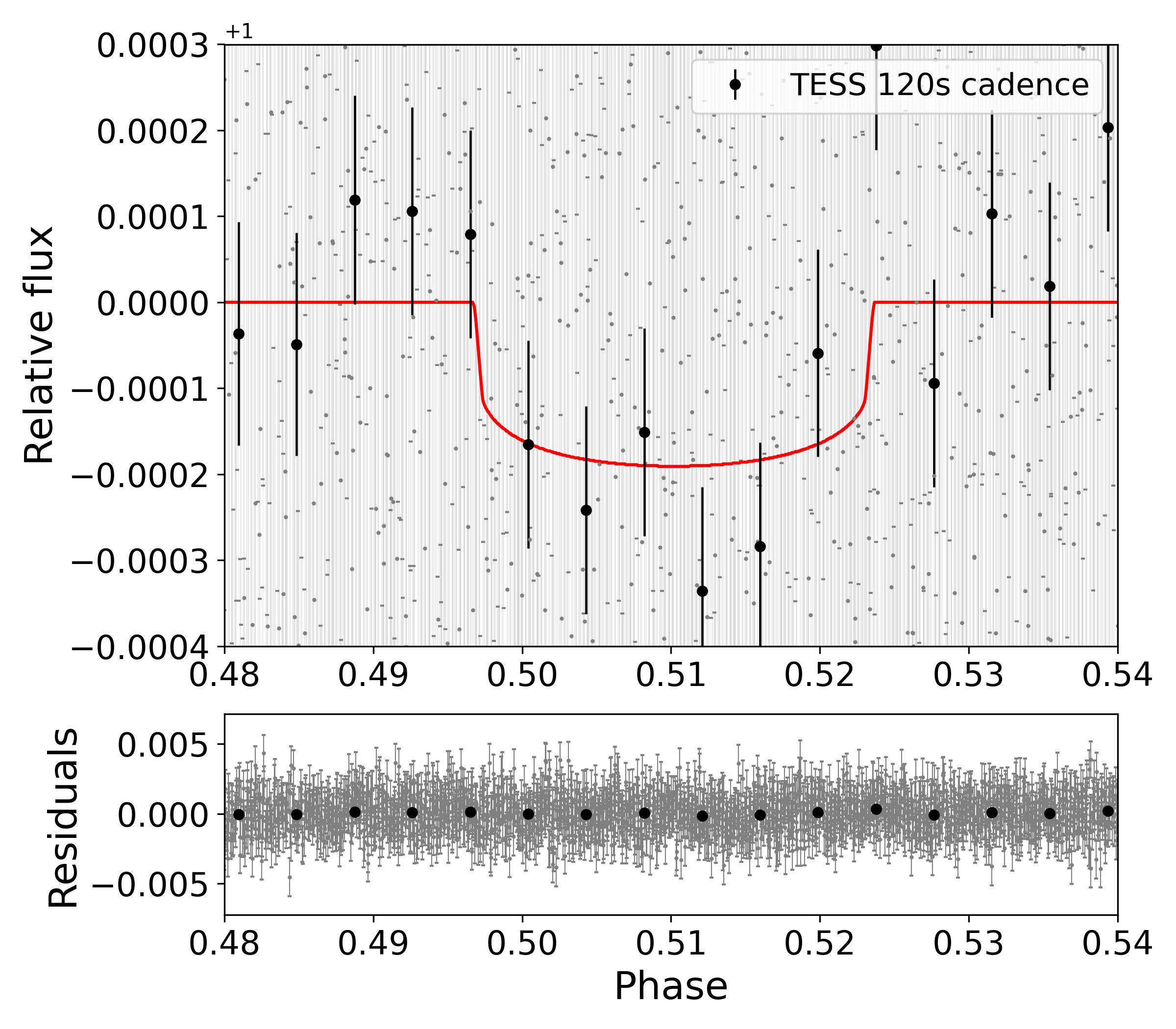}
    \includegraphics[width=0.33\linewidth]{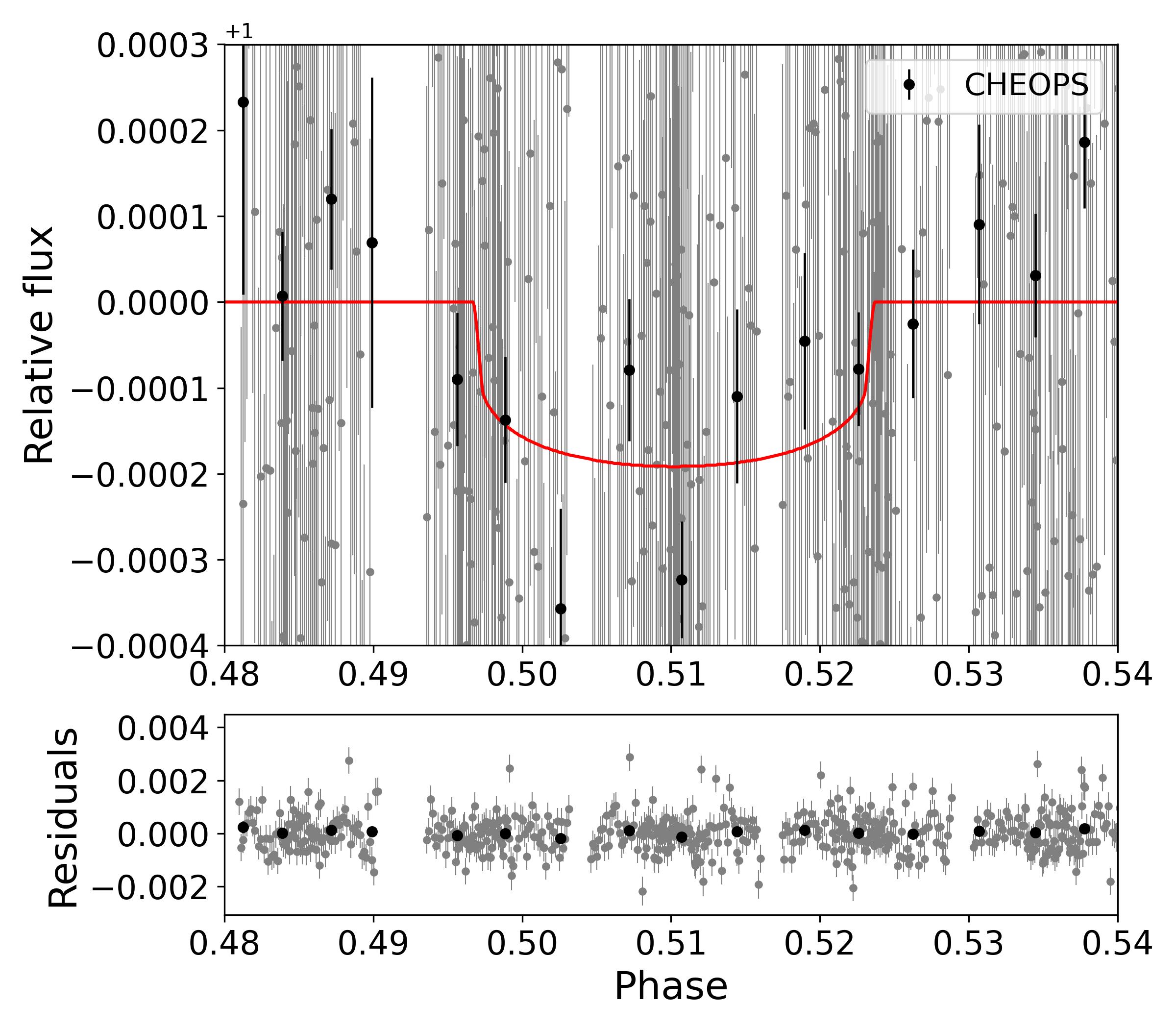}\\
    \includegraphics[width=0.33\linewidth]{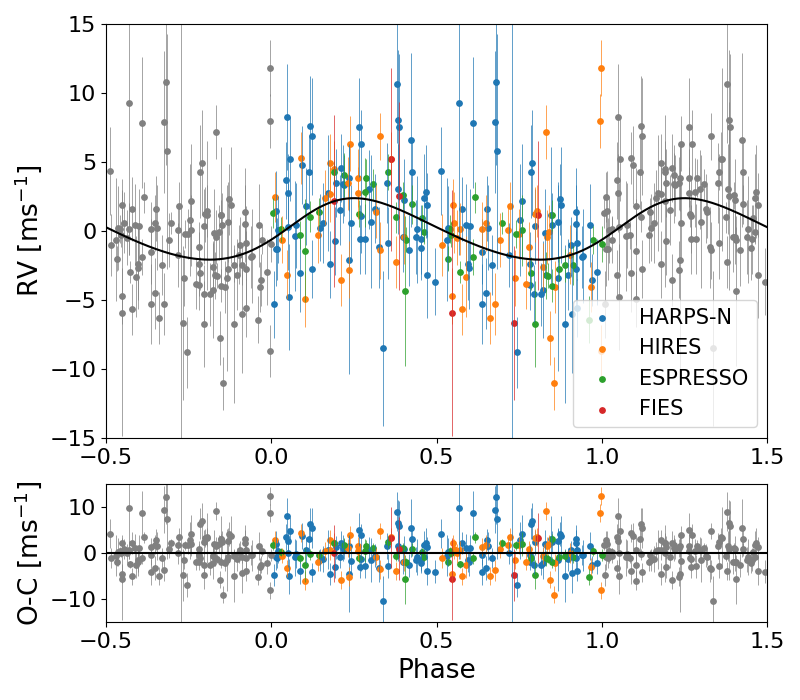}
    \includegraphics[width=0.33\linewidth]{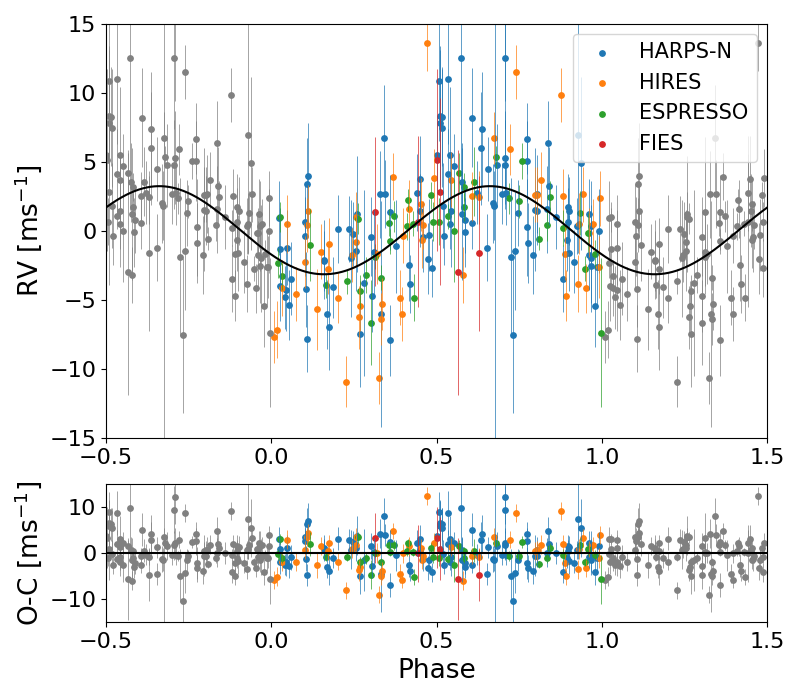}\\
    \includegraphics[width=0.5\linewidth]{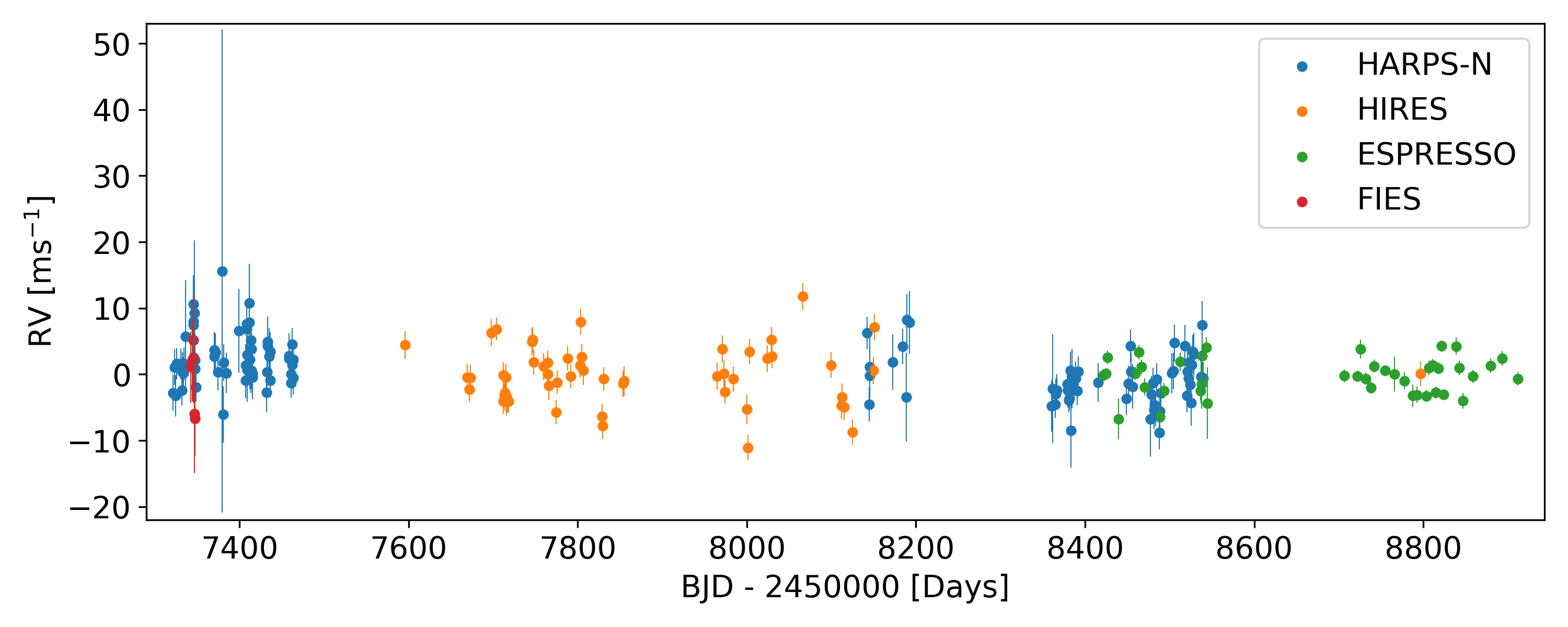}
    \caption{Combined fit results for K2-111\,b and RV fit results of K2-111\,c. Transit fit to the light curve data from \textit{K2} long cadence (\textbf{first row left}), \textit{TESS} long cadence (\textbf{first row centre}), and \textit{CHEOPS} (\textbf{first row right}). The cadence (29.4\,min for \textit{K2}, 120\,s for \textit{TESS}, and 60\,s for \textit{CHEOPS}) fluxes are plotted in grey, the fluxes binned every 30\,min are over-plotted in black, and the fitted transit is shown by the red solid line. The RV fit to the HARPS-N, HIRES, ESPRESSO, \& FIES  RVS are shown in the \textbf{second row left} for K2-111\,b and \textbf{second row right} for K2-111\,c. Coloured circular points show the phase folded RVs, grey points show these same values in subsequent phases, and the black line shows the best fit Keplerian model. The \textbf{third row} shows the RVs over the full $\sim4$\,yr baseline.}
    \label{fig:K2-111_transits}
\end{figure*}

\begin{figure*}
    \centering
    \includegraphics[width=0.33\linewidth]{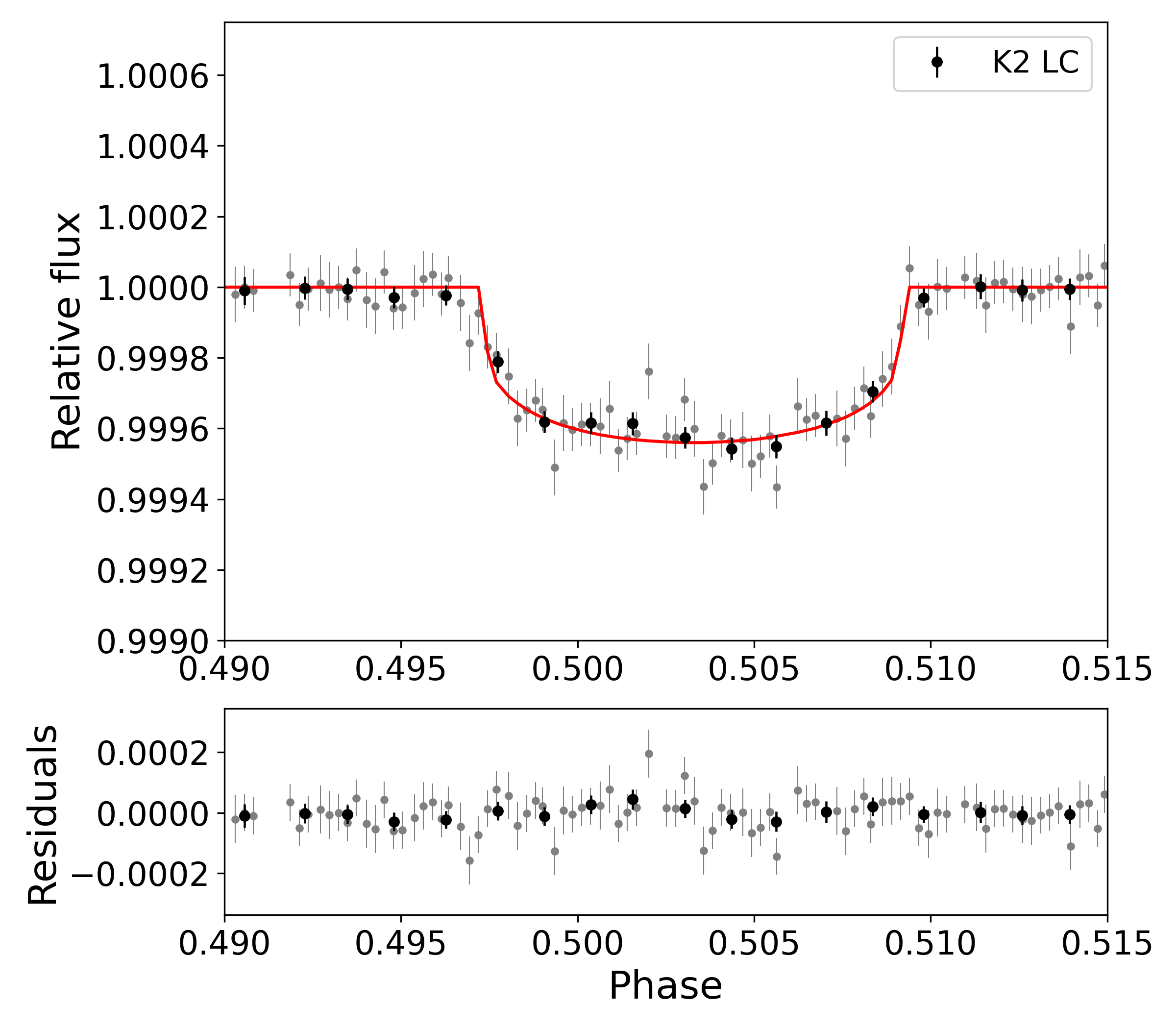}
    \includegraphics[width=0.33\linewidth]{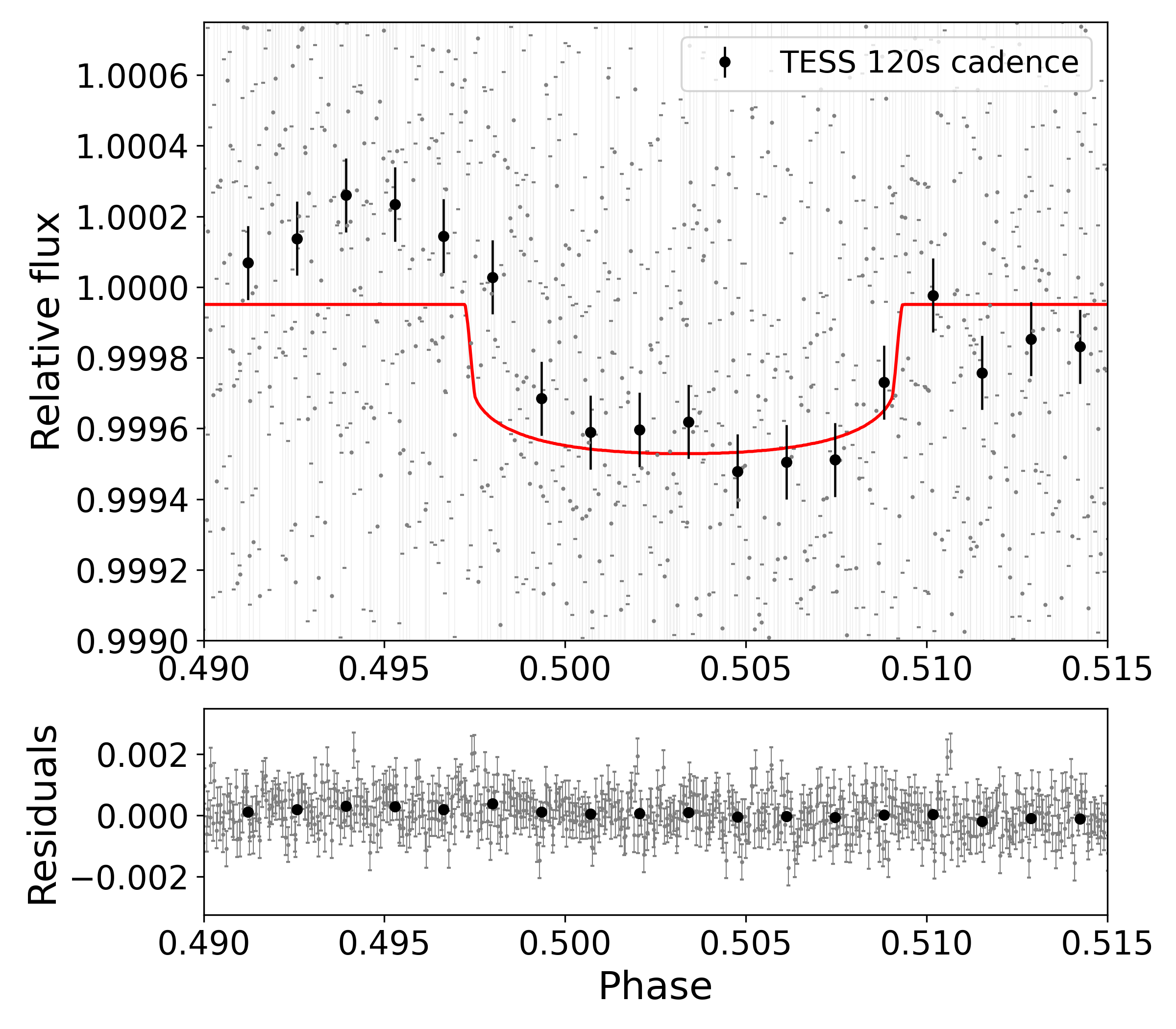}
    \includegraphics[width=0.33\linewidth]{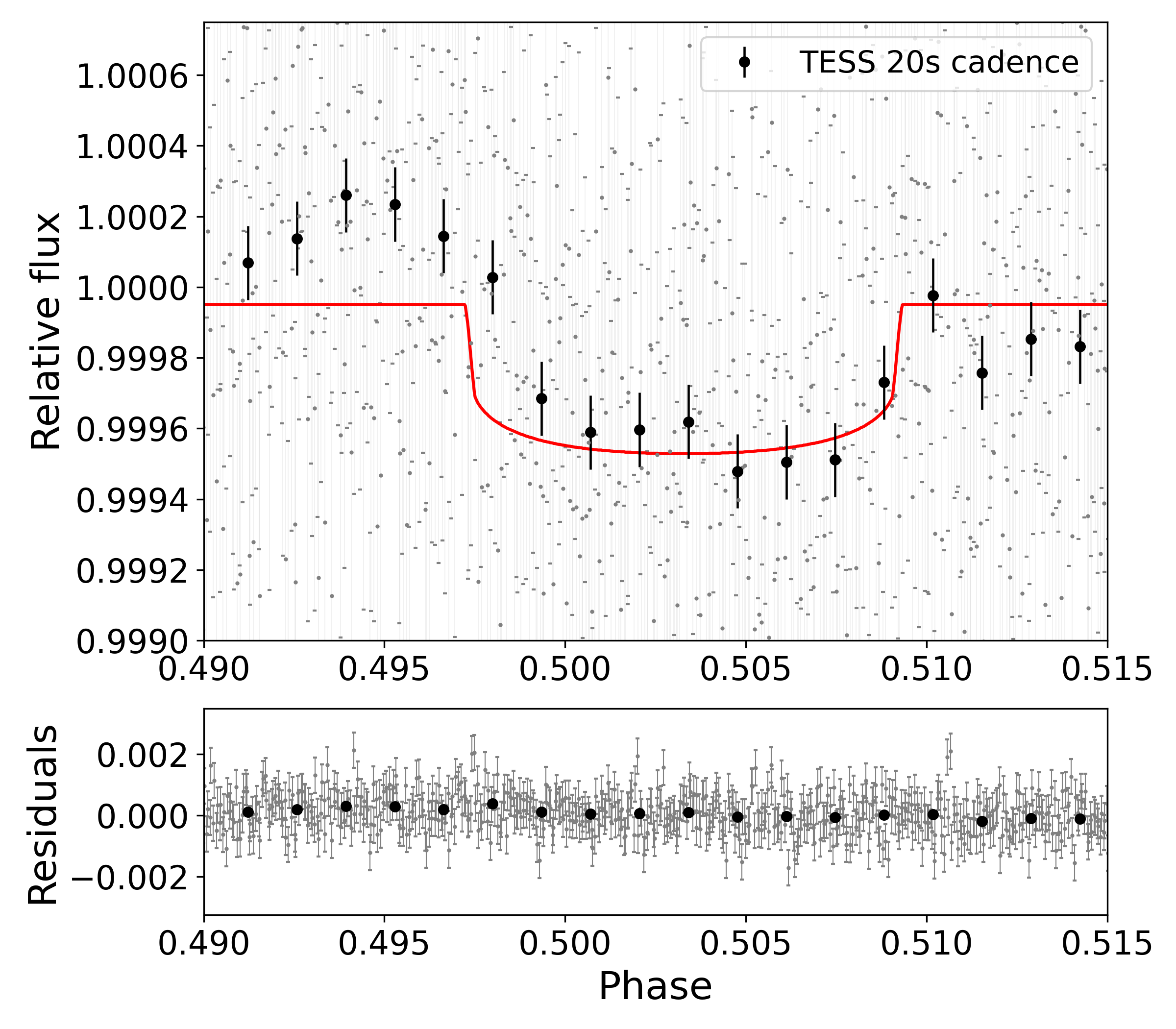}\\
    \includegraphics[width=0.33\linewidth]{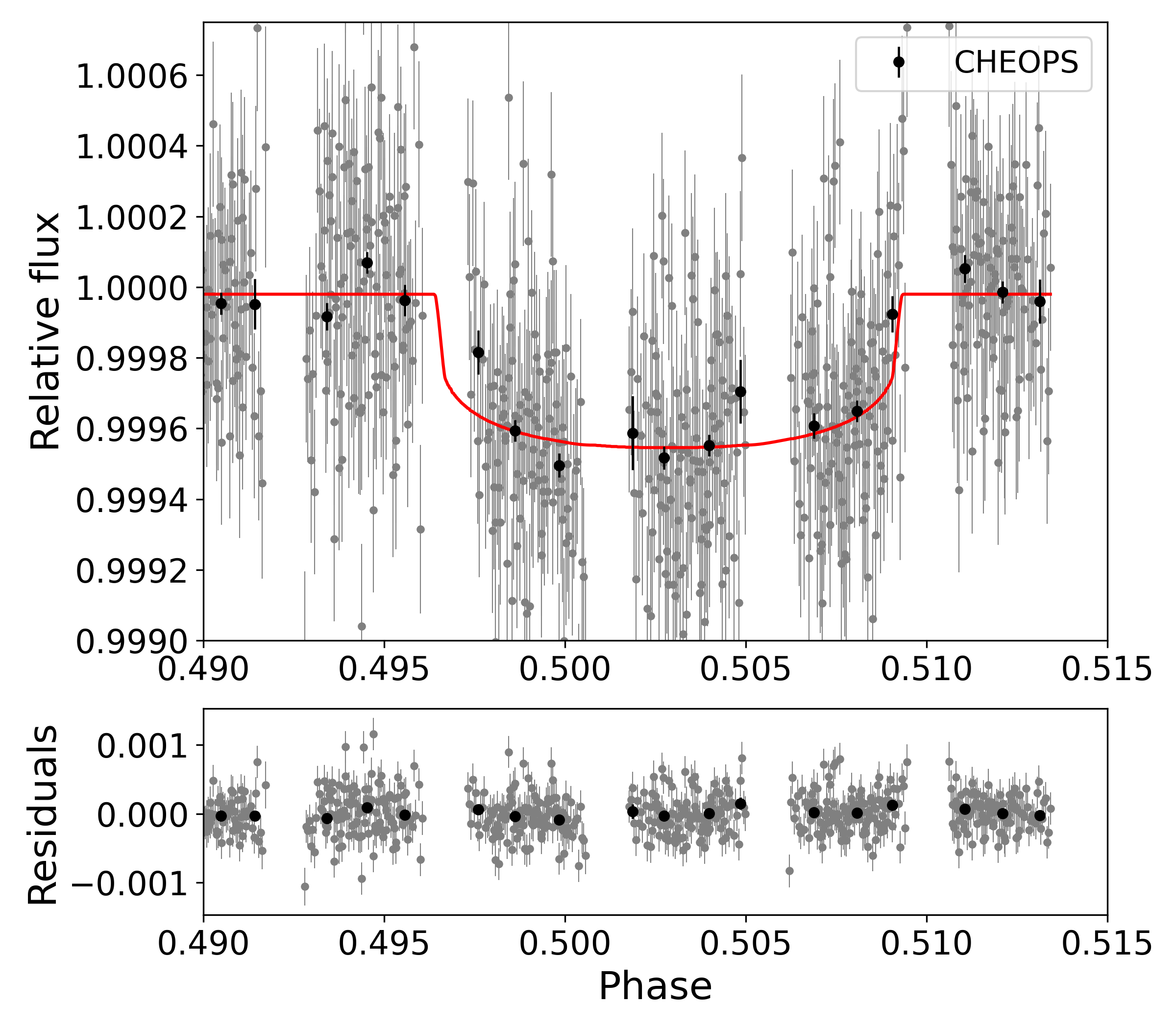}
    \includegraphics[width=0.33\linewidth]{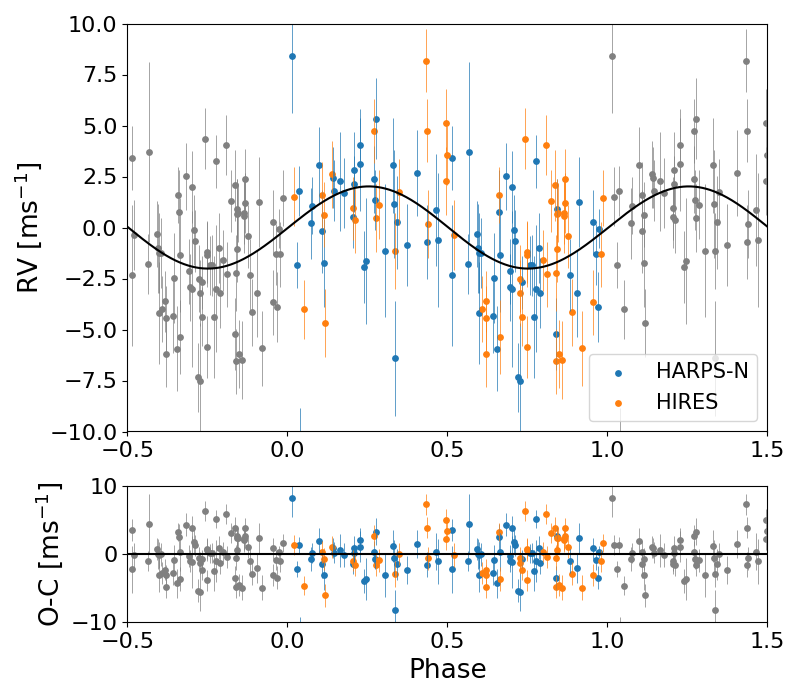}
    \includegraphics[width=0.33\linewidth]{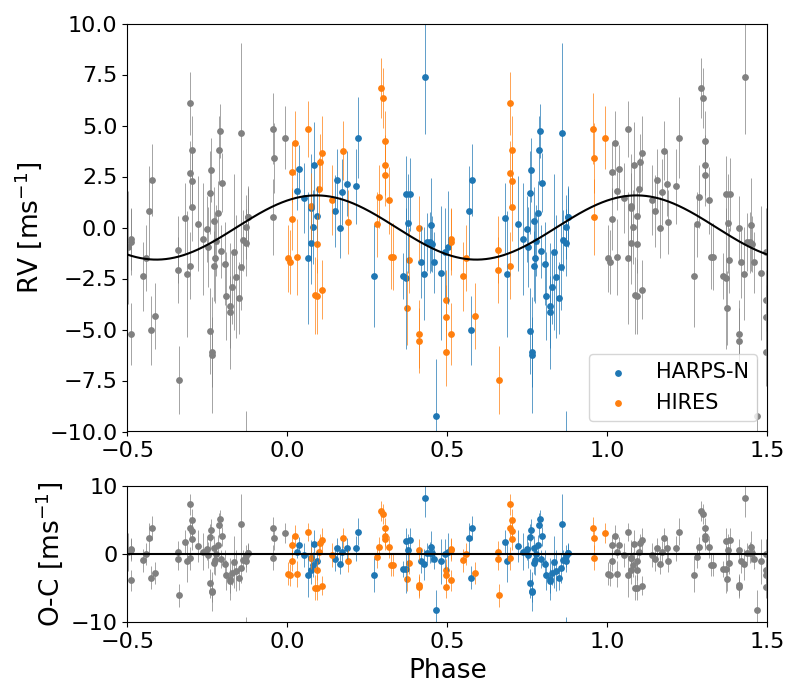}\\
    \includegraphics[width=0.5\linewidth]{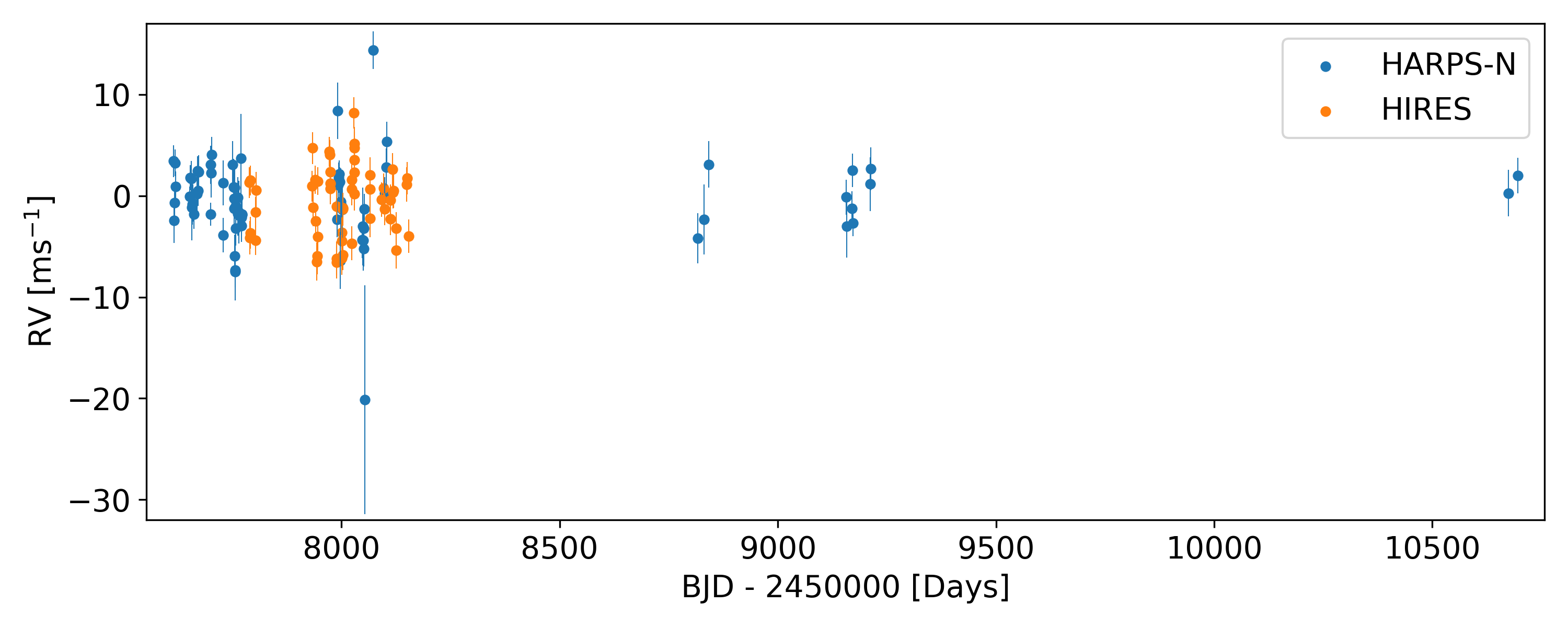}
    \caption{Combined fit results for K2-222\,b and RV fit results of K2-222\,c. Transit fit to the light curve data from \textit{K2} long cadence (\textbf{first row left}), \textit{TESS} long cadence (\textbf{first row centre}), \textit{TESS} short cadence (\textbf{first row right}), and \textit{CHEOPS} (\textbf{second row left}). The cadence (29.4\,min for \textit{K2} long cadence, 120\,s for \textit{TESS} long cadence, 20\,s for \textit{TESS} short cadence, and 60\,s for \textit{CHEOPS}) fluxes are plotted in grey, the fluxes binned every 30\,min are over-plotted in black, and the fitted transit is shown by the red solid line. The RV fit to the HARPS-N \& HIRES RVs are shown in the \textbf{second row centre} for K2-222\,b and \textbf{second row right} for K2-222\,c. Coloured circular points show the phase folded RVs, grey points show the values in subsequent phases, black line shows the best fit Keplerian model. The \textbf{third row} shows the RVs over the full $\sim7$\,yr baseline.}
    \label{fig:K2-222_transits}
\end{figure*}

\begin{figure*}
    \centering
    \includegraphics[width=0.33\linewidth]{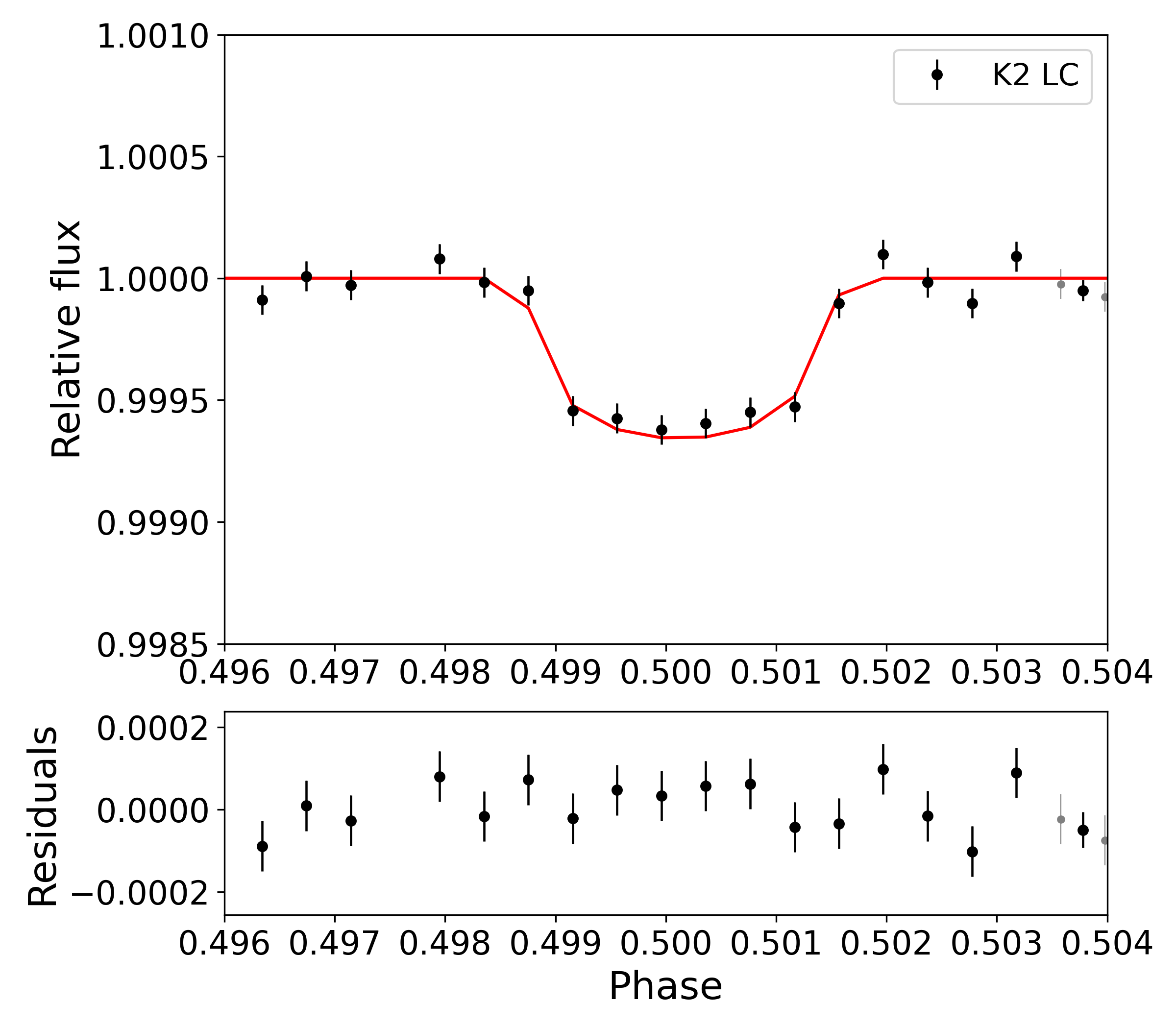}
    \includegraphics[width=0.33\linewidth]{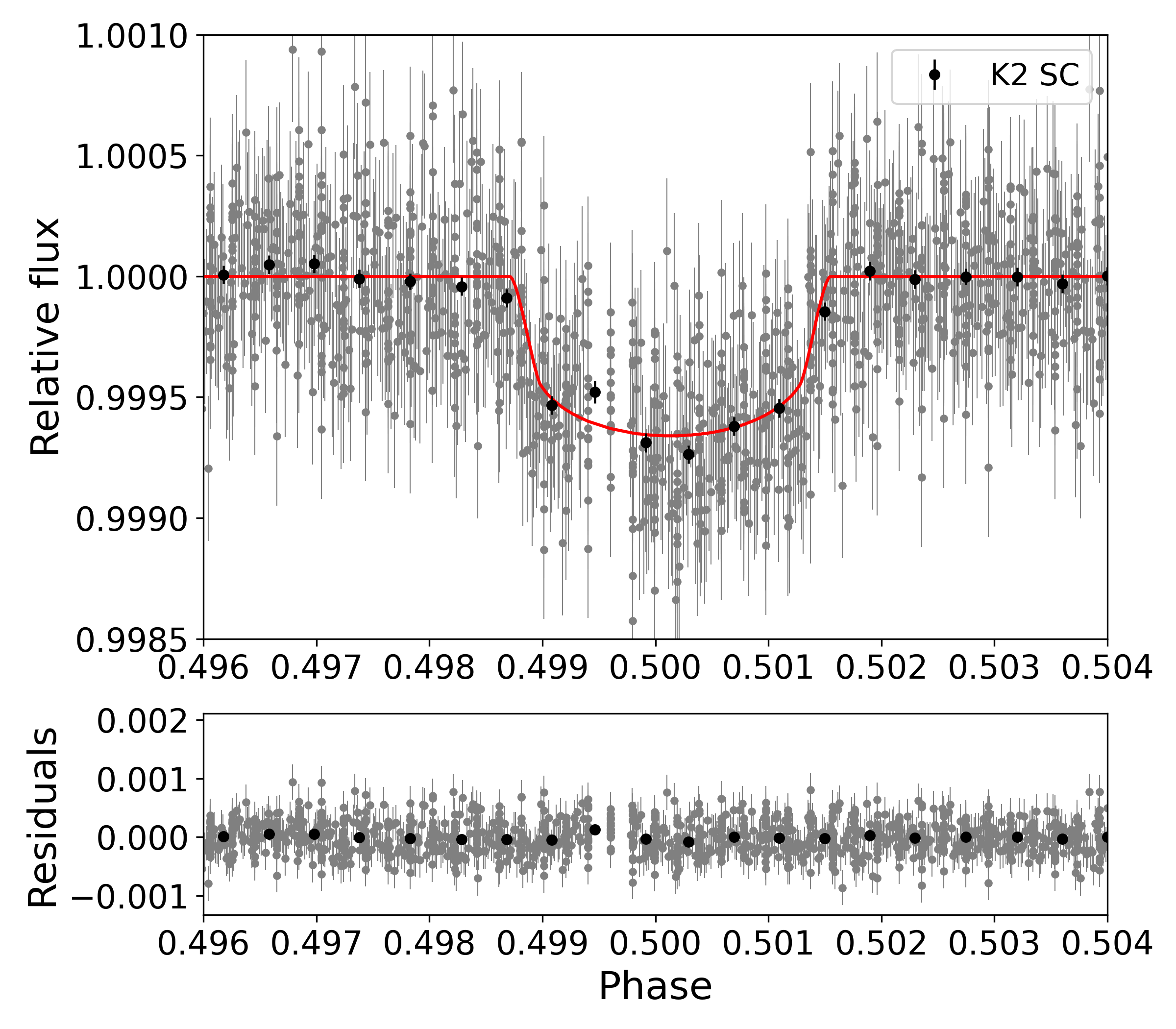}
    \includegraphics[width=0.33\linewidth]{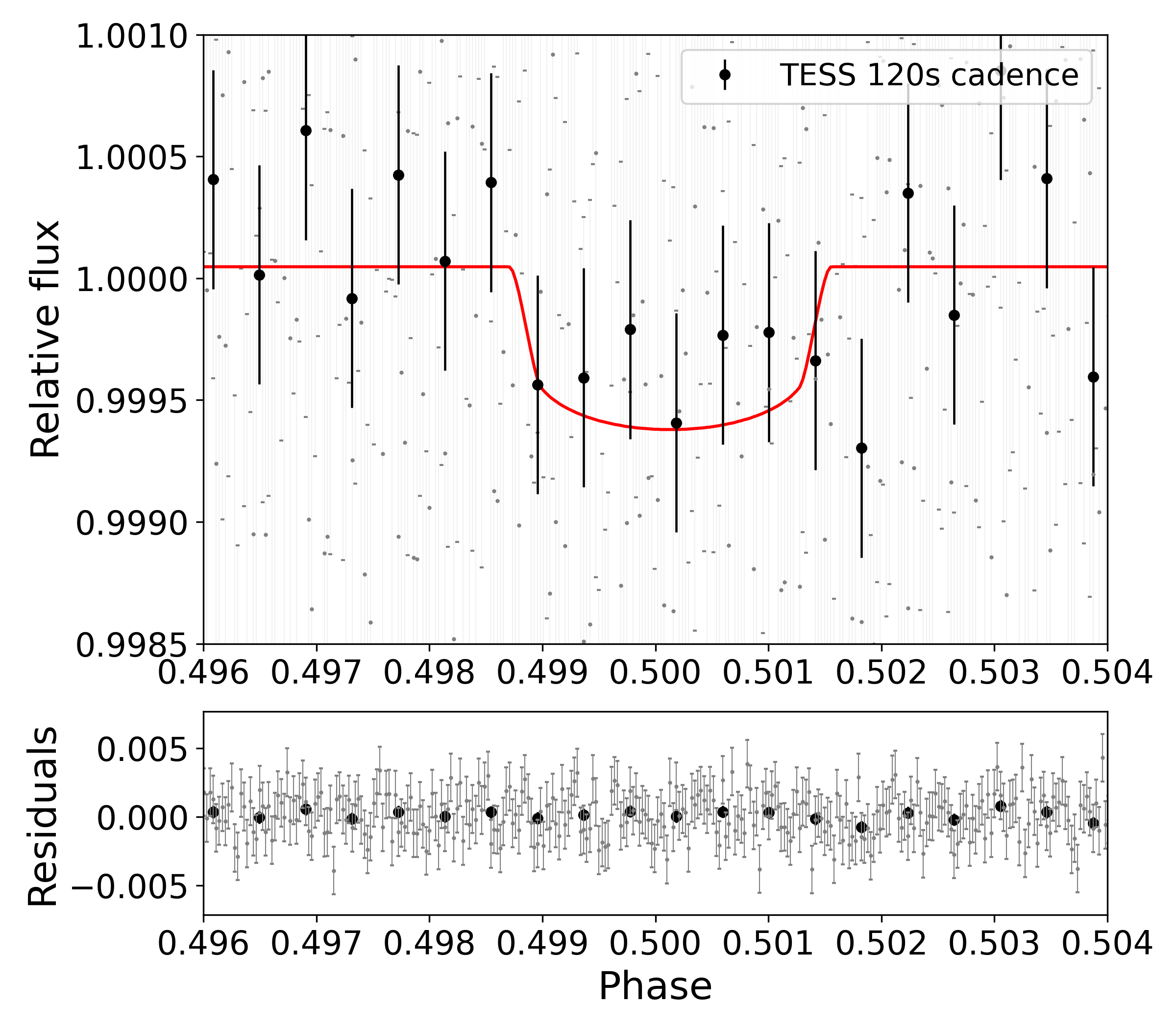}\\
    \includegraphics[width=0.33\linewidth]{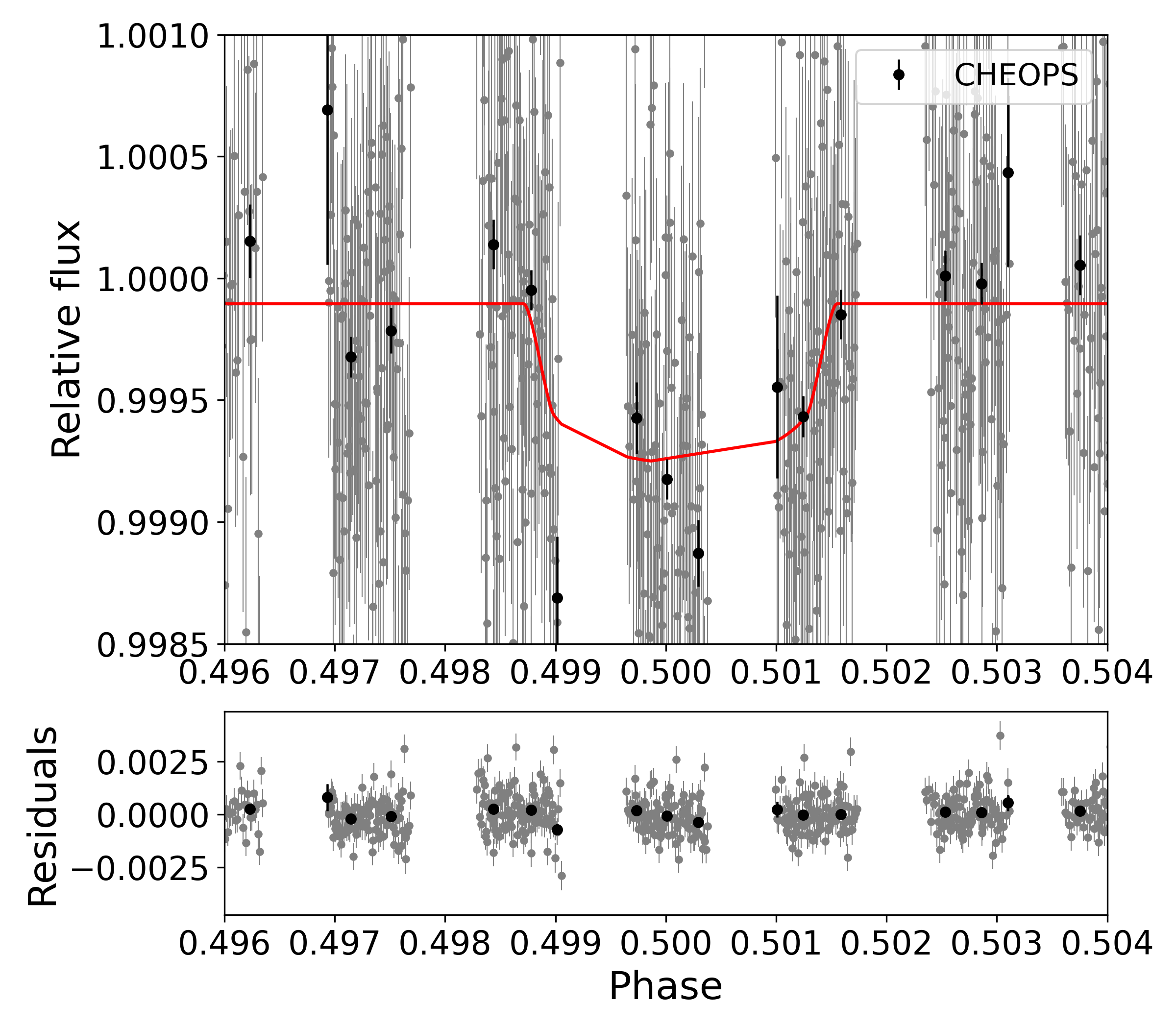}
    \includegraphics[width=0.33\linewidth]{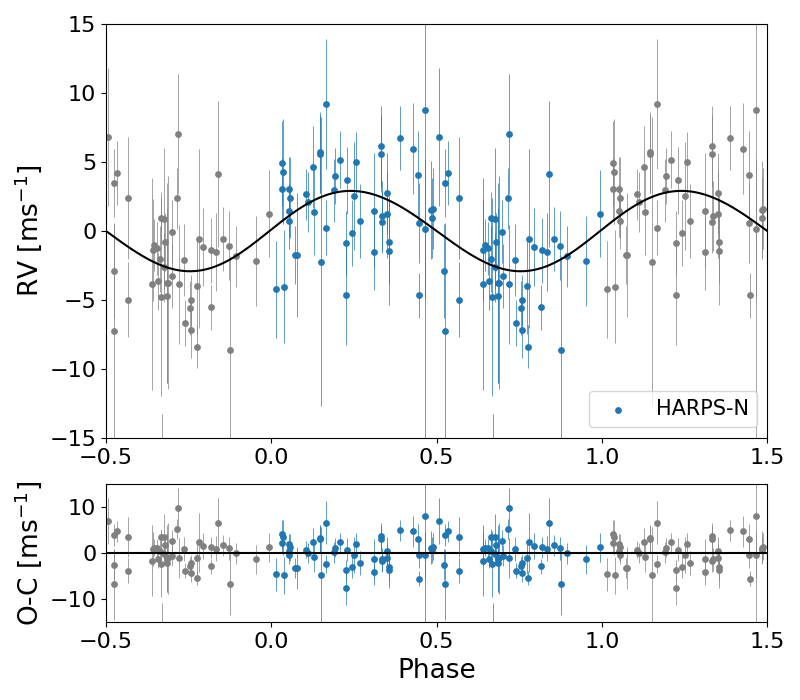}\\
    \includegraphics[width=0.5\linewidth]{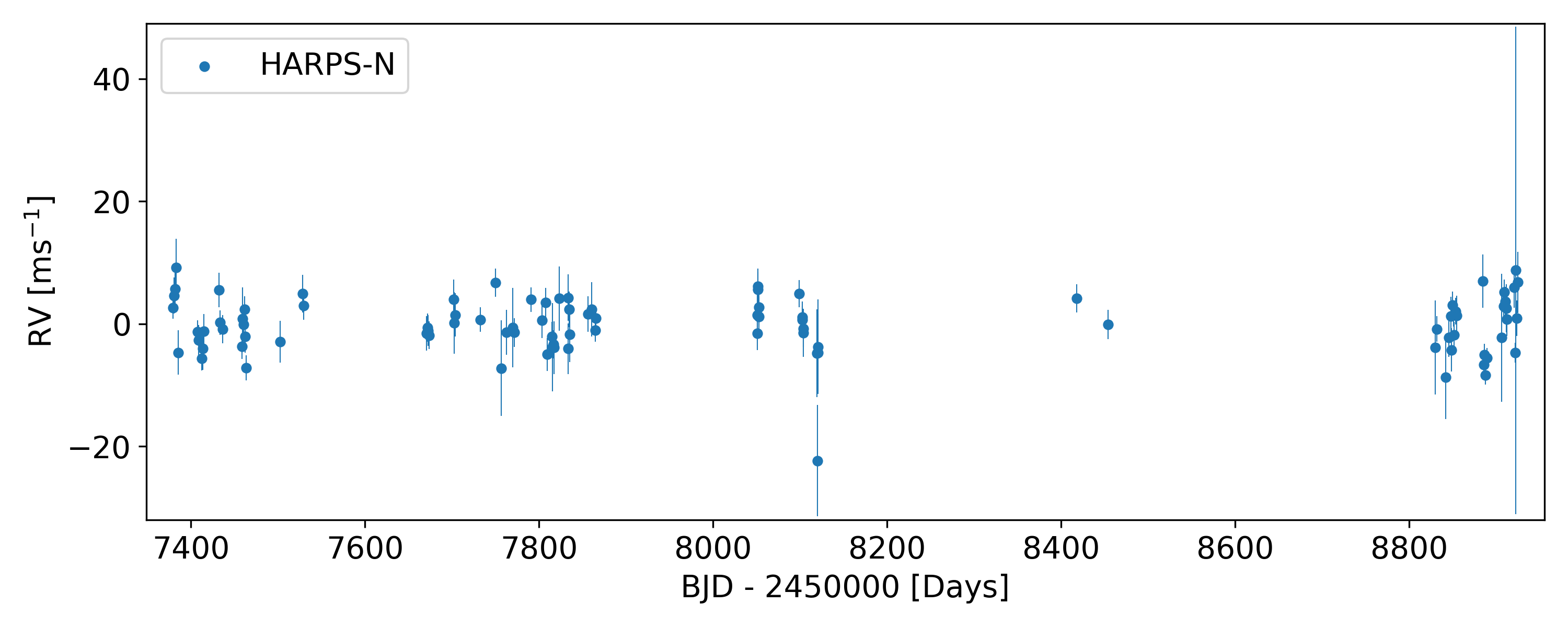}
    \caption{Combined fit results for K2-263\,b. Transit fit to the light curve data from \textit{K2} long cadence (\textbf{first row left}), \textit{K2} short cadence (\textbf{first row centre}), \textit{TESS} long cadence (\textbf{first row right}), and \textit{CHEOPS} (\textbf{second row left}). The short cadence (29.4\,min for \textit{K2} long cadence, 60\,s for \textit{K2} short cadence, 120\,s for \textit{TESS} long cadence, and 60\,s for \textit{CHEOPS}) fluxes are plotted in grey, the fluxes binned every 30\,min are over-plotted in black, and the fitted transit is shown by the red solid line. The RV fit to the HARPS-N RVs is shown in the \textbf{second row right} for K2-263\,b. Coloured circular points show the phase folded RVs, grey points show these same values in subsequent phases, and the black line shows the best fit Keplerian model. The \textbf{third row} show the HARPS-N RVs over the full $\sim4$\,yr baseline.}
    \label{fig:K2-263_transits}
\end{figure*}

\begin{figure*}
    \centering
    \includegraphics[width=0.33\linewidth]{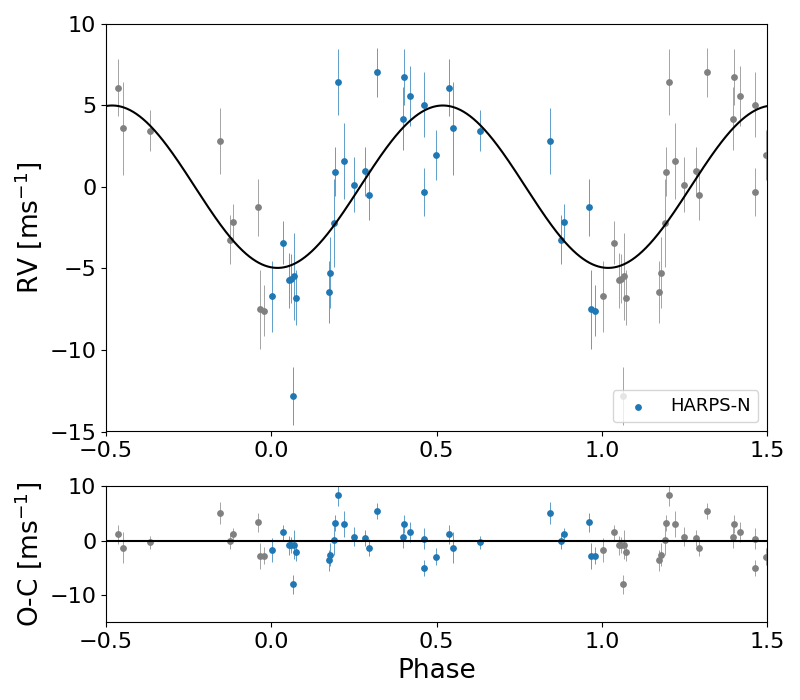}
    \includegraphics[width=0.33\linewidth]{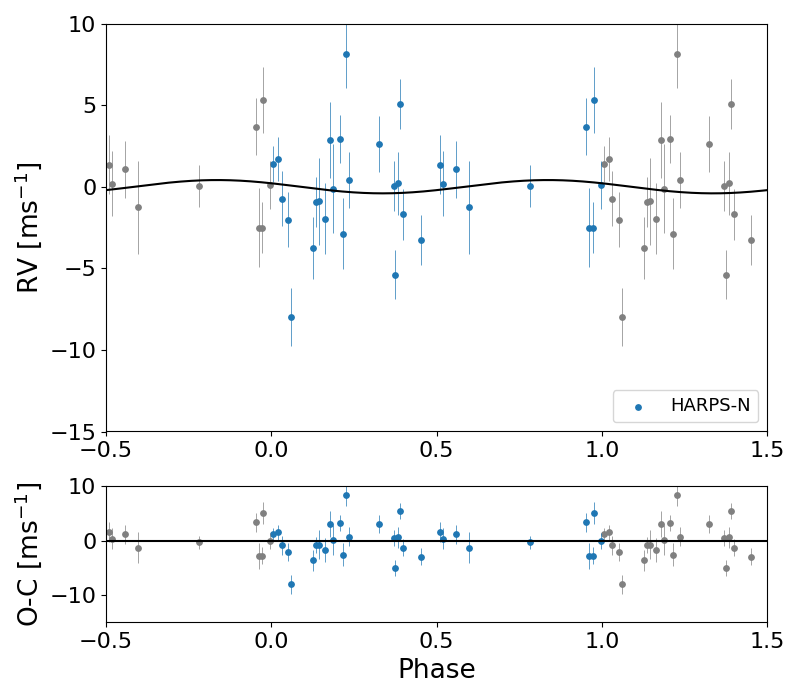}\\
    \includegraphics[width=0.5\linewidth]{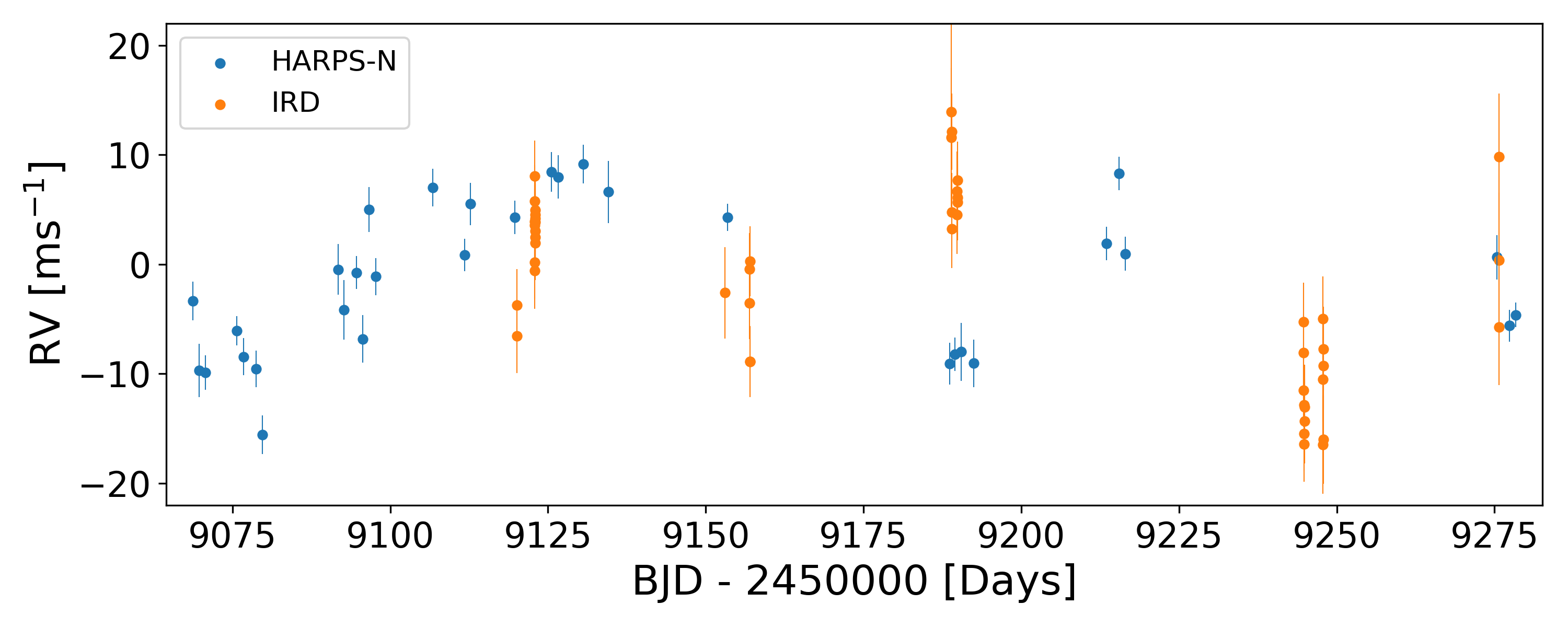}
    \caption{RV results for TOI-1634. The \textbf{top left} panel shows the phase-folded HARPS-N RVs of the confirmed planet TOI-1634\,b, with the best-fitting Keplerian model overplotted. The \textbf{top right} panel shows the HARPS-N RVs after subtraction of the signal of TOI-1634\,b, folded at the characteristic period ($P \sim 100$\,d) of a low-frequency modulation. The overplotted Keplerian curve in this panel is shown for illustrative purposes only and does not represent a confirmed planetary detection. Blue circular points show the phase-folded RVs, grey points show the same measurements repeated over multiple phases, and the black line indicates the corresponding model. The \textbf{bottom panel} shows the HARPS-N and IRD RVs as a function of time over the full $\sim$7 month baseline, highlighting the presence of additional low-frequency variability beyond the signal of TOI-1634\,b.}
    \label{fig:TOI-1634_RVs}
\end{figure*}

\begin{figure*}
    \centering
    \includegraphics[width=0.33\linewidth]{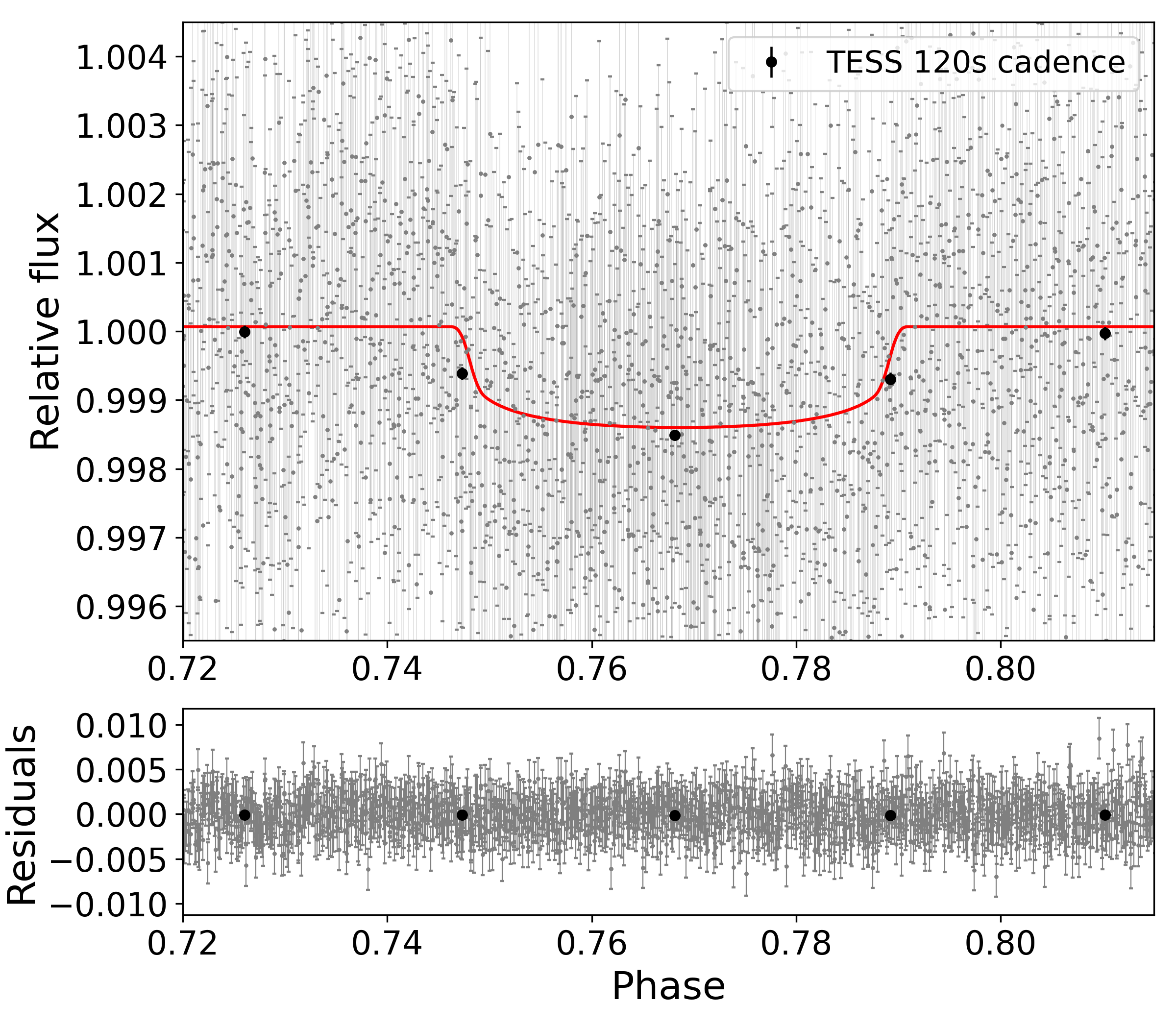}
    \includegraphics[width=0.33\linewidth]{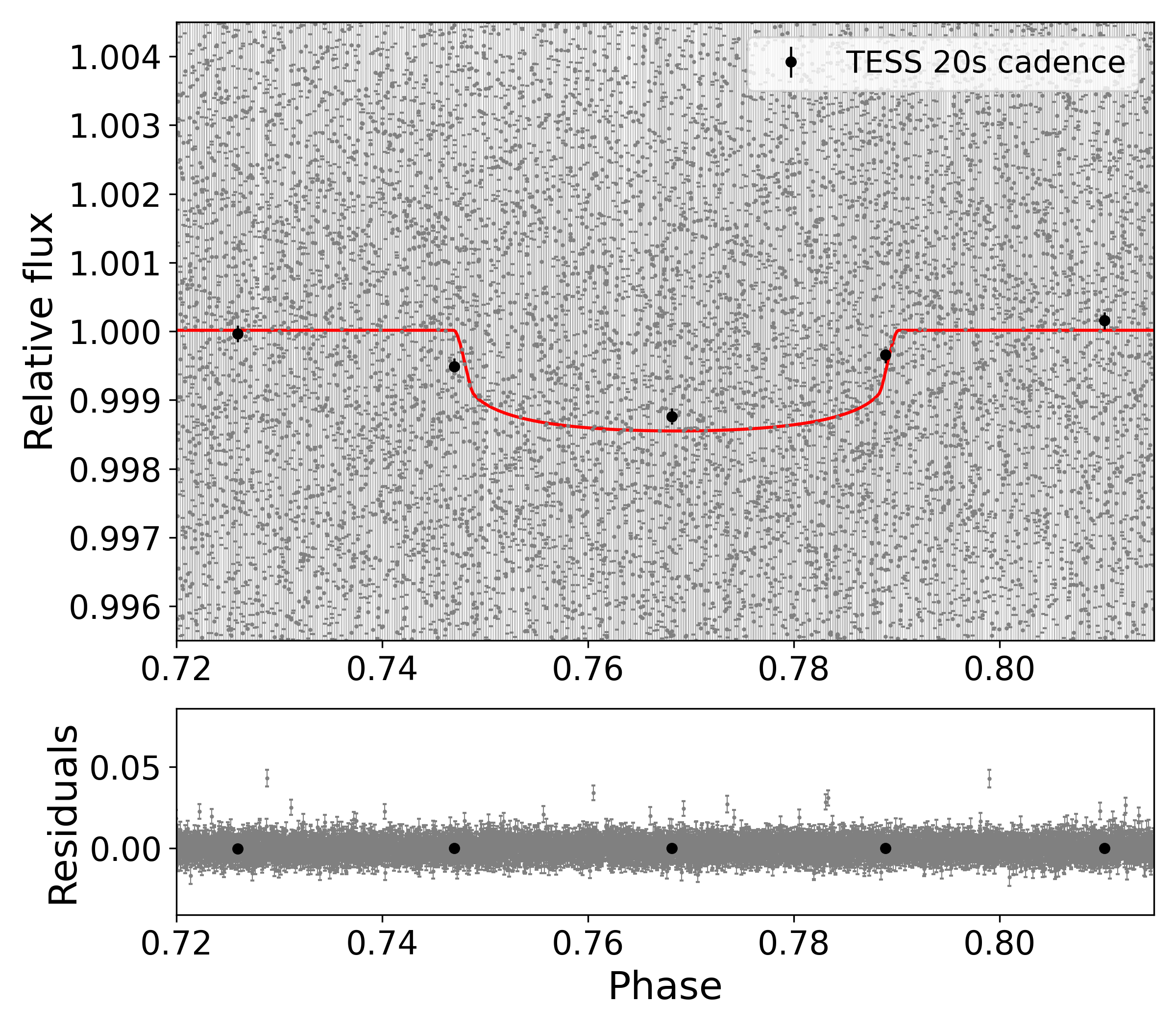}
    \includegraphics[width=0.33\linewidth]{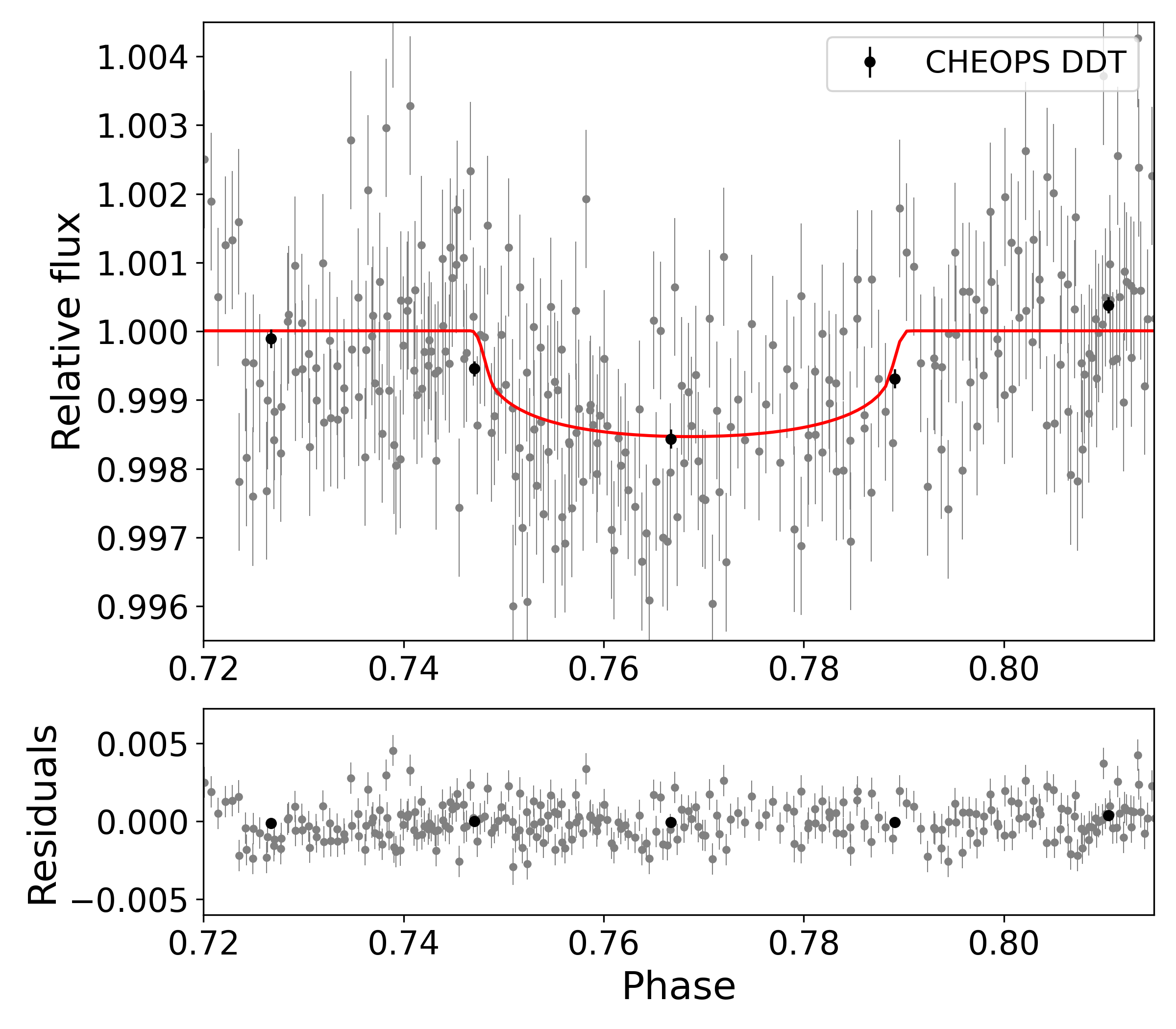}\\
    \includegraphics[width=0.33\linewidth]{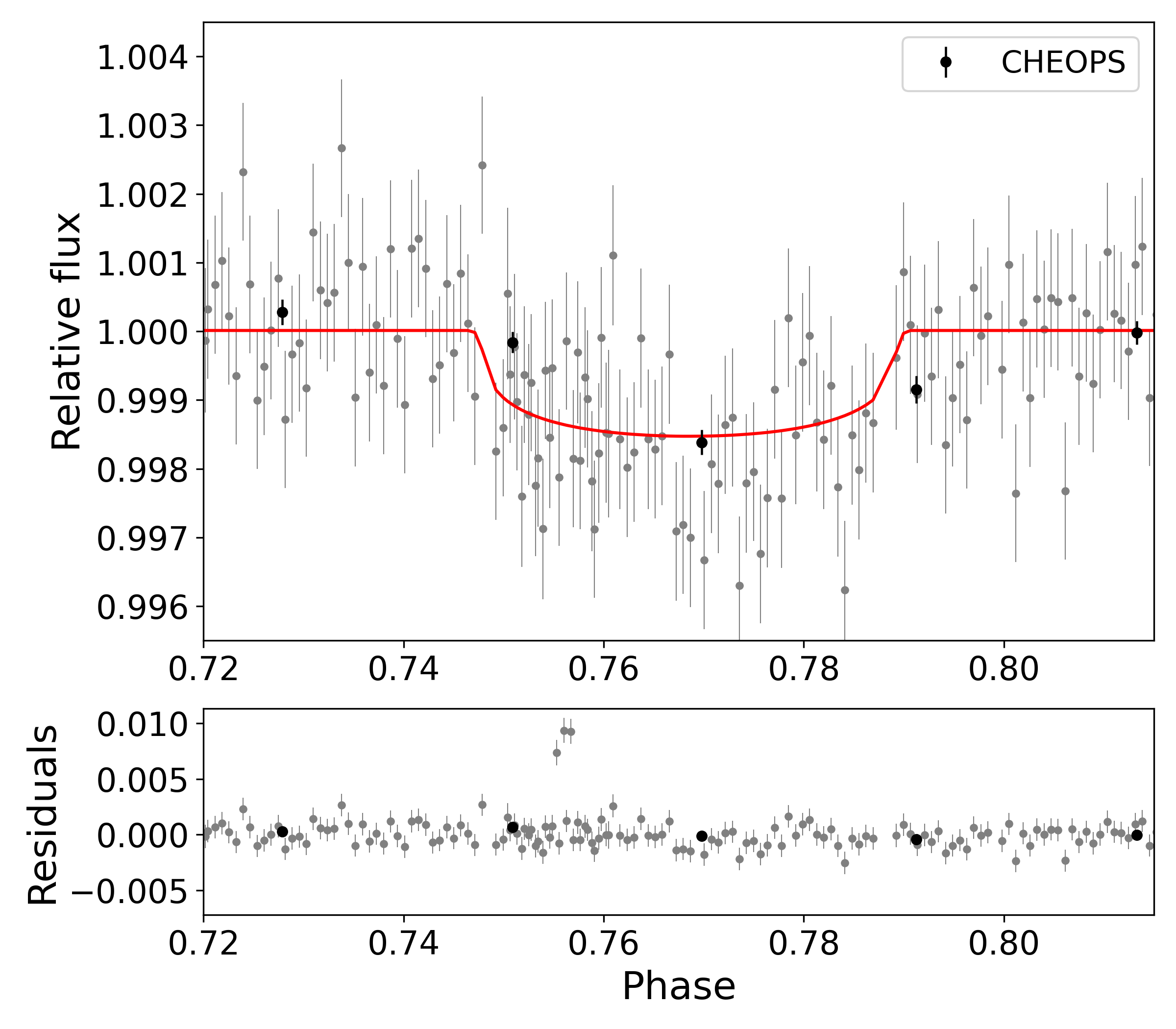}
    \includegraphics[width=0.33\linewidth]{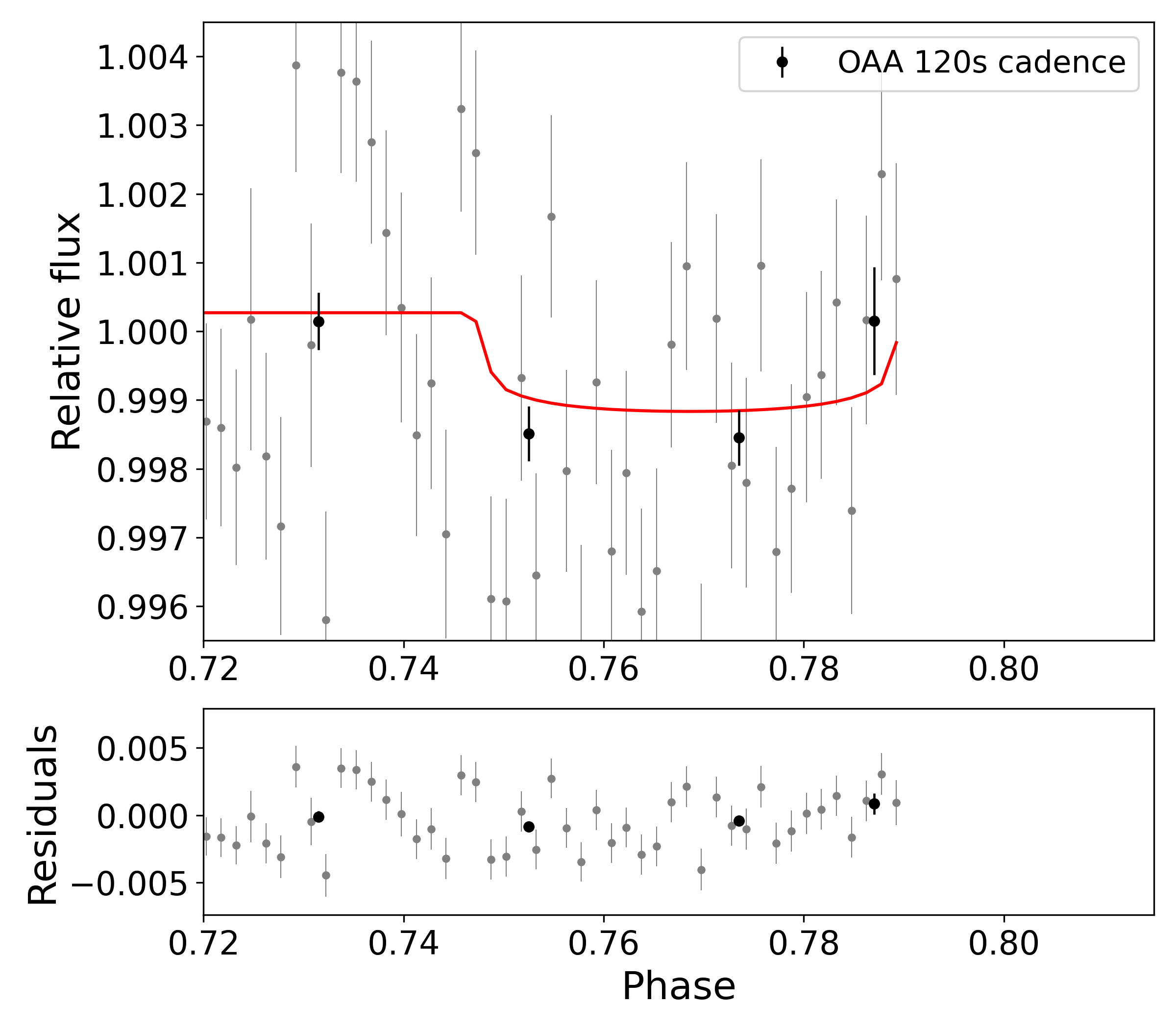}
    \includegraphics[width=0.33\linewidth]{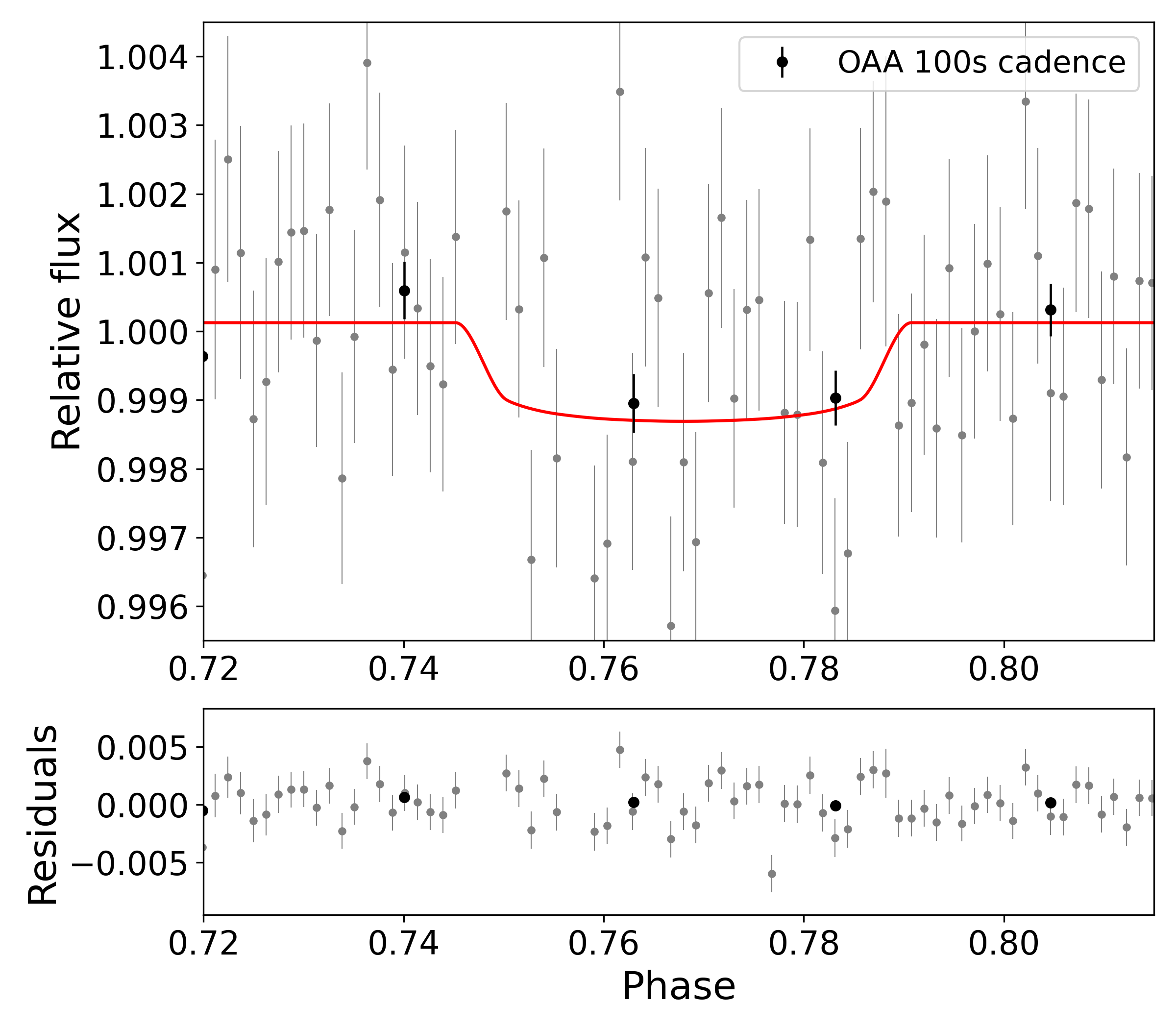}\\
    \includegraphics[width=0.33\linewidth]{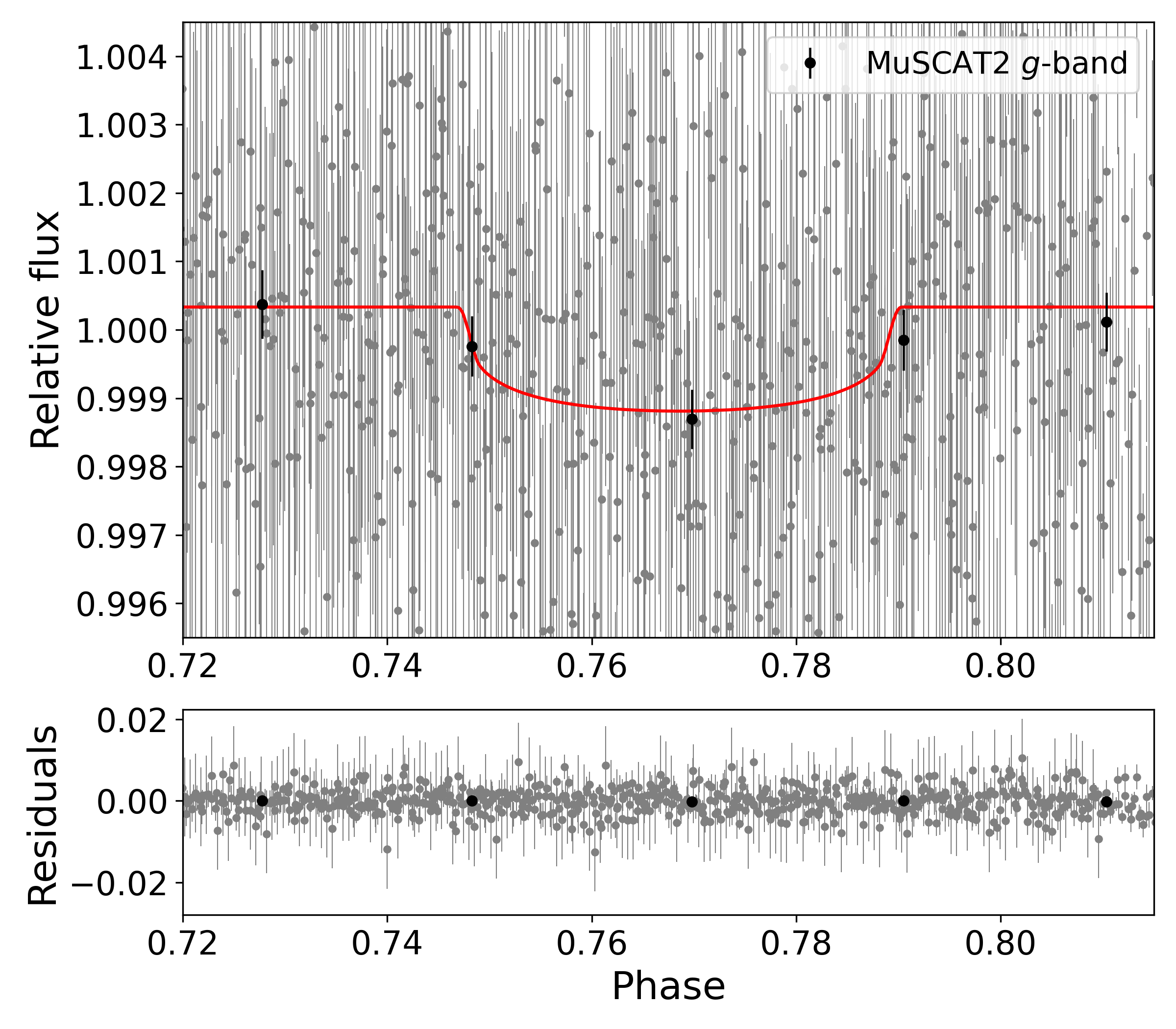}
    \includegraphics[width=0.33\linewidth]{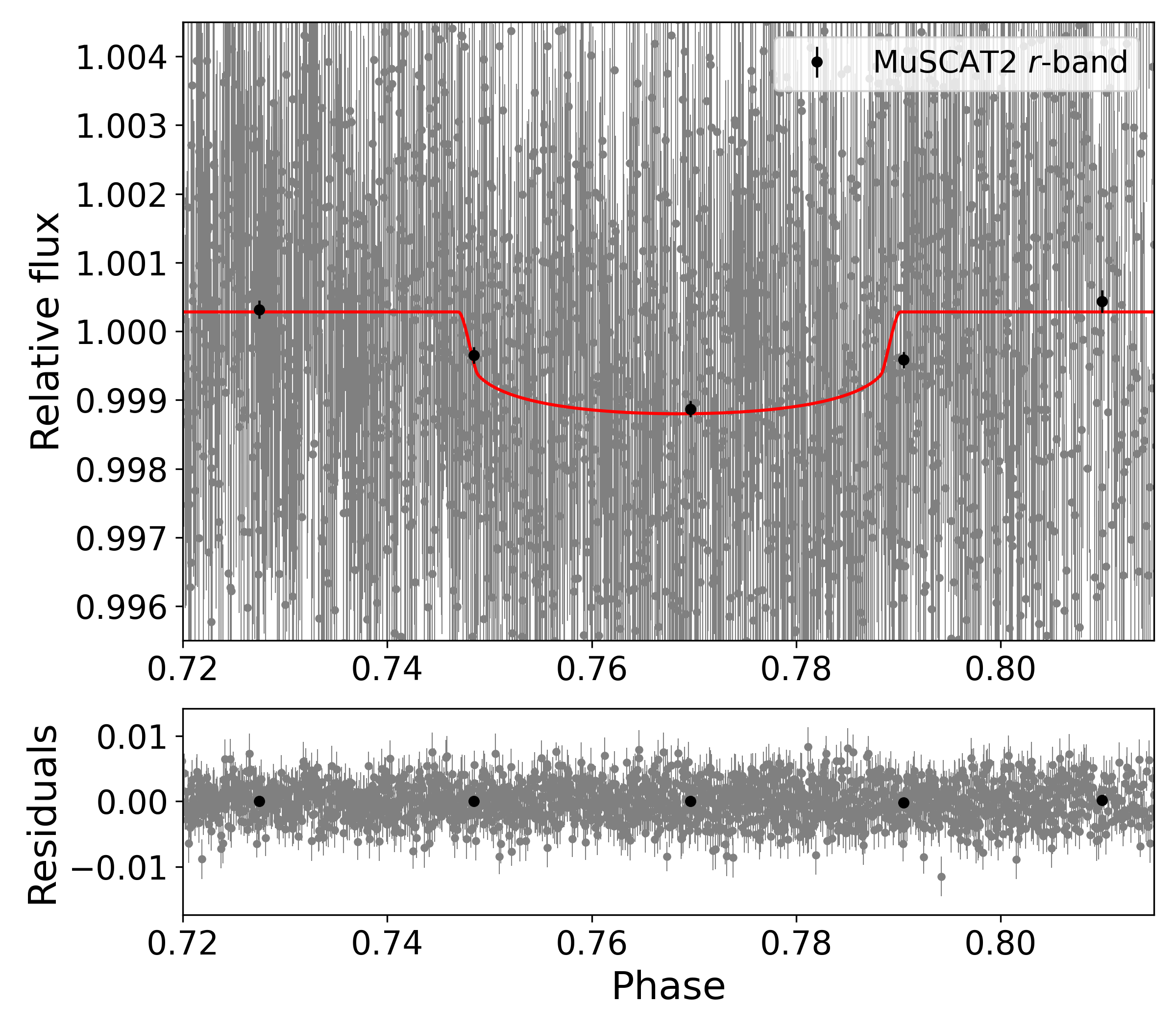}
    \includegraphics[width=0.33\linewidth]{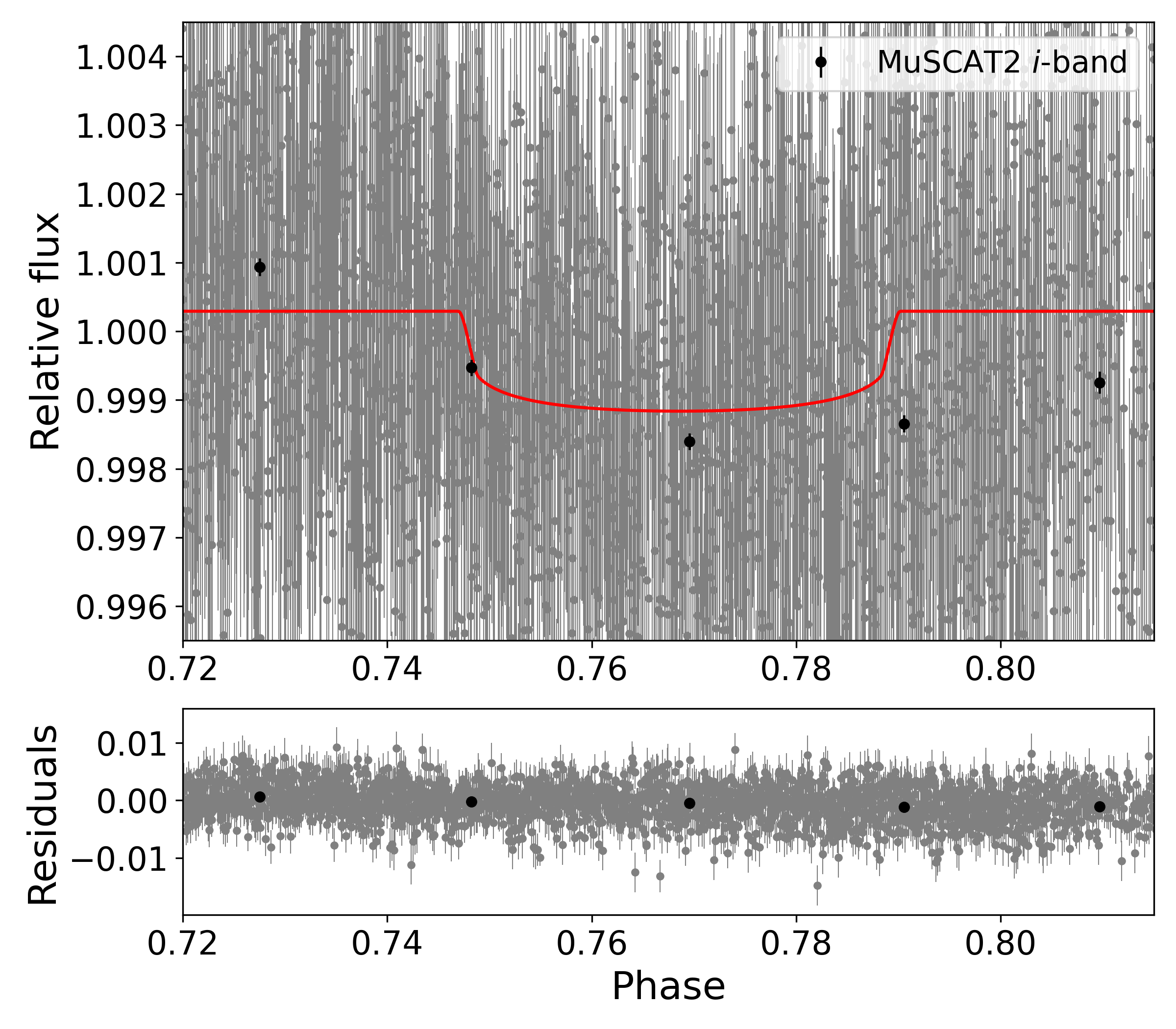}\\
    \includegraphics[width=0.33\linewidth]{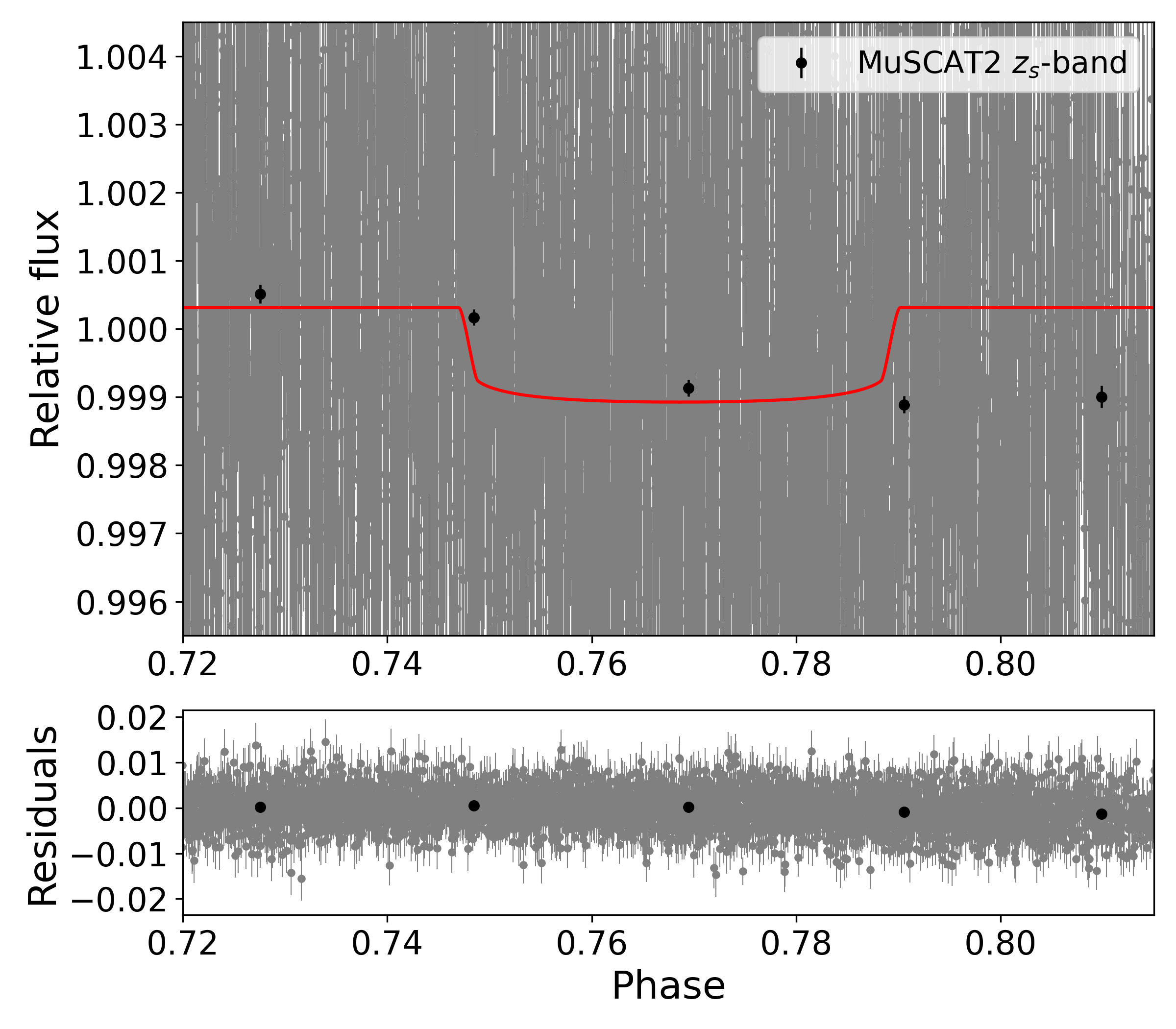}
    \includegraphics[width=0.33\linewidth]{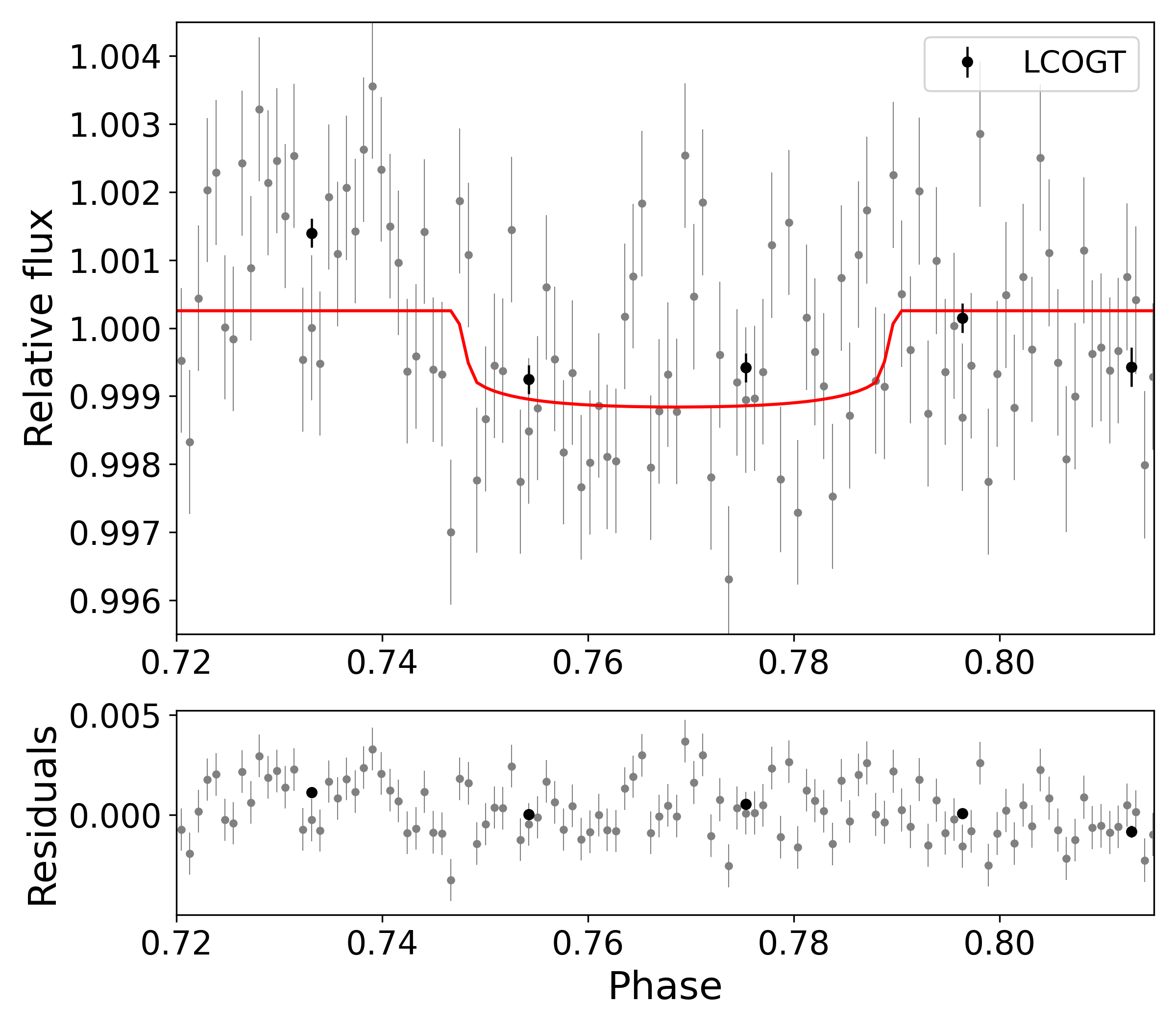}
    \includegraphics[width=0.33\linewidth]{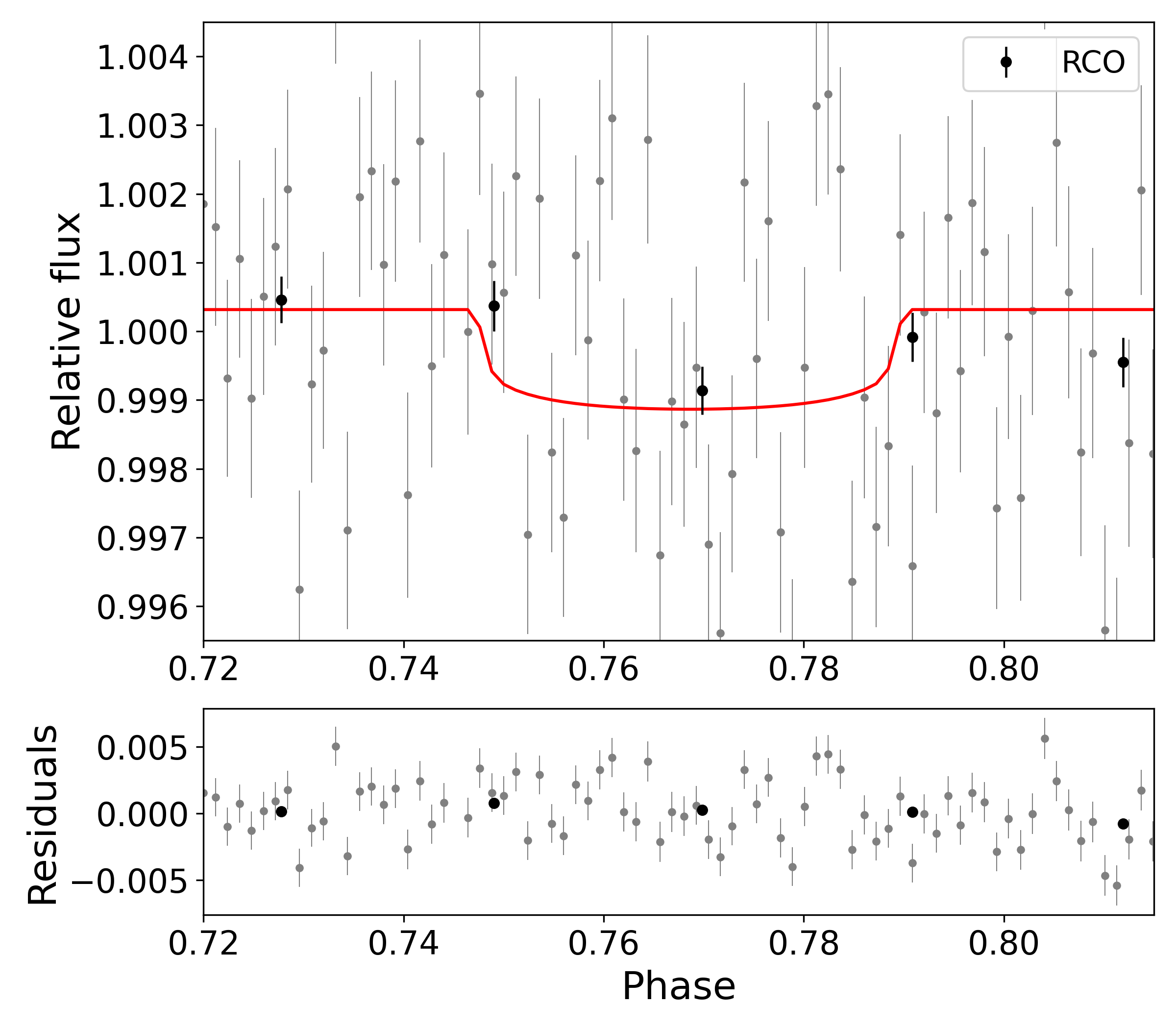}
    \caption{Transit fit results for TOI-1634\,b. Transit fit to the light curve data from \textit{TESS} long cadence (\textbf{first row left}), \textit{TESS} short cadence (\textbf{first row centre}), \textit{CHEOPS} DDT (\textbf{first row left}), \textit{CHEOPS} (\textbf{second row left}), \textit{OAA} long cadence (\textbf{second row centre}), \textit{OAA} short cadence (\textbf{second row right}), MuSCAT2 $g$-band (\textbf{third row left}), MuSCAT2 $r$-band (\textbf{third row centre}), MuSCAT2 $i$-band (\textbf{third row right}), MuSCAT2 $z_s$-band (\textbf{fourth row left}), LCOGT (\textbf{fourth row centre}), and RCO (\textbf{fourth row right}). The short/long cadence (120\,s for \textit{TESS} long cadence, 20\,s for \textit{TESS} short cadence, 60\,s for \textit{CHEOPS}, 120\,s for OAA long cadence, 100\,s for OAA short cadence, 40\,s for LCOGT, 90\,s for RCO, 46\,s for MuSCAT2 $g$-band, 21\,s for MuSCAT2 $i$- and $r$-bands, and 11\,s for MuSCAT2 $z_s$-band) fluxes are plotted in grey, the fluxes binned every 30\,min are over-plotted in black, and the fitted transit is shown by the red solid line.}
    \label{fig:TOI-1634_transits}
\end{figure*}


\bsp	
\label{lastpage}
\end{document}